\documentclass{aa}
\usepackage{graphicx}
\usepackage{txfonts}
\usepackage{natbib}
\usepackage{longtable}
\usepackage{multirow}
\usepackage{color}

\newcommand{\ha}{\ifmmode {\rm H}\alpha \else H$\alpha$\fi}
\newcommand{\hb}{\ifmmode {\rm H}\beta \else H$\beta$\fi}
\newcommand{\lya}{\ifmmode {\rm Ly}\alpha \else Ly$\alpha$\fi}
\def\micron{$\mu$m}

\def\msun{\ifmmode M_{\odot} \else $M_{\odot}$\fi}
\def\msunyr{\ifmmode M_{\odot} {\rm yr}^{-1} \else $M_{\odot}$ yr$^{-1}$\fi}
\def\zsun{\ifmmode Z_{\odot} \else Z$_{\odot}$\fi}
\def\lsun{\ifmmode L_{\odot} \else L$_{\odot}$\fi}
\def\hii{{\rm HII}}
\def\mstar{M$_{\star}$}

\begin{document}
\title{Properties of $z\sim3-6$ Lyman break galaxies. II. Impact of nebular emission at high redshift}

   \author{S. de Barros
          \inst{1,2}
	  \and
          D. Schaerer\inst{1,3}
          \and
          D.P. Stark\inst{4,5}}

   \institute{Geneva Observatory, University of Geneva, 51, Ch. des Maillettes, CH-1290 Versoix, Switzerland
   	\and
		Department of Physics and Astronomy, University of California, Riverside, 900 University Avenue, Riverside, CA 92521, USA
              \email{stephane.debarros@ucr.edu}
           \and
              CNRS, IRAP, 14 Avenue E. Belin, 31400 Toulouse, France
           \and
           	Kavli Institute of Cosmology and Institute of Astronomy, University of Cambridge, Madingley Road, Cambridge CB30HA, UK
	\and
		Steward Observatory, University of Arizona, 933 N Cherry Ave, Tucson, AZ 85721, USA
	}

   \date{Received 16 July 2012; accepted 18 November 2013}
   
   \authorrunning{} \titlerunning{Impact of nebular emission at high redshift}
    
   \abstract
   {To gain insight on the mass assembly and place constraints on the star formation history (SFH) of Lyman break galaxies (LBGs), it is important to accurately determine  
   their properties.}
  {We estimate how nebular emission and different SFHs affect parameter estimation of LBGs.}
   {We present a homogeneous, detailed analysis of the spectral energy distribution (SED) of $\sim$ 1700 LBGs from the GOODS-MUSIC catalogue with deep multi-wavelength photometry from $U$ band to 8 $\mu$m to determine stellar mass, age, dust attenuation, and star formation rate.
   Using our SED fitting tool, which takes into account nebular emission, we explore a wide parameter space. We also explore a set of different star formation histories.}
  {Nebular emission is found to significantly affect the determination of the physical parameters for the majority of $z \sim $ 3--6 LBGs.
 We identify two populations of galaxies by determining the importance of the contribution of emission lines to broadband fluxes. We find that $\sim$ 65\% of LBGs
 show detectable signs of emission lines, whereas $\sim$ 35 \% show weak or no emission lines. This distribution is found over the entire redshift
 range. We interpret these groups as actively star forming and more quiescent LBGs, respectively.
We find that it is necessary to considerer SED fits with very young ages ($<50$ Myr) to reproduce some colours affected by strong emission lines. Other arguments favouring episodic star formation and relatively short star formation
 timescales are also discussed.
 Considering nebular emission generally leads to a younger age, lower stellar mass, higher dust attenuation, higher star formation rate, and a large scatter in the SFR-$M_{\star}$ relation. Our analysis yields
  a trend of increasing specific star formation rate with redshift, as predicted by recent galaxy evolution models.}
   {The physical parameters of approximately two thirds of high redshift galaxies are significantly modified when we account for nebular emission.
   The SED models which include nebular emission shed new light on the properties of LBGs with numerous important implications.}
   
   \keywords{galaxies: starburst -- galaxies: high redshift -- galaxies: evolution -- galaxies: star formation}

   \maketitle
%
%________________________________________________________________

   \section{Introduction}
   \label{intro}
   
   For several years now, large multi-wavelength surveys undertaken with space and ground-based facilities, such as the {\em Hubble Space Telescope}, {\em Spitzer} or the {\em Very Large Telescope}, reach high redshifts and unveil properties of a growing number of objects.
The Lyman Break technique \citep[e.g.][]{steideletal1996} is currently used to detect star-forming galaxies at a redshift as high as $z\sim10$ \citep{oeschetal2013} and has been used to identify several thousands of galaxies at $2<z<8$ \citep[e.g.][]{shapleyetal2003,bouwensetal2013}. 
   
   Difficulties arise from the method to determine physical properties of these galaxies. A parameter known to be well constrained by the spectral energy distribution (SED) fitting method is the stellar mass, since the determination of this physical parameter is only slightly affected by change in assumptions \citep[e.g.][]{finlatoretal2007,yabeetal2009}. Determination of the star formation rate (SFR) at high redshift ($z>2$) is more difficult, since we usually do not have access to measurements of emission lines (e.g. H$\alpha$), which correlate well with this parameter \citep[since using Ly$\alpha$ line as a SFR tracer is still challenging, e.g.][]{ateketal2013}. Therefore, most studies rely on SFR estimated from UV luminosity \citep[e.g.][]{bouwensetal2009} using standard UV--SFR relation \citep{kennicutt1998,madauetal1998}. However, UV photons are strongly affected by dust attenuation, which requires to infer properly dust attenuation to estimate the SFR, which can be done statistically for large samples relying on the relation between dust attenuation and UV continuum slope \citep[$\beta$ slope, e.g.][]{dunlopetal2013,bouwensetal2012,bouwensetal2013,finkelsteinetal2012}.
   
   While there can be some discrepancies between studies that rely on the $\beta$ slope \citep[see][and references therein]{bouwensetal2013},  both methods, the UV--SFR conversion and $\beta$--dust attenuation relation, rely on several assumptions: a constant star formation history, age ($>100$ Myr), and an intrinsic $\beta$ slope \citep{meureretal1999}. Generally, the results provided by these relations are consistent with other SFR and dust attenuation tracers on average up to $z\sim3$ \citep[e.g.][]{reddy&steidel2004}. However, by using cosmological hydrodynamical simulations, \cite{wilkinsetal2013} show that the intrinsic $\beta$ slope is possibly lower at higher redshift ($z>4)$, even evolving with redshift, which leads to generally higher inferred dust attenuation than studies relying on the Meurer relation. 
   
   Furthermore, some studies relying on the SED fitting lead to results, which question several assumptions that are generally used to fit SEDs or  those used in SFR--UV conversion and the Meurer relation. These assumptions include age $>100$ Myr \citep[e.g.][]{vermaetal2007}, constant star formation history \citep{starketal2009}, or very low dust attenuation at high resdhift \citep[e.g.][]{yabeetal2009}. Since SED fitting suffers from several well known degeneracies (e.g.\ between dust attenuation and age), these results may be questioned too, but the SED fitting provides  parameters 
   (SFR, age, dust attenuation, etc), which are consistent among each other.  
  
On the other hand, theoretical studies can attempt to reproduce observables, such as the UV luminosity function, while several estimated parameters remain uncertain. The luminosity function provides the number of galaxies that emits light at a given redshift and luminosity. The UV luminosity function is the more commonly used, since the UV luminosity is a star formation tracer, as explained above, and UV wavelengths are the easiest to observe at high redshift \citep[e.g.][]{bouwensetal2007}. While stellar mass is not a direct observable, the insensitivity of this parameter to different assumptions used to infer it also allows us to compare theoretical predictions with inferred stellar mass functions, that is the number of galaxies at a given stellar mass \citep[e.g.][]{gonzalezetal2011} .

 Several theoretical studies are now able to reproduce these observed trends \citep[UV luminosity function, stellar mass function, e.g.][]{finlatoretal2007,boucheetal2010,finlatoretal2011,weinmannetal2011}, but these studies also predict a specific star formation (sSFR=SFR/\mstar)--redshift relation that rises with increasing redshift, which is not found by studies relying on a constant SFR (SFR=const) and an age $>100$ Myr (sSFR--z ``plateau"), while some studies rely on SED fitting that indeed finds rising sSFR \citep[][]{yabeetal2009,schaerer&debarros2009,starketal2013}. The finding of a rising sSFR is a consequence of a dust attenuation that is much greater than those inferred from the $\beta$ slope \citep{yabeetal2009}, or an effect of nebular emission \citep{schaerer&debarros2009,starketal2013}.
      
   Although nebular emission (i.e.\ emission lines and nebular continuous emission from \hii\ regions) is ubiquitous in regions of massive star formation, strong or dominant   in optical spectra of nearby star-forming galaxies and present in numerous types of galaxies, its impact on the determination of physical parameters of galaxies, in particular at high redshift, has been neglected until recently \citep[cf.\ overview in][]{schaerer&debarros2011}. Several spectral models  of galaxies have indeed included nebular emission in the past \citep[e.g.][]{charlot&longhetti2001,fioc&rocca1997,anders&fritz2003,zackrissonetal2008}; however, they had not been applied to the analysis of distant galaxies. \cite{zackrissonetal2008} show that nebular emission can significantly affect broadband photometry; the impact being stronger with increasing redshift, since the equivalent width (EW) of emission lines scales with ($z$+1). For the first time \cite{schaerer&debarros2009} include nebular emission to fit SEDs of a sample of Lyman break galaxies   at $z\sim6$ and show that nebular lines strongly affect age estimation, since some lines can mimick a Balmer break. Ages are strongly reduced, which can lead one to reconsider star formation rate estimations from UV luminosity, since the standard relation used to convert  UV luminosity into SFR, as previously explained, assumes a constant star formation activity during 100 Myr \citep{kennicutt1998,madauetal1998}.  The analysis of a sample of $z \sim$ 6--8 LBGs observed with {\it HST} and {\it Spitzer} further demonstrates the potential impact of nebular emission on the physical parameters as derived from SED fits of high-z galaxies \citep{schaerer&debarros2010}.
   
   It has now become clear \citep{schaerer&debarros2009,schaerer&debarros2010,onoetal2010,lidmanetal2012} 
that we must account for nebular emission (both lines and continuum emission) to interpret photometric measurements of the SEDs of star-forming galaxies, such
as Lyman-alpha emitters  and Lyman break galaxies, which are the dominant galaxy populations at high-z. Furthermore, as testified by the presence of \lya\ emission, a large and growing fraction of the currently known population of star-forming galaxies at high redshift shows emission lines \citep{ouchietal2008,starketal2010,schaereretal2011,curtislakeetal2012,schenkeretal2012,schenkeretal2013}.
   
   In parallel, diverse evidence of galaxies with strong emission lines and/or strong contributions 
of nebular emission to broadband fluxes has been found at different redshifts, e.g.\ by
\cite{shimetal2011,mclindenetal2011,ateketal2011,trumpetal2011,vanderweletal2011,labbeetal2012,starketal2013,smitetal2013}.
    
    Several studies have shown that the inclusion of nebular emission leads to modify parameter estimation from SED fitting, mainly reducing stellar mass and increasing SFR \citep[e.g.][]{onoetal2010,schaerer&debarros2010,acquavivaetal2012,mclureetal2011}. 
While analysis relying on standard SED models  show no or little evolution of the sSFR with redshift at $z>2$ \citep[e.g.][]{starketal2009,gonzalezetal2010,bouwensetal2012,reddyetal2012a}, the impact of nebular emission on physical parameter estimation  \citep[e.g.][]{debarros&schaerer2011,gonzalezetal2012,starketal2013} seems to lead to results more consistent with expectation from hydrodynamical simulations \citep[e.g.][]{boucheetal2010}. Furthermore, there is a possible trend of increasing dust attenuation with the stellar mass \citep{schaerer&debarros2010}, a trend already established at lower redshift \citep{brinchmannetal2004,reddyetal2010}.
    
      \cite{starketal2009} presents a first attempt to constrain the star formation history by studying the evolution of LBG samples that are uniformly selected among different bins of redshift. In this work, we use a similar approach with a large sample of LBGs that covers four bins of redshift between $z\sim3$ to $z\sim6$ by using an up-to-date photometric redshift and SED-fitting tool that treats the effects of nebular emission. This homogeneous analysis provides the main physical parameters, as star formation rate, stellar mass, age and reddening. We explore a large parameter space by using different assumptions on star formation history and nebular emission, which allows us to estimate the effects of these assumptions on parameter estimation.
            
Our paper is structured as follows. The observational data are described in Sect.\ \ref{data}, and the method used for SED modelling
is described in Sect.\ \ref{method}. The results are presented in Sect.\ \ref{results} and discussed in Sect.\ \ref{discussion}.
Section \ref{conclusions} summarises our main conclusions. 
We adopt a $\Lambda$-CDM cosmological model with $H_{0}$=70 km s$^{-1}$ Mpc$^{-1}$, $\Omega_{m}$=0.3 and $\Omega_{\Lambda}$=0.7. 
All magnitudes are expressed in the AB system \citep{oke&gunn1983}.

%__________________________________________________________________

  \section{Data}
  \label{data}
  
  \subsection{The GOODS Fields}

We focus our analysis on the data from the Great Observatories Origins
Deep Survey (GOODS).  Detailed descriptions of the datasets are
available in the literature \citep{giavaliscoetal2004,santinietal2009}, so we only provide
a brief summary here.  The GOODS-S and GOODS-N survey areas both cover
roughly 160 arcmin$^2$ and are  centered on the {\em Chandra} Deep
Field South (CDF-S; \citealt{giacconietal2002}) and the {\em Hubble} Deep
Field North (HDF-N; \citealt{williamsetal1996}).   Extensive multi-wavelength
observations have been conducted in each of these fields.  In this
paper, we utilize optical imaging from the Advanced Camera for Surveys
(ACS) onboard the Hubble Space Telescope (HST).  Observations with ACS
were conducted in F435W, F606W, F775W, and F850LP (hereafter
B, V, $i$, $z$) toward GOODS-S and
GOODS-N \citep{giavaliscoetal2004}.  The average 5$\sigma$ limiting
magnitudes in the v2 GOODS ACS data (0\farcs35 diameter photometric aperture) are
B=29.04, V=29.52, $i$=29.19, and $z$=28.54.
We also make use of U- and R-band observations of GOODS-S taken with 
the ESO Very Large Telescope (VLT) with the VIMOS wide field imager \citep{lefevreetal2003}, as provided by \cite{noninoetal2009}, with a 1$\sigma$ limiting magnitude (1\farcs\ radius aperture) reaching $U\approx29.8$.

In the near-infrared, we utilize publicly available deep J-, H- and K-band
observations of GOODS-S (PI: C. Cesarsky), using the ISAAC camera on
the  VLT.  The sensitivities vary across the
field  depending on the effective integration time and seeing FWHM.
Average 5$\sigma$ magnitude limits (corrected for the amount of flux
that falls outside of the 1\farcs0 diameter aperture) are
$J\simeq 25.2$, $H\simeq24.7$ and K$_s\simeq24.7$.

Deep {\it Spitzer} imaging is available toward both GOODS fields with
the Infrared Array Camera (IRAC) as part of the ``Super Deep'' Legacy
program (Dickinson et al. {\it in prep}).
Details of the observations have been described in detail elsewhere
\citep{eylesetal2005,yanetal2005,starketal2007} so we do not discuss them  further
here.  The 5$\sigma$ limiting magnitudes of the IRAC imaging are
$\simeq 26.3$ at 3.6$\mu$m and $\simeq 25.9$ at 4.5$\mu$m using
2\farcs4 diameter apertures and applying an aperture correction.  

In practice, we use the V2 GOODS-MUSIC catalogue from \cite{santinietal2009} with
optical, near-, and mid-infrared photometry. Non-detections are included in the SED fit with \textit{ Hyperz} by setting the flux in the corresponding filter to zero, and the error to 
the 1$\sigma$ upper limit.

For $U$, $B$, and $V$ drops, non-detections vary from 10 to 34\% in $J$, $H$, and $K$ bands from 6 to 10\% in the first three IRAC bands and from 26 to 34\% at 8$\mu$m, while  they vary from 26 to 57\% ($J$, $H$, and $K$ bands) and from 36 to 42\% in IRAC bands for $i$ drops.

\subsection{Dropout selection}

Galaxies at $z\simeq 3$, 4, 5, and 6 are selected via the presence of the
Lyman-break  as it is  redshifted through the U, B, V, 
and $i$ bandpasses, respectively.  Selection of Lyman break
galaxies  at these redshifts has now become routine
\citep{stanwayetal2003,giavaliscoetal2004,bunkeretal2004,beckwithetal2006,bouwensetal2007}.  To ensure a consistent comparison of our samples to these
previous samples, we adopt colour criteria, which are similar to those
used in \cite{beckwithetal2006}, and are very similar to those used by
\cite{bouwensetal2007}.  These criteria have  been developed to select
galaxies in the chosen redshift interval, while  minimizing
contamination from red galaxies likely to be at low redshift.

As mentioned in the previous section, galaxies are selected from the
GOODS-MUSIC catalog v2 \citep{santinietal2009}. It contains 14 999 objects which are selected in either the $z_{850}$ band or the $K_s$ band or at 4.5$\mu$m.

The selection criteria used in this work are identical to those from \cite{noninoetal2009} and \cite{starketal2009} with an additional constraint for $U$ dropouts (${\rm S/N}(U)<2$). This latter criteria helps greatly in removing contaminants, but it also removes galaxies at the low redshift tail of the $U$-drop selection. This selection leave us 440 $U$ drops, 859 $B$ drops, 277 $V$ drops, and 66 $i$ drops.

 \section{Method}
 \label{method}
  
 \subsection{SED fitting tool}
    
We use a recent, modified version of the Hyperz photometric redshift code of \cite{bolzonellaetal2000}, by taking into account nebular emission (lines and continua). We consider a large set of spectral templates \citep{bruzual&charlot2003}, which
covers different metallicities  and a wide range of star formation (SF) histories (exponentially decreasing, constant and rising SF),
and we add the effects of nebular emission following our method presented in \cite{schaerer&debarros2009,schaerer&debarros2010}.
We account for attenuation from the intergalactic and the interstellar medium and varying redshift. With these assumptions,
we fit the observed SEDs by straightforward least-square minimization.

In practice, we adopt spectral templates computed for a Salpeter IMF \citep{salpeter1955} from 0.1 to 100 $M_\odot$, and we 
properly treat the returned ISM mass from stars. 
Nebular emission from continuum processes and lines is added to the spectra predicted from the GALAXEV models, as described in 
\cite{schaerer&debarros2009}, proportionally to the Lyman continuum photon production. The relative line intensities of He
 and metals are taken from \cite{anders&fritz2003}, which includes galaxies grouped in three metallicity intervals by covering 
 $\sim$ 1/50--1  \zsun. Hydrogen lines from the Lyman to the Brackett series are included with relative intensities as given by case B.
For galactic attenuation, we use the Calzetti law \citep{calzettietal2000}. The IGM is treated following \cite{madau1995}.

To examine the effects of different star formation histories and for comparison with other studies, we define three sets of models:
\begin{itemize}
	\item {\em Reference model} (REF): this model has a constant star formation rate with a minimum age $t> 50$ Myr and solar metallicity.
	\item {\em Decreasing model} (DEC): this model has exponentially declining star formation histories (SFR $\propto \exp(-t/\tau)$) with
	variable timescales $\tau$. Metallicity and $\tau$ are free parameters.
	\item {\em Rising model} (RIS): this describes a rising star formation rate. We use the mean rising star-formation history from
the simulations of \citet[][their Fig.\ 1]{finlatoretal2011}. To describe this case, we assume that SF starts
at age=0 and grows by 2.5 dex during 0.8 Gyr by following their
functional dependence. After this period, we set SFR$=0$. 
	Metallicity is a free parameter. Overall, this SFH leads to similar parameters as for exponentially rising SF \citep{schaereretal2013}.
\end{itemize}

Furthermore, we define three options concerning the treatment of nebular emission:
\begin{itemize}
	\item No nebular emission.
	\item +NEB: this includes nebular continuum emission and lines except for Ly$\alpha$, since this line may be attenuated by radiation transfer 
	processes inside the galaxy or by the intervening intergalactic medium.
	\item +NEB+Ly$\alpha$: this includes nebular emission (all lines and continuum processes).
\end{itemize}

The {\em Reference model} is used here as a conservative model with a set of assumptions that are widespread in use to infer physical properties of high-redshift galaxies.

The comparison between the two latter models (NEB/NEB+Ly$\alpha$) is intended to determine the effect of the Ly$\alpha$ line on the parameter estimation. While several studies found inconsistencies between the SFR derived from SED fitting with a declining SFH and the SFR derived from UV+IR and/or H$\alpha$ (Reddy et al. 2012, Erb et al. 2006) at $z\sim2$, we use this SFH to reproduce episodic star formation, a scenario suggested by Stark et al. (2009).

The dynamical timescale, $t_{\mathrm{dyn}} \simeq 2r_{\mathrm{hl}}/\sigma$ with $r_{\mathrm{hl}}$ defined as the half-light radii and $\sigma$ as the velocity dispersion, should be a lower limit for age estimation of a galaxy for reasons of causality. Using different studies \citep{bouwensetal2004,fergusonetal2004,douglasetal2010} that provide $r_{\mathrm{hl}}$ values and the study by \cite{forsterschreiberetal2009} that provides velocity dispersion at $z\sim3$, and with the assumption of no evolution of $\sigma$ with redshift, we estimate that $t_{\mathrm{dyn}}$ evolves from 24$^{+39}_{-17}$ Myr at $z\sim3$ to 12$^{+19}_{-8.5}$ Myr at $z\sim6$. The dynamical timescale is also found to scale with $(1+z)^{\frac{-3}{2}}$ \citep[e.g.][]{wyithe&loeb2011}, which implies a variation of a factor $\sim3$ between redshift 3 to 6. For age, we use a lower limit of 50 Myr with the reference model to illustrate the effect of this physical limit on SED fitting, and parameter estimation. The value of 50 Myr corresponds to a high estimate of typical $t_{\mathrm{dyn}}$ at z$\sim$3.
 
The SED fits to all galaxies have been computed for each of the above model sets and nebular options, for nine different combinations.
This allows us to examine the impact that these assumptions/options have on the derived physical parameters
and to compare also their fit quality.
In detail, for the B-drop sample, we have also tested the effect of the SMC extinction law of \cite{prevotetal1984}. 

In general, the free parameters of the SED fits are the redshift $z$, the metallicity $Z$ (of stars and gas), the star 
formation history as parametrised by $\tau$, the age $t$ defined since the onset of star formation, and the attenuation $A_V$.
For the reference model set, the SFH and metallicity are fixed and the age limited to a minimum.
For the RIS model set, the SFH is also fixed.
In all cases, we consider $z \in [0, 10]$ in steps of 0.1, $A_V =$ 0--4 mag in steps of 0.1, and
51 age steps from 0 to the age of the Universe \citep[see][]{bolzonellaetal2000}.
The SFH of the decreasing models is sampled with $\tau=$ (10, 30, 50, 70, 100, 300, 500, 700, 1000, 3000, $\infty$) Myr.

For all the above combinations, we compute the $\chi^2$ and the scaling factor of the template, which provides 
information about the SFR and stellar mass ($M_{\star}$) from the fit to the observed SED. Minimisation over the entire parameter space yields the best fit 
parameters.
To determine both confidence intervals (68\%) and medians for all the parameters, we ran 1000 Monte Carlo simulations for each object by perturbing the input broadband photometry assuming the photometric uncertainties are Gaussian. This procedure yields the one or two dimensional probability distribution functions of the physical parameters
of interest both for each source and for the ensemble of sources.

 \begin{figure}[htbf]
      	\centering
	       \includegraphics[width=10.5cm,trim=3.75cm 6.5cm 1.5cm 7.5cm,clip=true]{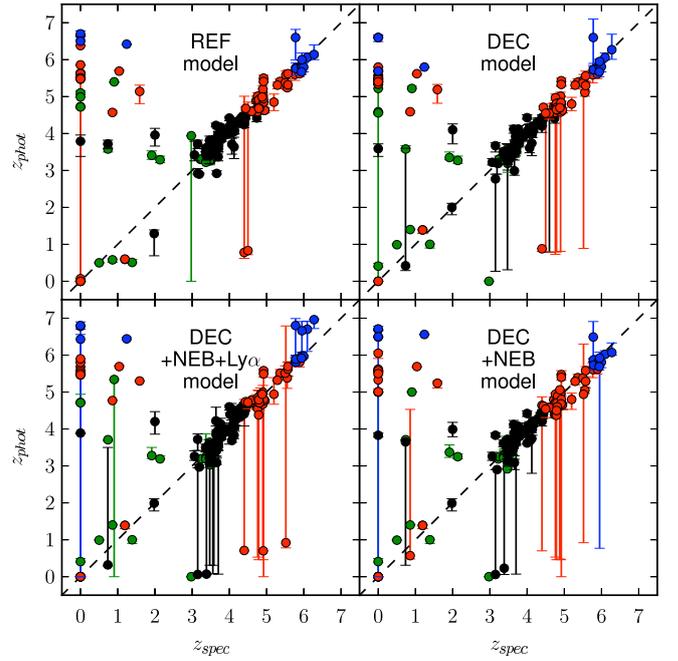}
        \caption{Comparison between photometric redshift and spectroscopic redshifts with
	 green, black, red, and blue the $U$, $B$, $V$, $i$-dropout.
	 Disagreements within a 68\% confidence limit of $U$, $B$, $V$ and $i$-dropout affect 9 objects with a very good/good spectroscopic redshift, 6 uncertain and 5 unreliable.  
         Large error bars are due to sources with maxima probability at low and high redshift.}
        \label{our_redshift}
      \end{figure}
      
      \begin{figure}[htbf]
      	\centering
	        \includegraphics[width=8.75cm,trim=1.25cm 0.5cm 1.75cm 1.25cm,clip=true]{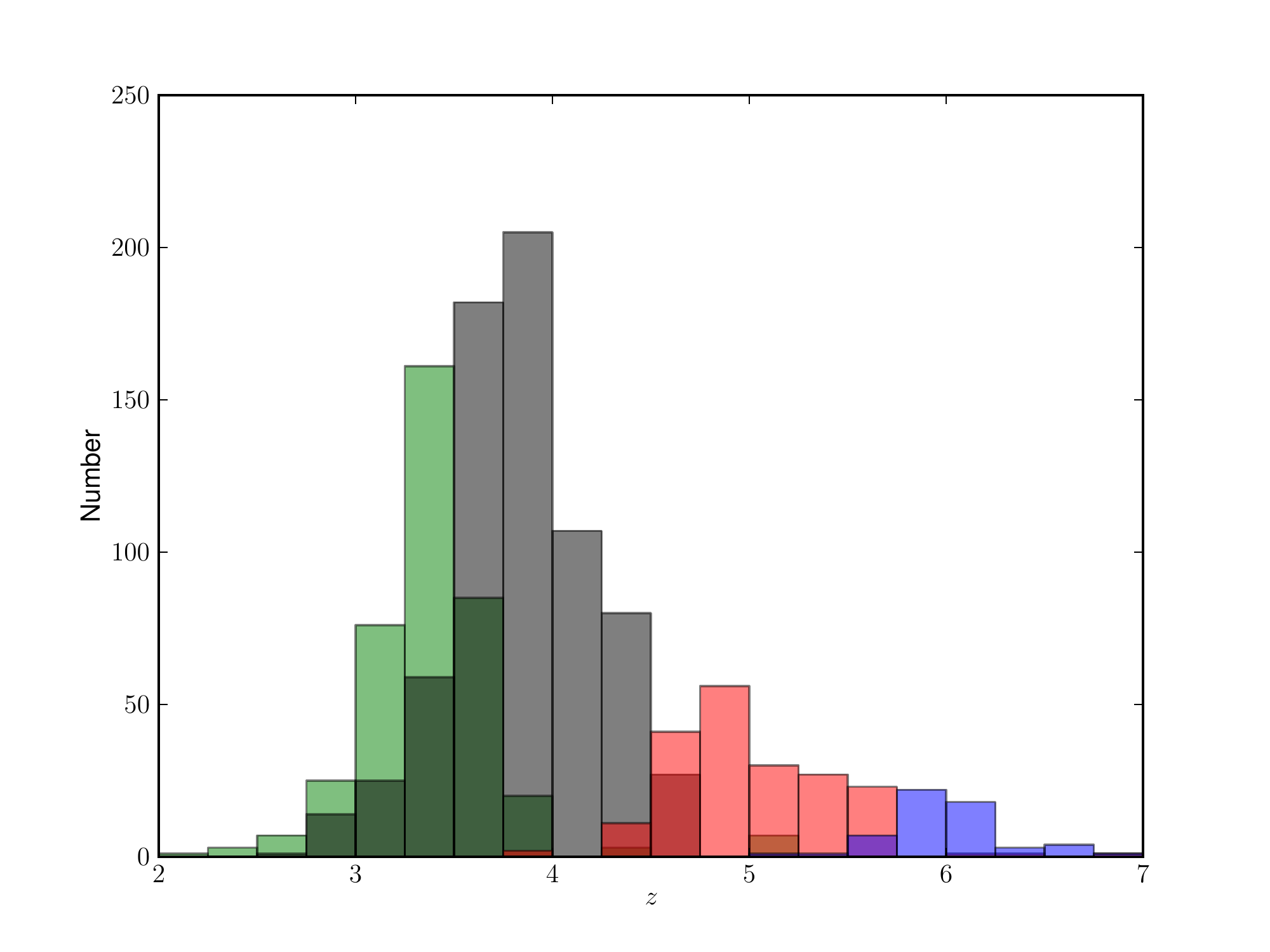}
        \caption{$U$, $B$, $V$, and $i$-drops redshift distribution after the redshift ``cleaning'' with respective green, black, red, and blue histograms.}
        \label{distrib_redshift}
      \end{figure}
      
\subsection{Redshift selection}

We have compared our photometric redshifts against objects with a known spectroscopic redshift that is taken from the literature\citep{vanzellaetal2005,vanzellaetal2006,
  vanzellaetal2008,mignolietal2005,szokolyetal2004,lefevreetal2004,dohertyetal2005,wolfetal2004,daddietal2005,cristianietal2000,strolgeretal2004,
  vanderweletal2005,rocheetal2006,ravikumaretal2007,teplitzetal2007,xuetal2007,gronwalletal2007,nilssonetal2007,hathietal2008,finkelsteinetal2008,
  yangetal2008,popessoetal2009}.
The result is shown in Figure \ref{our_redshift}, respectively for the $U$, $B$, $V$, and $i$ -dropout samples,
for which 42, 72, 50, and 14 spectroscopic redshifts are available. To estimate our photometric redshift performance, we compute the median $\Delta z /(1+z_{spec})$ with $\Delta z$, which we define as the difference between the median $z_{phot}$ and $z_{spec}$. For each sample and model, we obtain values from $-0.02$ to 0.09 with no significant differences among models. We also compute the median absolute deviation, and we find a typical value of $\sigma_{\mathrm{MAD}}=0.03$. These results show that we recover redshift with a good accuracy, which is consistent with typical values found in other studies \citep{wuytsetal2009,hildebrandtetal2010}.
	
	Combining the results of the DEC, DEC+NEB+Ly$\alpha$, and DEC+NEB models, we find 20 objects among all samples, whose
      median photometric redshifts are inconsistent with the spectroscopic redshift within the 68\% confidence limit.
The GOODS MUSIC catalog provides quality flags for the 
      spectroscopic redshift; among the objects with inconsistent redshifts, six are very good, three good, six uncertain, and five unreliable spectroscopic redshifts, 
leading to an estimated 5--11\% of outliers in our samples.
      We also obtain objects with large error bars, which is
      due to a double-peaked redshift probability distribution function with maxima at low and high redshift. 

	To eliminate low redshift contaminants and to have the most reliable sample at each redshift, we proceed with a conservative cut: for $U$, $B$,
      $V$ and $i$-dropouts, we take a lower limit for the median photometric redshift of $z>2$, $z>3$, $z>4$, and $z>5$ respectively, 
as derived from the DEC, DEC+NEB+Ly$\alpha$, and DEC+NEB models. We obtain  389 ($\sim 88\%$), 705 ($\sim 82\%$),
      199 ($\sim 72\%$) and 60 ($\sim 91\%$) objects (Figure \ref{distrib_redshift}) respectively.  Similar criteria applied with REF models leads to larger samples (5 to 13\%) and with RIS models to  similar samples with a maximum variation
      %of the size of the sample
      of $\-4$\%. We notice that there is a significant overlap  between the $U$ and $B$ dropout with 96 objects in both samples, with this final selection.
            
      Table \ref{zmedian} shows the median redshifts and 68\% confidence limits for REF, DEC, and RIS model. Accounting for nebular emission (+NEB+Ly$\alpha$ and +NEB models), median redshifts do not vary more 0.1.
      \begin{table}[htbf]
         \centering
         \caption{Median redshift values and 68\% confidence limits of final samples.}
         \begin{tabular}{cccc}
           \hline
           & REF & DEC & RIS \\
           \hline
           \\
	   $U$-dropout           & $3.32_{-0.13}^{+0.25}$  & $3.33_{-0.14}^{+0.26}$ & $3.30_{-0.13}^{+0.25}$\\
           $B$-dropout           & $3.88_{-0.38}^{+0.41}$  & $3.79_{-0.39}^{+0.44}$   & $3.78_{-0.38}^{+0.42}$\\
           $V$-dropout           & $4.94_{-0.33}^{+0.51}$  & $4.81_{-0.23}^{+0.59}$   & $4.81_{-0.25}^{+0.59}$\\
           $i$-dropout           & $6.00_{-0.29}^{+0.48}$  & $6.00_{-0.32}^{+0.46}$     & $6.00_{-0.33}^{+0.56}$\\
           \\
           \hline
         \end{tabular}
         \label{zmedian}
      \end{table}
     
     \begin{figure}[htbf]
  \centering
    \includegraphics[width=10cm,trim=1.5cm 0cm 0cm 14.cm,clip=true]{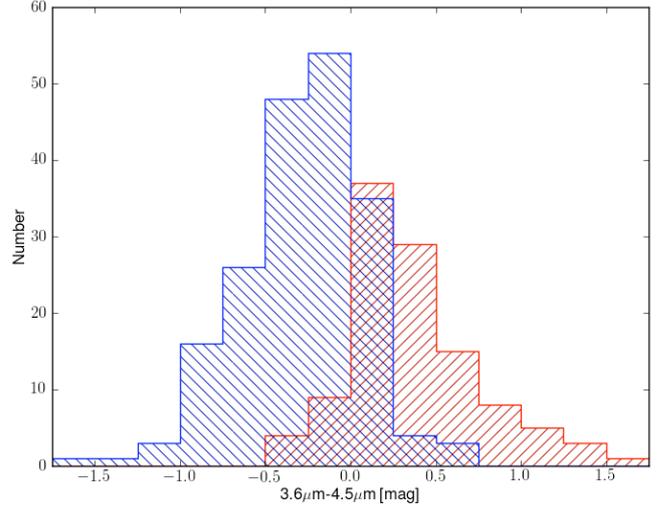}
  \caption{3.6$\mu$m-4.5$\mu$m colour histogram for a subsample of $z$  $\epsilon \left[3.8,5\right]$ objects. In blue, we show objects that are best fit with nebular emission
   and in red, we show  objects that are best fit without nebular emission. Both for a decreasing SFH.}
     \label{histcolor}
\end{figure}
      
\section{Results}
\label{results}

\subsection{Two LBG categories revealed}
  \label{fitq}

As this is the first time large samples of LBGs are analysed with SED fits that include the effects of nebular
emission, we have examined if this leads to better fits and by how much. 
Our main result from this comparison is that typically
60--70\% of the galaxies are better fit with nebular emission (option +NEB or +NEB+Ly$\alpha$) than without. That is, $\chi^2$ values associated with each SED fit have lower values when our models include nebular emission. Using the Akaike Information Criterion (AIC) with the sample {for which we find better fits} with nebular emission (i.e. 60-70\% of the objects), we find that models that includes nebular emission (+NEB/+NEB+Ly$\alpha$) have a relative probability to be the best model 5 to 10 times higher than models without nebular emission, for a given SFH, while these latter models are about twice as likely as models +NEB/+NEB+Ly$\alpha$ for the rest of the sample (30-40\%). Furthermore, models +NEB+Ly$\alpha$ have an increasing probability with redshift to be the best model in comparison with +NEB models, which is up to 5-10 times more likely at $z\sim6$.
This is found independently of the adopted SFH and for all samples, which is from $z \sim$ 3 to 6.
In other words, for $\sim$35\% of the objects, the best fit is found {\em without} taking account of nebular emission.
This fraction is independent of properties, such as the absolute UV magnitude $M_{1500}$ or the number of filters 
available.
Furthermore, all SF histories (REF, DEC, RIS model sets) yield approximately the same percentages (30\%-39\%), and all models lead to almost identical samples. That is, an object identified as a ``strong" (``weak") emitter with one SFH is generally identified as a ``strong" (``weak") emitter with any other SFH. Thus, ``strong" and ``weak" samples are similar at 68\% for $U$ drops and up to 94\% for $V$ drops.
Finally, this is not only a statistical property, but the vast majority of objects can be assigned 
to such a category.

Since this distinction in two groups is fairly model- and redshift- independent, there must be a physical explanation for it.
The easiest and most natural explanation is found when considering a subsample of objects over a restricted
redshift interval. Indeed, since H$\alpha$ is a strong line at 656.4 nm (rest-frame) and few strong lines are found longward of it,
this line must affect the 3.6-4.5 \micron\ colour for objects between $z$=3.8 and $z$=5 \citep[cf.][]{shimetal2011}. 
We therefore selected
$B$-dropout objects with available 3.6$\mu$m and 4.5$\mu$m data (excluding non-detections) and with a median redshift between 3.8 and 5. 
We obtain a subsample of 303 objects for which again $\sim$ 35\% of the objects are best fit when nebular emission is not taken into account.
This should thus be a representative subsample of all galaxies studied here.
Figure \ref{histcolor} shows that the objects best fit with nebular emission have a systematically bluer 3.6$\mu$m-4.5$\mu$m colour than those
better fit without nebular effects. This shows that objects better fit with models which account for nebular emission 
do indeed show strong \ha\ emission lines. This is not a trivial finding, since these models also allow ages/SF histories,
where nebular emission is absent/insignificant.
We therefore conclude that the objects best fit with models accounting for nebular emission ($\sim$ 60--70\%) correspond to galaxies
with ``strong" emission lines, 
whereas the rest shows few or no discernible signs of emission lines (``weak" emission lines). Median H$\alpha$ equivalent widths for these two categories are shown in Table \ref{ewha}.

\begin{table}[htbf]
         \centering
         \caption{Median H$\alpha$ equivalent width (\AA) in the rest-frame 
         for ``strong" and ``weak" nebular emitters with different SFH.}
         \begin{tabular}{ccccccc}
           \hline
           & \multicolumn{2}{c}{REF+NEB} & \multicolumn{2}{c}{DEC+NEB} & \multicolumn{2}{c}{RIS+NEB} \\
%           \hline
           & weak & strong & weak & strong & weak & strong \\
           \hline
           \\
           $U$ drops & 175 & 309 & 130 & 544 & 264 & 722 \\
           $B$ drops & 234 & 203 & 175 & 787 & 463 & 1292 \\
           $V$ drops & 384$^{\mathrm{b}}$ & 384 & 236 & 832 & 722 & 1345 \\
           $i$ drops$^{\mathrm{a}}$ & 384 & 384 & 147 & 2755 & 575 & 3088 \\
           \\
           \hline
         \end{tabular}
         \label{ewha}
         \begin{list}{}{}
\item[$^{\mathrm{a}}$ For the $i$-dropouts, we use +NEB+Ly$\alpha$ models. See Section \ref{sfr}.]
\item[$^{\mathrm{b}}$ Maximum possible value for the REF+NEB model.]
\end{list}
      \end{table}

It is interesting to note that both \cite{yabeetal2009} and \cite{shimetal2011} also found $\sim 70\%$ of a sample of LBGs at $z \sim5$, and 4, respectively,
with a 3.6 $\mu$m excess, which they explained by the presence of \ha\ emission.
While their samples include $\sim$ 100 (70) such galaxies, we have $\sim$ 300 galaxies at $z \approx$ 3.8--5 with direct
empirical evidence of strong \ha\ emission. In addition, our SED fits suggest that a similar percentage of objects with ``strong"
emission lines exist over the entire examined redshift range ($z \sim$ 3--6).  
Below, we  show that this interpretation is perfectly consistent with the differences found for the physical
parameters of these galaxies, and we propose a physical explanation for the existence of these two groups of LBGs.

\begin{figure}[htbf]
  \centering
    \includegraphics[width=11.5cm,trim=3cm 6.5cm 0cm 7.5cm,clip=true]{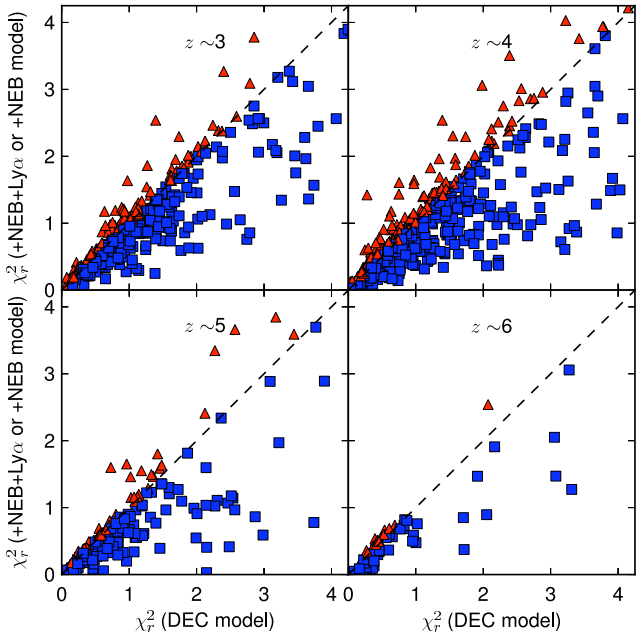}
  \caption{Comparison between DEC model $\chi^2_r$ and DEC+NEB+Ly$\alpha$/+NEB model $\chi^2_r$. Red triangle: DEC model best fit and blue square:
   DEC+NEB+Ly$\alpha$/+NEB best fit. Blue squares and red triangles underlie the two LBGs categories, respectively ``strong" nebular emitters and ``weak" nebular emitters.}
  \label{chi2_nebvsstd}
\end{figure}

\subsection{Fit quality and constraints on star formation histories}
\label{s_sfh}

To compare the fit quality of the different models, we compare values of $\chi^2_r$ ($\chi^2$ value divided by the number of filters minus 1). At each redshift, the SEDs are
systematically better fit with RIS and DEC model sets (considering or not nebular emission) in comparison with REF model sets. The $\chi^2_r$ values
are on average 20--40\% lower for RIS model sets and 25--40\% lower for DEC model sets, which also show that DEC model sets fit slightly better than RIS model sets. 
As shown in Figure \ref{chi2_nebvsstd} for declining SF and all redshifts, 
``strong'' nebular emitters show a large improvement in $\chi^2_r$ when they are fit with models that consider nebular emission 
($\chi^2_r$ $\sim$30 to 55\% lower on average). At the opposite, ``weak'' nebular emitters show a slight improvement of their $\chi^2_r$ when they are fit without nebular emission (10 to 30\% on average). Models with nebular emission are able to provide significantly better fits for ``strong" nebular emitters and a more or less similar fit quality for ``weak" nebular emitters. To provide a fair comparison among models, we use the AIC with models of the same number of free parameters again, since models based on rising and declining SFHs respectively have one (metallicity) and two (metallicity and $\tau$) additional free parameters in comparison with models based on a constant SFH. Since the SED fitting is generally insensitive to changes in
metallicity \citep[while it can be different when accounting for nebular emission,][]{schaerer&debarros2009}, and since our results with DEC models show that short timescale are preferred (see Section \ref{s_tau}), we compute relative probabilities using the same number of free parameters for all the models, running SED fits with solar metallicity for RIS models and DEC models and a fixed timescale ($\tau=10$ Myr) for DEC models. Under these assumptions (which do not strongly affect our results), RIS+NEB model is 1.5 times more likely to be the best model than the REF model, while the DEC model is 10 times more likely than REF model. The comparison with REF+NEB model leads to similar results.

\begin{figure}[htbf]
  \centering
    \includegraphics[width=9.5cm,trim=1.5cm 6.5cm 2cm 7.5cm,clip=true]{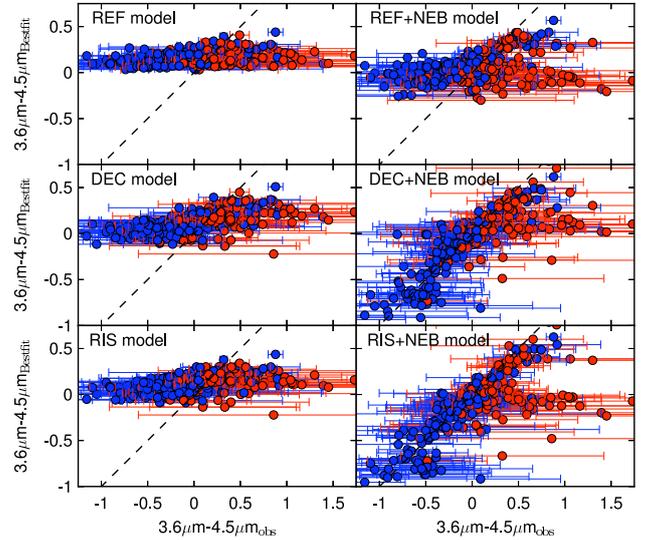}
    \caption{Comparison between observed 3.6$\mu$m-4.5$\mu$m colour and best fit colour for a subsample of $z$  $\epsilon \left[3.8,5\right]$ objects for different models with and without nebular emission. Blue dots show objects identified as ``strong" nebular emitters, and red dots are ``weak" nebular emitters.}
  \label{colortest}
\end{figure}
	
For the subsample of objects with redshifts $z \sim$ 3.8--5 as discussed above, the observed 3.6$\mu$m-4.5$\mu$m colour
provides also an interesting constraint on star formation history and nebular emission. 
Indeed, models without nebular emission are unable to reproduce the range of observed 3.6$\mu$m-4.5$\mu$m colours at $z\sim4$,  as shown in Fig.\ \ref{colortest}. The REF+NEB model (constant star formation) is also unable to reproduce the observations, while the DEC+NEB and RIS+NEB models provide fits fully consistent with this colour (+NEB+Ly$\alpha$ option leads to similar results) with DEC+NEB providing even better results than RIS+NEB. Abandoning the age limitation ($\mathrm{age}>50$ Myr) for the REF+NEB model would allow this model to provide fits consistent with observations. We need to have very large H$\alpha$ equivalent widths to reproduce the colour of ``strong" nebular emitters. This can be obtained for any SFH with young ages (median of $\sim20$ Myr for DEC+NEB and RIS+NEB). On the other hand, small EW(H$\alpha$) for ``weak" nebular emitters  are necessary to reproduce their red colour. Since no physical process is able to suppress effectively nebular emission for both rising and constant SFs, DEC+NEB(+Ly$\alpha$) is the model which best fits both LBG categories. However, RIS+NEB(+Ly$\alpha$) and REF+NEB(+Ly$\alpha$) also provide acceptable fits for ``weak" nebular emitters, considering errors in colour estimation.
Further constraints and results on the timescales of the exponentially declining star formation histories are discussed
in Sect.\ \ref{s_tau}.

Finally, we have found a shift between
those best fit with or without \lya\ (i.e.\ between +NEB and +NEB+\lya\ option) with redshift, among objects better fit with nebular emission. In this 
sense the higher-$z$ galaxies favour a larger fraction of objects with \lya\ emission.
This shows that SED fitting is also sensitive to \lya\ emission, which is a finding we have demonstrated and 
discussed in detail in  \cite{schaereretal2011}.

  \subsection{Physical properties of the LBGs}
  \label{phys}
  
  We now turn to discuss the main physical properties (stellar mass, SFR, age, attenuation, and the star formation timescale
where appropriate) of the LBGs and their dependence on model assumptions.
The median values and uncertainties of the physical parameters derived in several bins of UV magnitude $M_{1500}$ for all our samples and by 
using all nine combinations of model assumptions are listed in Tables \ref{tabmagall_ref}, \ref{tabmagall_dec}, and \ref{tabmagall_ris}.
For each physical parameter, we now describe the median properties and their model dependence, explain their origin,
and compare the behaviour of the individual values. Furthermore, we examine possible correlations between derived and observed parameters.
To do this, we choose the largest subsample, here consisting of 705 $B$-drop ($z \sim 4$) galaxies, since the same trends/differences overall 
are found  at all redshifts (except stated otherwise).
In Sect.\ \ref{evoz}, we then discuss the redshift evolution of the physical parameters and their model dependence. 

	 \subsubsection{Absolute UV magnitude}

In what follows, the absolute UV magnitude $M_{1500}$ refers to the absolute magnitude at 1500 $\AA$. To determine it for each object, we use the integrated SED flux in an artificial filter of  200 \AA\ width centered on 1500 $\AA$.
	 Using the $V$-band magnitude for $U$-drop, $i$-band for $B$-drop, and $z$-band for $V$- and $i$-drop samples and spectroscopic redshift when availaible to estimate the UV magnitude \citep{starketal2009} leads to no significant difference on $M_{1500}$, except for one $B$-drop, two $V$-drops, and one $i$-drop galaxy. These are objects with a spectroscopic redshift identification at low redshift, which passes our selection. As they represent less than 1\% (slightly more for $i$-drop) of each sample, we consider that they can not alter significantly our conclusions. 
In passing, we note that the number of objects in each UV magnitude bin listed in Tables \ref{tabmagall_ref}, \ref{tabmagall_dec} and \ref{tabmagall_ris} can change from one model to another, mostly due to small differences in photometric redshifts.

	\begin{figure}[htbf]
	\centering
		\includegraphics[width=7.9cm,trim=7cm 6.75cm 7.5cm 7.5cm,clip=true]{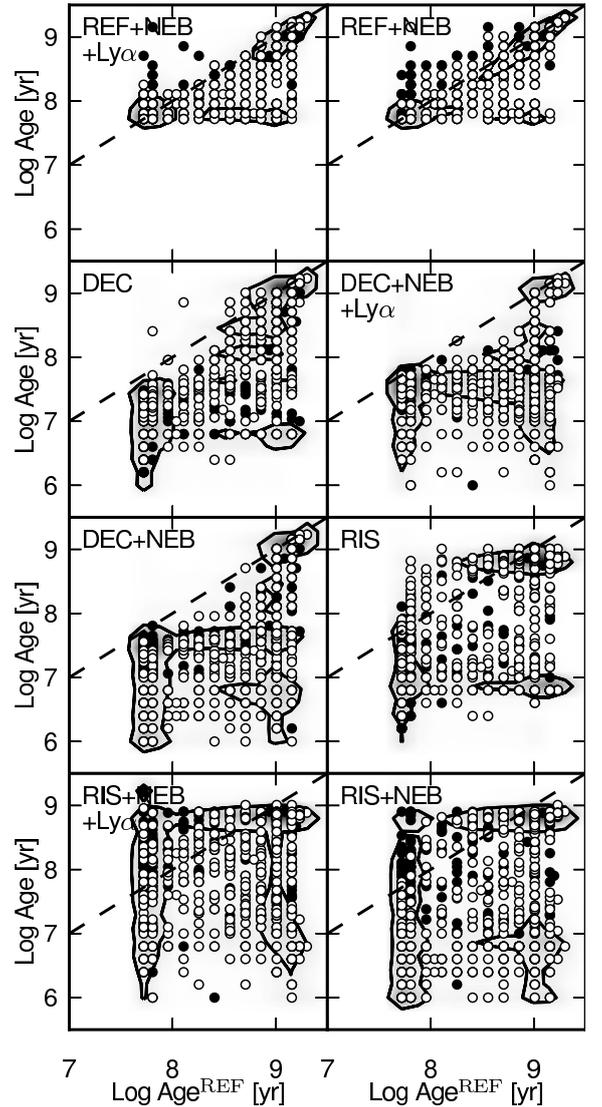}
	\caption{Composite probability distribution of age for REF model and age for all other models at $z\sim4$. The points overlaid show the median value properties for each galaxy in the sample. Black dots represent ``weak" nebular emitters and white dots ``strong" nebular emitters. The overlaid contour indicates the 68\% integrated probabilities on the ensemble properties measured from the centroid of the distribution.}
	\label{agecomp}
	\end{figure}  
		   
  	\subsubsection{Age}
	\label{age}

	Overall, the ages of individual $z \sim 4$ LBGs derived from the different models span a wide range, typically from $\sim$ 4 Myr (if no lower limit
is specified) to $\sim 1.5$ Gyr, which is the maximum age at this redshift (see Fig.\ \ref{agecomp}). A wide age range is found for all nine model sets.
     
     The individual ages and the resulting median age of the sample depend strongly on the model assumptions.
As can be seen in Tables \ref{tabmagall_ref}--\ref{tabmagall_ris} and by assuming declining (DEC) or rising (RIS) star formation
histories leads for models without nebular emission to median ages younger by a factor 5--10 as compared to constant star formation (REF).
These differences are much larger than the typical age uncertainty ($\sim0.15$ dex) found for the REF model.
       The reason is that galaxies keep a high UV rest-frame flux with a constant SF and if the observed SED shows the presence of a Balmer break, older ages are necessary for the population to reproduce the observed break.
         For a declining or rising SF, much younger ages are enough to obtain a suitable evolved population.
Indeed, the observed median ratio of the optical/UV flux is high and consistent with a Balmer break at each redshift \citep[see][]{gonzalezetal2010,leeetal2011}.
            
            \begin{figure}[htbf]
     \centering
          \includegraphics[width=7.9cm,trim=7cm 6.75cm 7.5cm 7.5cm,clip=true]{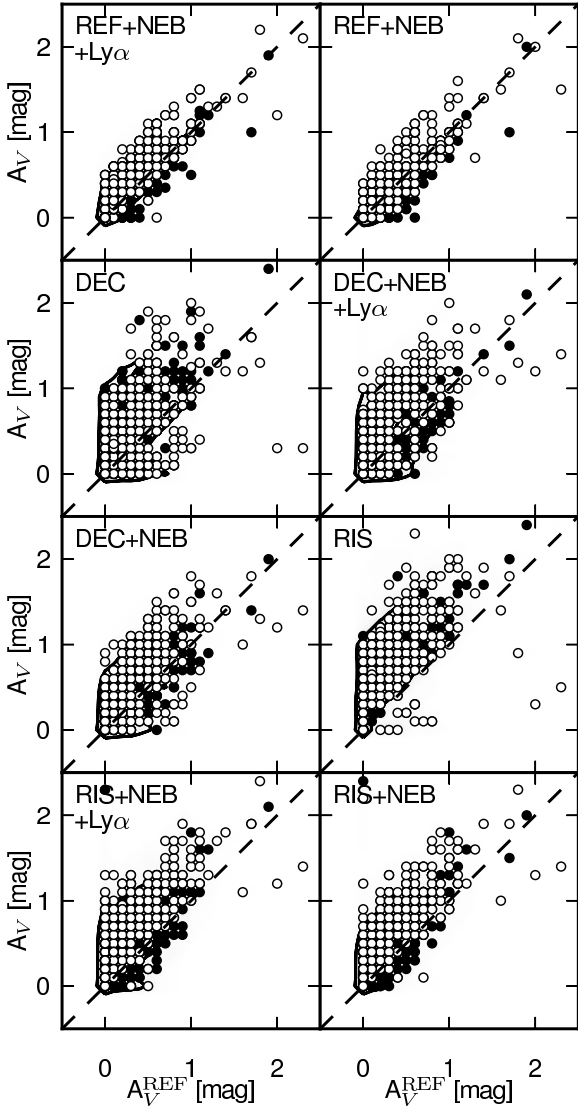}
     \caption{Same as Figure \ref{agecomp} for $A_V$.}
     \label{avcomp}
     \end{figure}  
            
            The effect of different assumptions on age estimation is more easily understood if we examine the two populations, ``weak" and ``strong" emitters, separately as defined in Section \ref{fitq}. For ``strong" nebular emitters, median ages are decreased for all SFHs and for all redshift, which is an expected result \citep{schaerer&debarros2009,schaerer&debarros2010}, since emission lines (mostly [O~{\sc III}] $\lambda\lambda$4959, 5007 and H$\beta$) can mimic a Balmer break. This effect is strong enough (e.g.\ by a factor 2--10 for the REF model; see Fig.\ \ref{agecomp}) to lead to significant differences between models, even considering uncertainties. Uncertainties are increased when nebular emission is taken into account: from 0.15 to 0.27 dex for REF models, from 0.39 to 0.55 dex for RIS models. These increased uncertainties are due to double peaked age probability distribution functions \citep{schaerer&debarros2010}, since these objects can be fit with young population (and strong lines) or old population (Balmer break). For declining SFH, there is no such effect, since there are already large uncertainties on age ($\sim0.4$ dex, even without nebular emission) because of the strong degeneracy between age and timescale.
            
            For LBGs with ``weak" lines, considering nebular emission leads to {\it older} ages, a result that can be easily explained if we assume that these objects have truly weak lines: the only way to decrease emission lines strength is by adopting old ages at least with our assumptions and SFHs. While the trend is the same for all SFHs, there are quantitative differences with older ages by a factor 2--5 for REF and DEC models (at $z\sim4$) and a factor 10--20 for rising SF.

In absolute terms, the derived ages of ``weak" and ``strong'' nebular emitters are fairly similar when REF models are assumed.
For decreasing and rising SF, the ages of ``strong" emitters are systematically decreased when considering nebular emission, while they are increased for ``weak" emitters. ``Strong" emitters are younger by a factor $\sim2$ (DEC+NEB model) to $\sim5$ (RIS+NEB model) in comparison with ``weak" emitters.
In any case, the age differences seem to confirm an intrinsic difference between these two LBG categories, and the strength of the lines seems to be the main driver for the age determination at high redshift when nebular emission is taken into account.

     \begin{figure}[htbf]
     \centering
          \includegraphics[width=8.5cm,trim=1cm 0.5cm 2cm 1cm,clip=true]{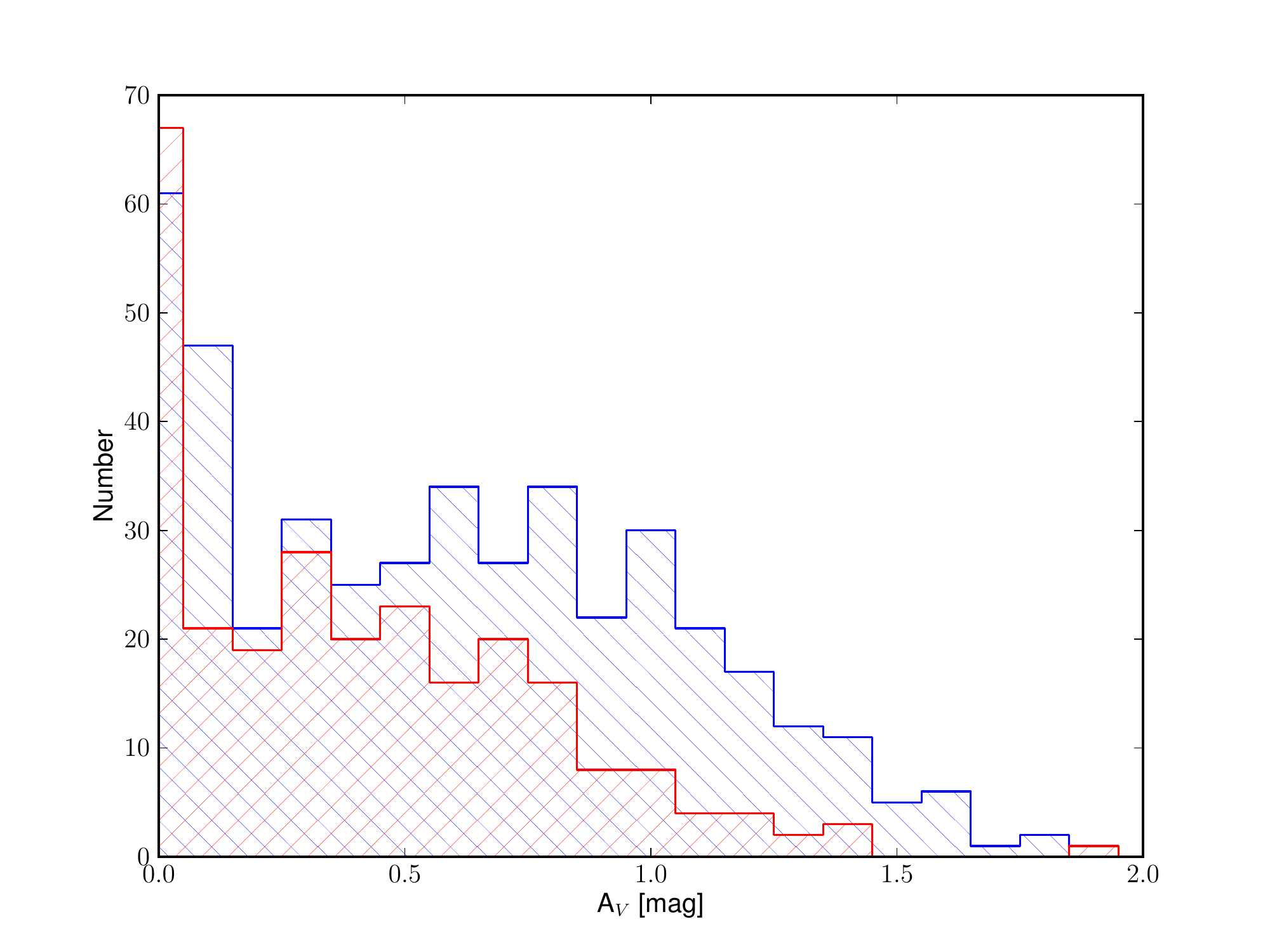}
     \caption{A$_V$ distribution at $z\sim4$ (DEC+NEB model) for ``strong" (blue) and ``weak" emitters (red).}
     \label{avdiff}
     \end{figure}  
    
     \subsubsection{Reddening}
     \label{av}

     The attenuation of individual $z \sim 4$ LBGs derived from the SED fits range from $A_V=0$ to a maximum of $\sim$ 1.5--2 mag for a few objects,
as shown in Fig.\ \ref{avcomp}. Although all model sets yield a similar range of attenuation, relatively large systematic
differences, which we discuss, are found between them.
         
	Whether one considers nebular emission or not for a given SFH a variation no larger than 0.2 mag can be determined. Comparing REF model to other models (without nebular emission), the median $A_V$ is higher for the DEC model (+0.2 mag) and for the RIS model (+0.5 mag), which can be partially explained by the well-known degeneracy between reddening and age, since DEC and RIS models lead to younger median ages than constant SF (cf.\ above).
          Furthermore, a higher attenuation is required to
fit the observations, compared to models with constant SF, as already pointed out by \citet{SP05}, since young stars always dominate the UV flux for rising star formation histories.
Considering this, nebular emission leads  to variations no larger than 0.1 mag in reddening for the REF model on average;
a slight increase is found when the \lya\ line is included (REF+NEB+Ly$\alpha$ model, assuming the maximum case B intensity for \lya),  which is explained by the additional flux from the Ly$\alpha$ line.

 \begin{figure*}[htbf]
     \centering
          \includegraphics[width=19.25cm,trim=2.5cm 7cm 2cm 7cm,clip=true]{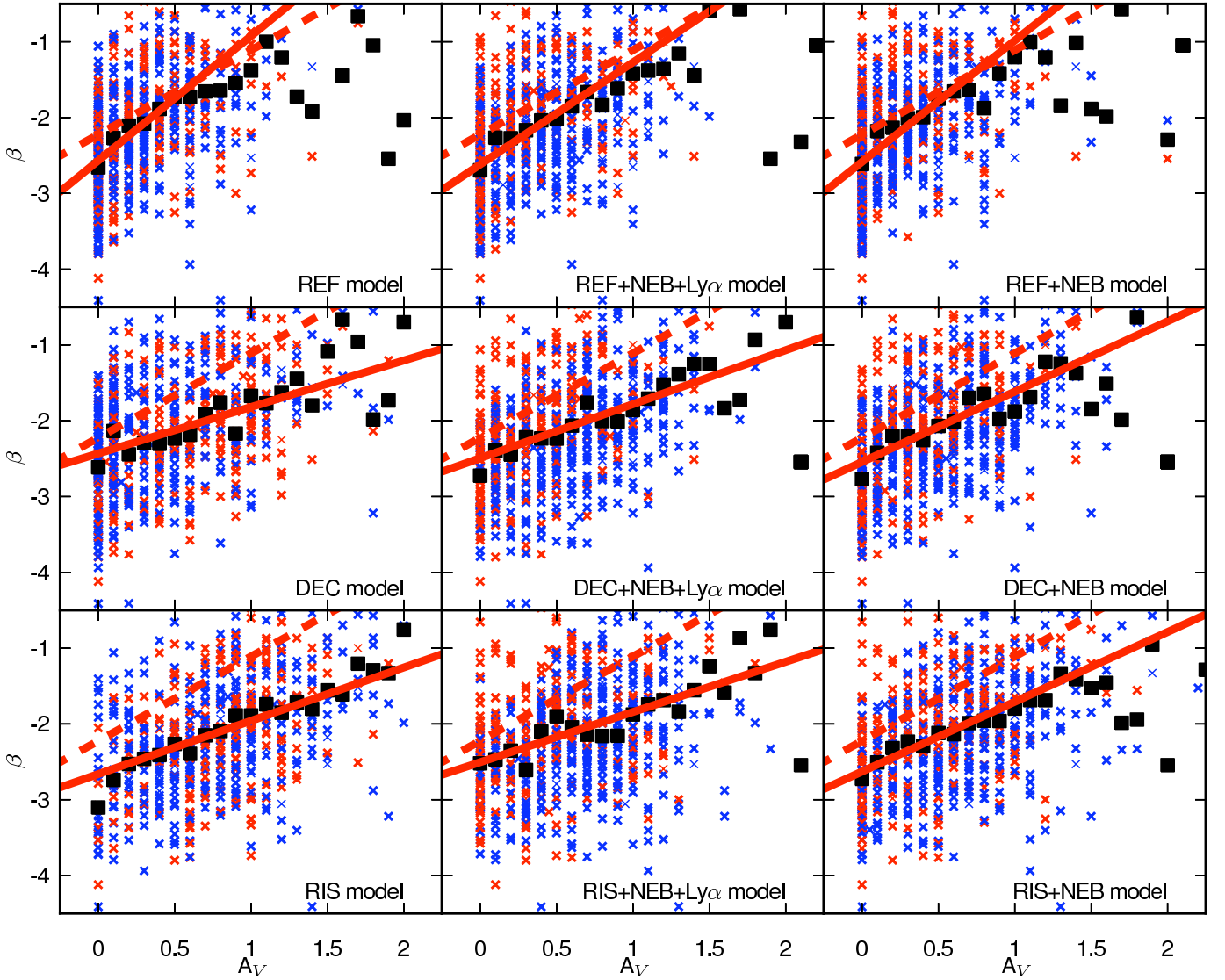}
     \caption{UV-continuum slope $\beta$ vs $A_V$ for all models at z$\sim4$. Each cross shows the median value properties for each galaxy: blue for ``strong" nebular emitters and red for ``weak" nebular emitters. Black squares are median values by bins of 0.1 $A_V$ mag. The red dashed line is the relation between extinction \citep{calzettietal2000} to a given $\beta$ if the base spectrum is a young star-forming galaxy of constant star formation \citep{bouwensetal2009}. The red line is a linear fit among the whole composite probability distribution function.}
     \label{beta_av}
     \end{figure*} 
     
On the other hand, nebular emission leads to a systematically lower median $A_V$ for rising star formation history, since the contribution of strong nebular lines in the optical and infrared (rest-frame) leads to redder SEDs, because of extremely large EWs (see Table \ref{ewha}). This effect is seen for both ``weak'' and ``strong'' categories. These results have to be taken with some caution since typical errors on $A_V$ for individual objects are 0.1 mag for constant SF and $\sim0.2$ mag for decreasing and rising SF, which does not allow a clear distinction for example between REF and DEC models.

This latter result can seem surprising since previous studies at $z\sim2$ \citep[e.g.][]{shapleyetal2005,erbetal2006b,reddyetal2012a} found that a declining SFH usually leads  to lower dust attenuation than a constant SFH due to the best fits with $t/\tau>1$. This implies that there is a significant population of old stars that explained the redness of UV continuum. This also leads to a discrepancy between the SFR inferred from the SED fitting and other SF indicators \citep{erbetal2006b,reddyetal2012a} with a declining SFH which leads to an underestimated SFR. Our SED modelling results differ from these previous studies on the $t/\tau$ ratio, for which we find that  40--50\% of our objects (at all redshift) have $t/\tau<1$. This condition should provide similar results between declining and constant SFH in terms of dust attenuation and SFR. Since we allow younger ages for DEC models than for REF models, we find typically higher dust attenuation and SFR.
By comparing our SED modelling with those from \cite{shapleyetal2005}, \cite{erbetal2006b}, and \cite{reddyetal2012a}, we find that these three studies rely only on solar metallicity, when we use three different metallicities (0.02, 0.2, and 1Z$_{\odot}$). \cite{papovichetal2001} shows in their Figure 10 that the impact of subsolar metallicity on confidence intervals for a composite probability distribution function between age and timescale. While the range of possible values is large at solar metallicity, it is increased at lower metallicity, mainly in the $t/\tau<1$ area. Indeed, if we consider our models with a declining SFH in our three metallicity bins, we find a trend consistent with this result: objects with $t/\tau<1$ are 25--30\% at Z=Z$_{\odot}$ up to 60--80\% at Z=0.02Z$_{\odot}$. This shows how some parameters can be sensitive to assumption on metallicity. We are reminded that we infer median physical parameters of each object and sample through MC simulations with a marginalization over the parameter space. As shown in Tables \ref{tabmagall_dec} and \ref{tabmagall_ris}, metallicity is bracketed between 0.2 and 1 Z$_{\odot}$ at $z\sim3-5$, which is consistent with metallicity inferred at $z\sim2$ \citep{erbetal2006c}. Extreme subsolar metallicity (0.02Z$_{\odot}$) is preferred only at $z\sim6$. We provide full comparison with several other studies in Section \ref{compothst}.

The inclusion of nebular emission with a constant star formation history does not lead to any significant change in dust reddening for ``weak" nebular emitters. For declining and rising SF, the extinction decreases strongly (by $\sim$ -0.2 to -0.5 mag in $A_V$) when we consider nebular emission. Since these objects seems to have intrinsically no discernible signs of emission lines, models include a nebular emission fit by minimizing equivalent widths, which is achieved by minimizing SFR and UV flux. Model sets based on constant star formation history (REF/REF+NEB/REF+NEB+Ly$\alpha$ models) do not allow sufficient variations to modify the reddening estimation. 
For ``strong" nebular emitters, the median $A_V$ increases (by +0.2 to +0.5 mag) between $z\sim3$ and $\sim5$ when we consider declining star formation, while it decreases for rising SF (-0.1 to -0.2 mag), as already explained. The increased dust attenuation with DEC+NEB model can easily be explained by the effect of emission lines on age,  leading generally to younger ages and thus to a bluer slope.
     
     At $z\sim6$,  effects of modelling with nebular emission are different:  for decreasing and rising model sets, the consideration of nebular emission leads to a decrease of the median $A_V$ for both ``weak" and ``strong" nebular emitters, respectively with $\sim-0.6$ and $\sim-0.2$ mag. The SED fits with nebular emission lead to an important contribution of nebular lines longward UV, and so an additional amount of dust attenuation is required, since strong lines are associated with strong UV flux.
     
     Overall, ``strong" emitters are more dusty than ``weak" emitters. Figure \ref{avdiff} illustrates this for the DEC+NEB model, and a KS-test shows that  the A$_V$ distributions are drawn from different populations if we consider any model that accounts for nebular emission (at $z\sim4$, $p<10^{-5}$), even REF+NEB model (constant SF and age$>50$ Myr). In contrast, $p=0.72$ for the REF model.

     \subsubsection{Reddening and UV-slope}
     \label{red_uvslope}

     Since the observed UV slope $\beta$ is often used to measure the attenuation in LBGs, it is interesting to examine how the attenuation
derived from the SED fits are based on the different model assumptions that compare with $\beta$. Such a comparison is shown in Fig.\ \ref{beta_av}
for the nine model sets applied to the $B$-drop sample. The UV slope has been determined using the same filters and relations as \cite{bouwensetal2009}. 
Figure \ref{beta_av} shows that there is a significant trend of increasing $\beta$ with $A_V$, as expected, albeit with a large scatter for individual objects.
We have done linear fits to the 2D composite probability distribution function, which yields the mean relations indicated in the plot
by red lines. For comparison, the ``standard'' relation between $\beta$ and $A_V$ taken here from \cite{bouwensetal2009} is also shown.
As expected, our relations agree well with the ``standard'' one for models assuming constant star formation and ages $>50$ Myr (REF model sets), since this corresponds to the main assumptions made to derive the standard $\beta$--reddening relation.
For a given SF history, differences between the three options with/without nebular emission 
can be explained by the behaviour of $A_V$ discussed above.
Since all models with declining and rising star formation histories yield higher reddening on average (cf.\ above),
a relation shallower than the standard one is found. Since the relations obtained are fairly similar,
we can combine them to obtain the following mean relations between $\beta$ and $A_V$ for the three cases:
 (\ref{betaavnoneb}) modeling without nebular emission, (\ref{betaavneb}) +NEB+Ly$\alpha$ and (\ref{betaavwlya}) +NEB. This is shown as 

\begin{equation}\label{betaavnoneb}
     A_V = 1.54 \times (\beta+2.54)
     \end{equation}

     \begin{equation}\label{betaavneb}
     A_V = 1.47 \times (\beta+2.49)
     \end{equation}
     
     \begin{equation}\label{betaavwlya}
     A_V = 1.09 \times (\beta+2.58).
     \end{equation}
     
     The last relation (Eq.\ \ref{betaavwlya}) is probably the most appropriate one, since it combines
the models which best fit the data \citep[i.e.\ models including nebular emission but no strong \lya\ for the majority of the galaxies, cf.][]{schaereretal2011}.
It should be reminded that these relations assume a Calzetti attenuation law. To translate this into the colour
excess one has $E(B-V)= A_V/R_V$, where $R_V=4.05$.
In short,  we find that LBGs with a given UV slope have higher attenuation 
than derived from the commonly used $\beta$--$A_V$ relation, from our new relations derived from a subsample of 705 $B$-drop galaxies using
various star formation histories. For typical UV slopes of 
$\beta \sim -2.2$ ($-1.7$) found for faint (bright) $z \sim 4$ galaxies, this translates
to an increase in the UV attenuation by a factor $\sim 3$.
     
While implications of our different reddening estimation are discussed in \cite{schaereretal2013}, we note that our preferred $\beta$--A$_V$ relation differs from the one found at $z\sim2-3$ with radio, X-ray and IR data \citep{reddy&steidel2004,reddyetal2010,reddyetal2012b}, which is consistent with the $\beta$--A$_V$ relation established by \cite{meureretal1999}. However, this relation relies on several assumptions, such as $\beta_0$ (UV continuum slope in the absence of dust absorbtion) being dependent from the SFH, metallicity, and IMF \citep{leitherer&heckman1995}. The value of $\beta_0=-2.23$  is obtained with a constant SFH lasting for 100 Myr, solar metallicity, and a Salpeter IMF. Since we obtain significant fractions of galaxies that have an age $<100$ Myr, and subsolar metallicities, we find $\beta_0\sim-2.6$, which is a value consistent with prediction from \cite{leitherer&heckman1995} under similar assumptions.

\begin{figure}[htbf]
     \centering
     \includegraphics[width=8.5cm,trim=1.5cm 6.5cm 2cm 7.5cm,clip=true]{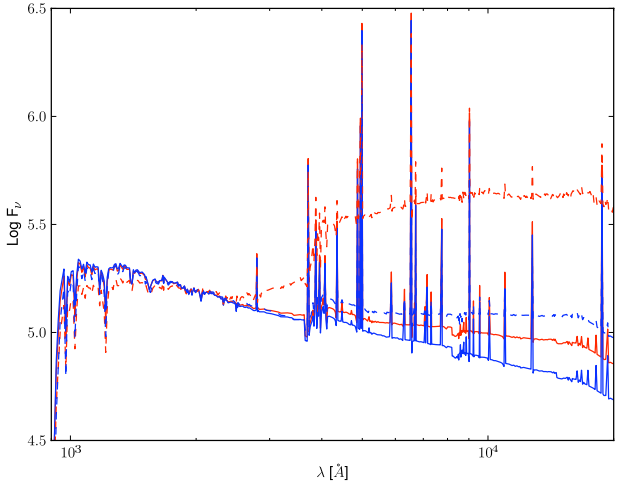}
     \caption{Theoretical SEDs in the rest-frame, which are normalized at $2000 \AA$ for models with $\mathrm{EW}(\mathrm{H}\alpha)=500 \AA$ (solid lines) and $100 \AA$ (dashed lines) and 
    different star formation timescales (constant SF in red and $\tau=10$ Myr in blue). For  the models with $\mathrm{EW}(\mathrm{H}\alpha)=500 \AA$,  the ages are 52 Myr and 16 Myr respectively.
    For $\mathrm{EW}(\mathrm{H}\alpha)=100 \AA$, 2.1 Gyr and 35 Myr.}
     \label{theosed}
     \end{figure}

     \begin{figure}[htbf]
     \centering
     \includegraphics[width=9cm,trim=1.25cm 6.5cm 1.25cm 7cm,clip=true]{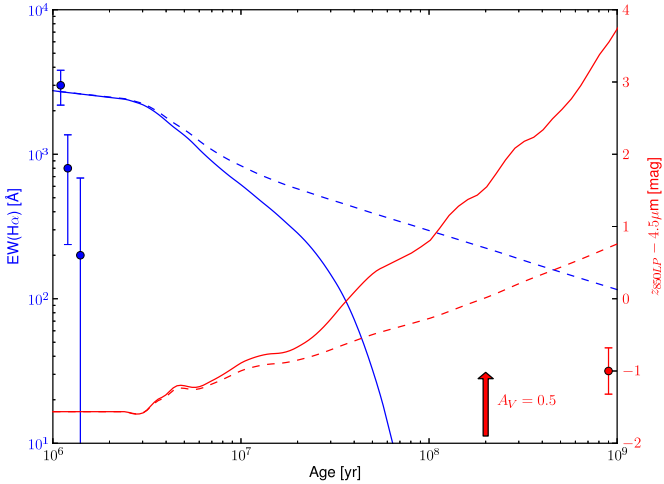}
     \caption{Evolution of EW(H$\alpha$) (blue), and $z_{850LP}-4.5\mu$m colour (red) with age for $\tau=10$ Myr (solid lines) 
and for $\tau=\infty$ (dashed lines).
Typical error bars, as shown on the left for EW(H$\alpha$) and on the right for the $z_{850LP}-4.5\mu$m colour, have been derived from error estimation on measured fluxes at $z\sim4$. 
The effect of redddening ($A_V=0.5$) on $z_{850LP}-4.5\mu$m is shown with the red arrow.}
     \label{color_ewha}
     \end{figure} 

     \begin{figure}[htbf]
      	\centering
        \includegraphics[width=6cm,trim=7.5cm 0.5cm 9cm 14cm,clip=true]{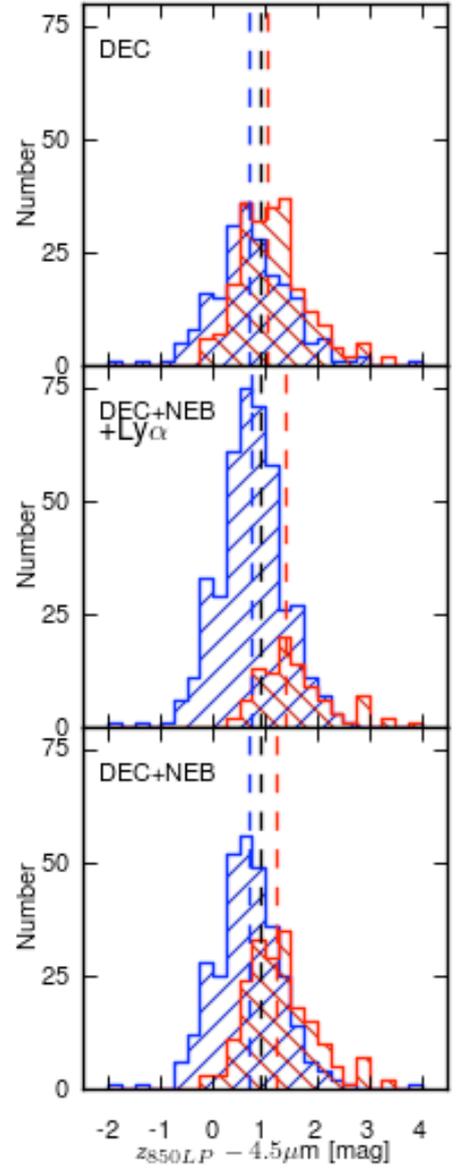}
        \caption{$z_{850LP}-4.5\mu$m observed colour histogram at $z\sim4$. In red, we have best fit objects with $\tau \geq 100$ Myr,
   and in blue, best fit objects with $\tau=10$ Myr for the three models with decreasing SF. The blue, red, and black dashed lines show respectively a median colour for objects that are best fit with $\tau=10$ Myr, with $\tau\geq100$ Myr and for the whole sample. For the two subsamples, KS test for the three models gives $p=7.8\times10^{-8}$, $p=2.6\times10^{-17}$, and $p=1.3\times10^{-16}$, respectively, for DEC, DEC+NEB+Ly$\alpha$, and DEC+NEB models, showing that the two subsamples are not drawn from the same population.}
        \label{hist_colorz}
      \end{figure}
     
\subsubsection{Star formation timescale}
     \label{s_tau}

     For the models with exponentially declining star formation histories (DEC models), which are found to provide the best fits for
the majority of objects (i.e. lower $\chi^2_r$) but not always by large margins it is of interest to examine the resulting timescales $\tau$
and the uncertainties on this quantity. As a reminder, the DEC model set considers 10 different star formation
timescales $\tau \in [10,3000]$ Myr plus the limiting case of $\tau=\infty$ that corresponds to constant star formation.
          
For all options with/without nebular emission (DEC, DEC+NEB+Ly$\alpha$ and DEC+NEB), we find median values of $\tau$ between 10 and 300 Myr
in the different UV magnitude bins (cf.\ Table \ref{tabmagall_dec}). Models including nebular emission favour  shorter timescales on average than those without.
Although the timescales found are relatively short compared to the dynamical timescale
at values of $t_{\rm dyn}=21^{+34}_{-15}$ Myr at $z\sim4$ \citep{bouwensetal2004,fergusonetal2004,douglasetal2010,forsterschreiberetal2009} the uncertainties on $\tau$ are large.

What constrains the SF timescales and why are short timescales preferred? 
Since SFR$\propto \exp(-t/\tau)$, two (or more) observational constraints are needed to determine both the age $t$ and timescale $\tau$.
Lets us first consider the case of models that include nebular emission and examine galaxies with $z \sim$ 3.8--5. 
In this case, we find that $t$ and $\tau$ are mostly constrained by a combination of the (3.6-4.5) \micron\ colour tracing EW(H$\alpha)$
and by a UV/optical (rest-frame) colour. This works as follows.
As already discussed above, the 3.6-4.5 \micron\ colour of galaxies at $z \sim$ 3.8--5 reflects the H$\alpha$ equivalent width.
However, it is well known that a given equivalent width can be obtained with different values of $t/\tau$ (cf.\ Fig.\ \ref{color_ewha}).
To illustrate this, we show the predicted SEDs for galaxies with the same EW(H$\alpha)=$ 100 (500) \AA\
but different SF timescales ($\tau=10$ Myr and $\infty$) in Fig.\ \ref{theosed}. It is obvious that the main feature allowing to lift this degeneracy is the 
ratio of the UV/optical flux. At $z \sim 4$, this ratio is reflected by the $z_{850LP}-4.5\mu$m colour, whose evolution with
$t$ and $\tau$ is also shown in Fig.\ \ref{color_ewha}.
Therefore, these two colours provide good constraints on $t$ and $\tau$, assuming they are not strongly affected by reddening (cf.\ below). Within the typical 68\% error bars, these two extreme SFHs can be 
discriminated by their colour and EW(H$\alpha$) for $t \ga$ 20--30 Myr. For younger ages, the uncertainties in both EW(H$\alpha$) 
and colour do not allow a clear separation.
A posteriori, we can verify that the objects best fit with ``long'' timescales do indeed statistically differ from those
with ``short'' timescales. Figure\ \ref{hist_colorz} shows a statistically significant difference between 
galaxies best fit with $\tau=10$ Myr, which are bluer in  $z_{850LP}-4.5\mu$m and $\tau > 100$ Myr galaxies
showing redder colours.

 \begin{figure}[!htbf]
     \centering
          \includegraphics[width=7.9cm,trim=7cm 6.75cm 7.5cm 7.5cm,clip=true]{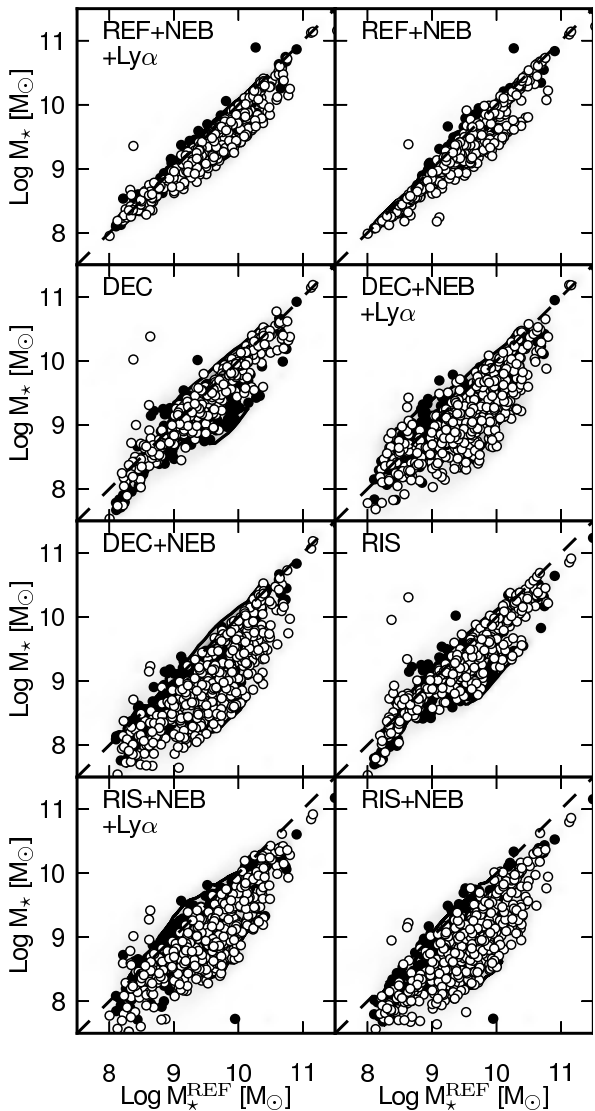}
     \caption{Same as Figure \ref{agecomp} for $M_{\star}$.}
     \label{smcomp}
     \end{figure}

     We also have to consider dust attenuation, which can increase $z_{850LP}-4.5\mu$m, as shown in Fig.\ \ref{color_ewha}, and introduce a degeneracy between SFHs.
However, considering the observed median colour and dust attenuation for the REF and DEC model sets, we expect to have higher extinction for REF model sets than for DEC model sets, but we find the opposite with a median $A_V$ that is larger by 0.1--0.2 mag for DEC models. Furthermore, Figure \ref{color_ewha} shows that it is more difficult to reproduce a given red colour with a constant SF than with a declining SF. This shows that dust attenuation seems to be only weakly correlated with $z_{850LP}-4.5\mu$m colour at $z\sim4$, which allows us to conclude that the ratio of UV/near-IR flux and equivalent widths of different emission lines (mainly Ly$\alpha$, O{\sc iii} and H$\alpha$) drive the choice of $\tau$ for models with declining star formation.
Despite the possibility to discriminate two extreme SFHs like constant SF and decreasing SF with $\tau=10$ Myr, large uncertainties on $\tau$ estimates are found (from 0.7 dex for DEC+NEB+Ly$\alpha$ model to $\sim2$ dex for DEC+NEB model), which prevents us to derive strong constraints on $\tau$.  These uncertainties come mainly from reddening; indeed, fixing reddening to an arbitrary value ($\mathrm{A}_V=0$) also lead to a low median $\tau$ ($<300$ Myr) but with lower typical uncertainties from $\sim0.4$ dex for the DEC+NEB+Ly$\alpha$ model to $\sim0.6$ dex for the DEC model.

Interestingly, our models with declining SF histories indicate a possible increase of the timescale $\tau$ with UV luminosity also with
stellar mass but only for ``strong" nebular emitters. It is tempting to suggest that this could be due to a decrease of the feedback efficiency with increasing galaxy mass, since
the star formation timescale is likely related to the dynamical one and modulated by feedback \citep[e.g.][]{wyithe&loeb2011}. We do not see any evolution of the timescale for the ``weak" nebular emitters, which is easily explained by weaker constraints due to the absence of strong distinctive features. This explains why the trend of $\tau$ with UV magnitude 
cannot be seen in Table \ref{tabmagall_dec}, where the combined data for the entire sample is listed.

           \begin{figure*}[htbf]
     \centering
     \includegraphics[width=18cm,trim=3cm 7cm 4cm 7cm,clip=true]{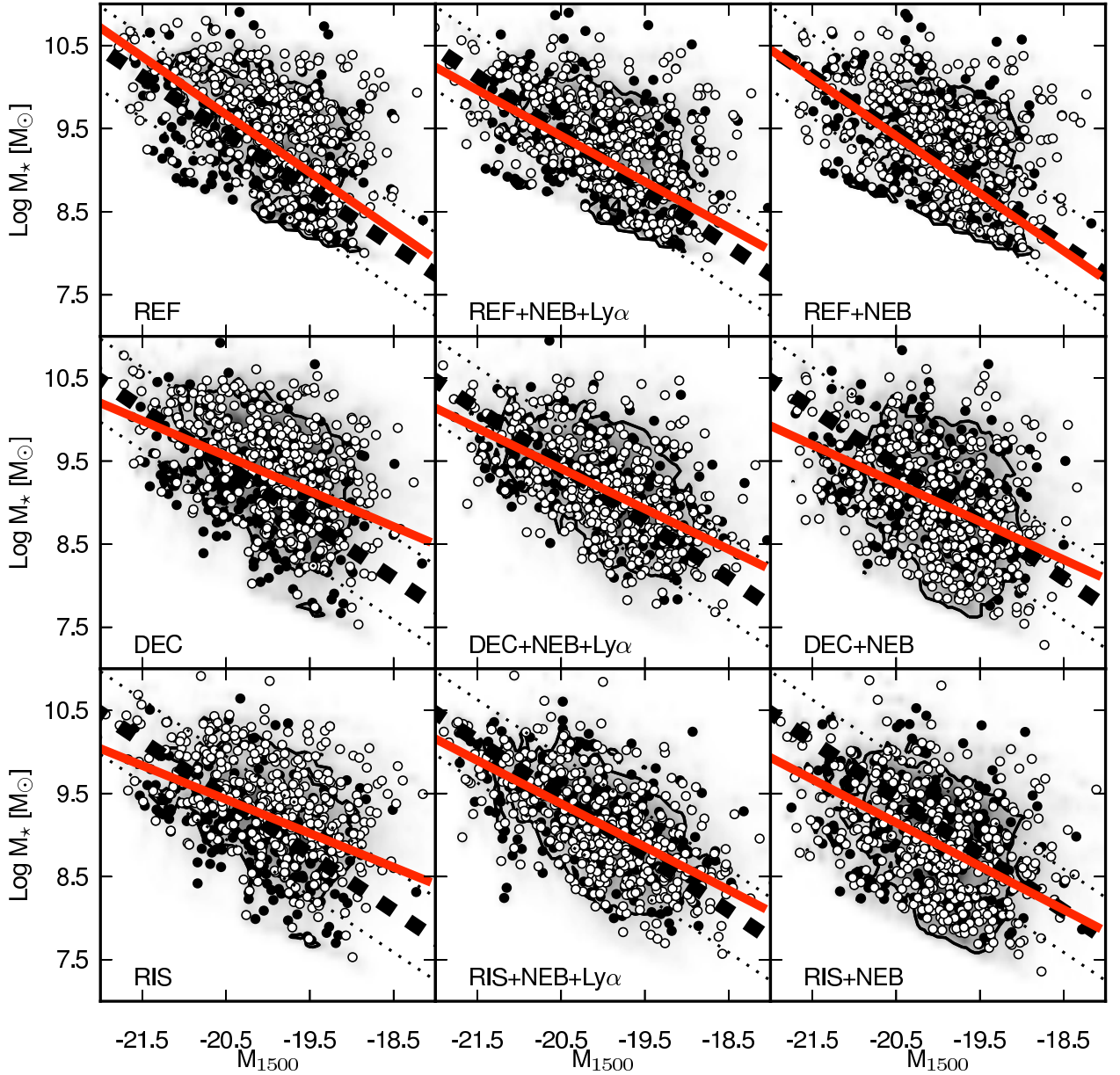}
     \caption{Composite probability distribution of $M_{1500}$ and $M_{\star}$ for all the models at $z \sim 4$. The black dashed line represents the M${\star}$-$M_{1500}$ trend found by \cite{gonzalezetal2011}, and the black dotted lines show a scatter of $\pm$0.5 dex. The solid red line shows a linear fit established by considering the whole composite probability distribution.
     The points overlaid show the median value properties for each object in the sample, black dots for ``weak" nebular emitters and white dots for ``strong" nebular emitters. The overlaid contour indicates the 68\% integrated probabilities on the ensemble properties measured from the centroid of the distribution.}
     \label{sm_muv_z4}
     \end{figure*}  
     
      \begin{figure*}[htbf]
     \centering
          \includegraphics[width=15cm,trim=3cm 0cm 3.5cm 1.5cm,clip=true]{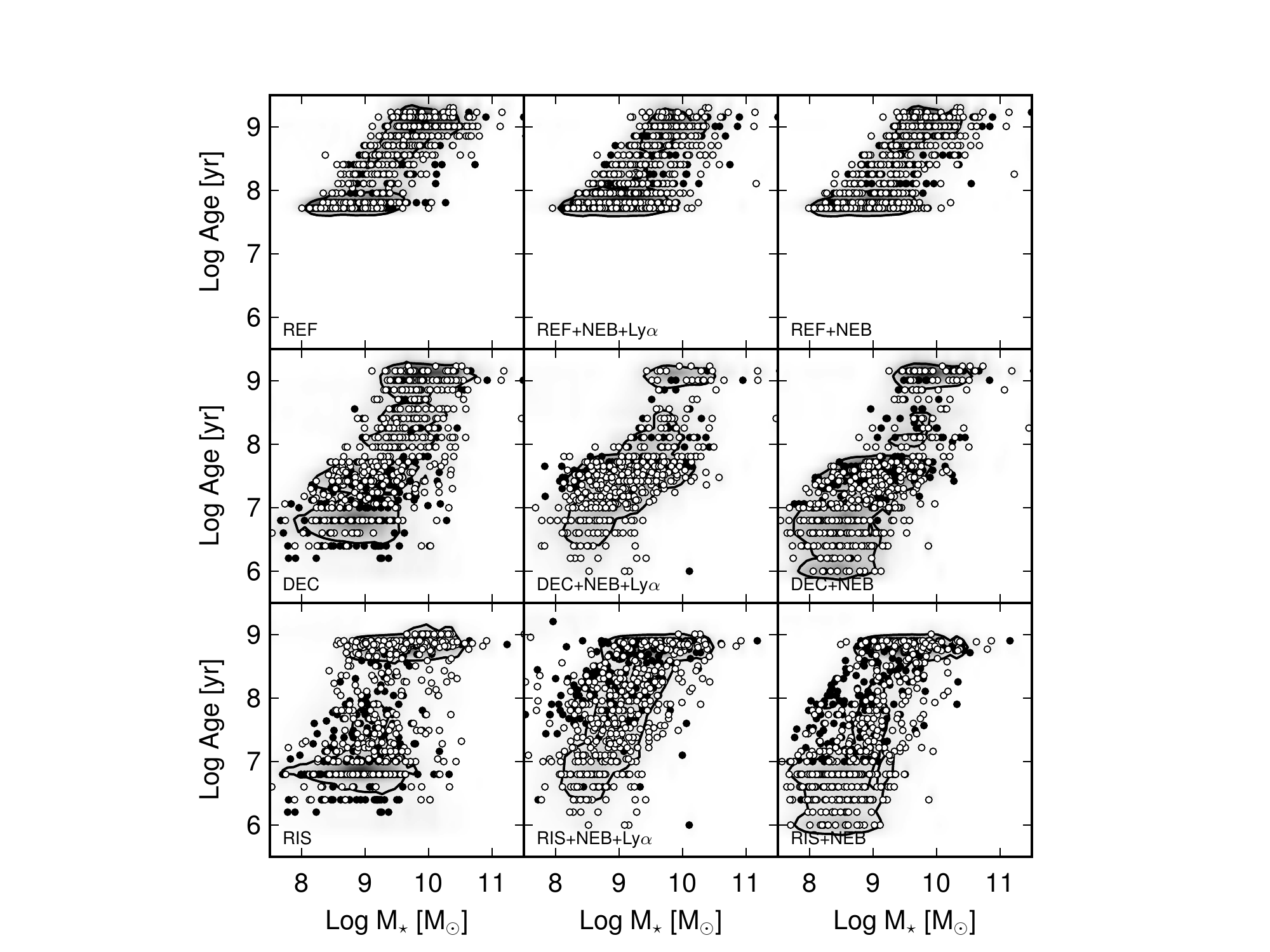}
     \caption{Composite probability distribution of $M_{\star}$ and age at $z\sim4$. The points overlaid show the median value properties for each object in the sample, black dots for ``weak" nebular emitters and white dots for ``strong" nebular emitters. The overlaid contour indicates the 68\% integrated probabilities on the ensemble properties measured from the centroid of the distribution.}
     \label{agesm}
     \end{figure*}

           \begin{figure*}[htbf]
     \centering
     \includegraphics[width=17cm,trim=1cm 0cm 2cm 1.5cm,clip=true]{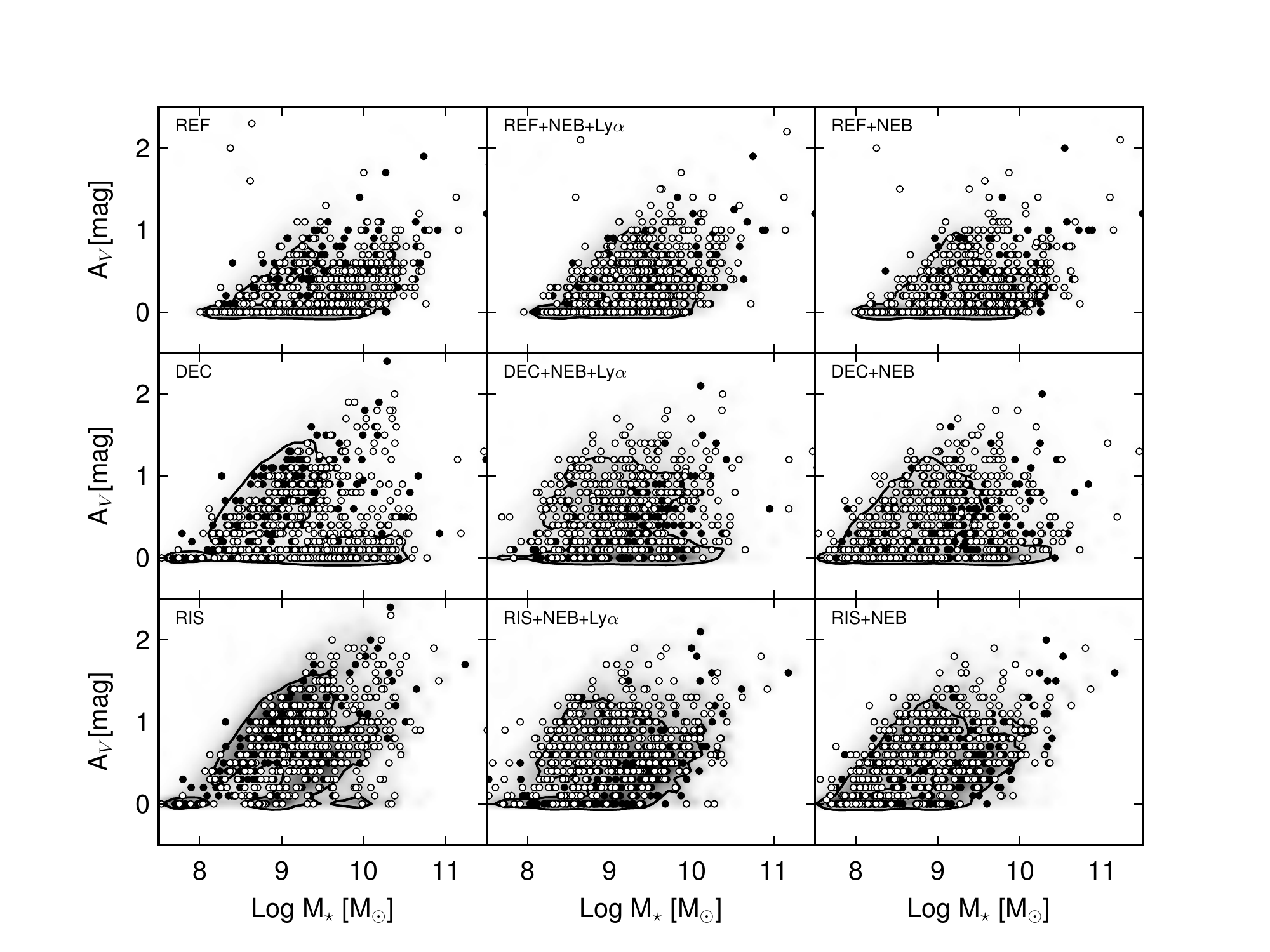}
	\caption{Same as Figure \ref{agesm} for $A_\mathrm{V}$ and $M_\star$.}
     \label{avsm}
     \end{figure*}

     \subsubsection{Stellar mass}
     \label{sm}
                  
     Stellar mass is generally considered as the most reliable parameter that is estimated by SED fitting, since relatively small differences are found when varying assumptions like the star formation
     history or dust extinction.
     \cite{finlatoretal2007} estimates that differences due to different assumptions on SFHs are typically not higher than 0.3 dex, and \cite{yabeetal2009} who adds effects of metallicity and extinction law, estimates differences to be not higher than $\sim0.6$ dex. 
Our stellar mass comparisons based on different model assumptions are shown in Fig.\ \ref{smcomp} and in Tables \ref{tabmagall_ref}--\ref{tabmagall_ris}.
Our results confirm the earlier results about the stellar mass dependence on the assumed SFH. Indeed, median stellar masses do not differ by more than $\sim0.3$ dex among
 different star formation histories from $z\sim3$ to $z\sim6$, when we do not account for nebular emission, even using metallicity as a free parameter.
With respect to models with constant star formation and without nebular emission (REF), all other models and options (+NEB or +NEB+\lya) lead systematically to
lower stellar masses with differences larger than the typical uncertainty of $\sim0.15$ dex found with the REF model.
The REF+NEB/+NEB+Ly$\alpha$ models (constant SF, nebular emission and age$>50\mathrm{Myr}$) lead to stellar mass differences of the same order as typical uncertainty, not larger than $\sim 0.2$ dex. 
For DEC and RIS models and the accounted nebular emission, we find stellar masses that are lower by $\sim0.4$ dex on average compared with REF model.

Differences in stellar mass found between ``strong" and ``weak" emitters are again consistent with an intrinsic difference between these two categories. When we consider nebular emission, stellar mass estimation of ``strong" emitters are more affected than for ``weak" emitters. Typically, stellar masses decrease by $\sim$ 0.4--0.9 (0.2) dex for ``strong" (``weak") nebular emitters in comparison to stellar mass estimates from the REF model. Furthermore, when we account for nebular emission, ``strong" emitters are slightly less massive than ``weak" emitters for any SFH ($\sim0.1$ dex), while both categories overall  span the same range of stellar mass and M$_{UV}$.

This is due to the less extended range of possible EW, since the impact of nebular emission on stellar mass estimation is correlated with this quantity. Indeed, as shown in Figure \ref{color_ewha}, declining SFH allows EW(H$\alpha$) variations up to 3 dex, while our REF+NEB+Ly$\alpha$/+NEB model shows possible variation of EW(H$\alpha$) by a factor $\sim5$. For these latter models, contribution of emission lines to broadband photometry is roughly similar for any objects, and thus, by considering nebular emission, does not introduce large variation on stellar mass estimation.

 In Figure \ref{sm_muv_z4}, we show the stellar mass--$M_{1500}$ relation  found for all our models at $z\sim4$. For constant star formation with or without nebular emission, we find, as expected, a relation in good agreement with the one found in \cite{gonzalezetal2011} within a scatter of $\pm$0.5 dex. Indeed, our REF model is based on assumptions similar to those of \cite{gonzalezetal2011}, except for the metallicity (We assume $Z=Z_\odot$ when they assume $Z=0.2Z_{\odot}$.) and for the minimum age (We assume 50 Myr when they assume 10 Myr.) Considering that the solar metallicity leads to $\sim0.06$ dex of increase in mass in comparison with 0.2 Z$_\odot$ and a higher minimal age
 increasing the lower bound in $M_\star$, the slight offset of our stellar mass-$M_{1500\AA}$ relation is easily explained. 
Although differences exist in the $M_{\star}$--$M_{1500}$ relation obtained among different model sets, our relation remains overall (within the scatter of $\pm 0.5$ dex) 
fairly similar to the relation derived in \cite{gonzalezetal2011}.

          A correlation between stellar mass and age is found for all the models, as shown in Figure \ref{agesm}. Since age estimation depends on the mass to light ratio, this relation is trivial if we describe the star formation history with a monotonic function. Indeed, luminosity is fixed by both measured fluxes and redshift, and the NIR data putting strong constrains on the stellar mass estimation. The stellar mass--age relation simply reflects the increase in the mass to light ratio with age.
As previously explained, different assumptions on the SFH lead to different trends between ``weak" and ``strong" nebular emitters when analysed with
SEDs that include nebular emission: 
for variable star formation histories (both rising or declining) ``weak" nebular emitters are found to be older and more massive on average than ``strong" nebular emitters.
In contrast,  physical properties of the two populations do not differ when constant star formation and an age$>50 \mathrm{Myr}$ is assumed (REF model).

\begin{figure}[htbf]
     \centering
     \includegraphics[width=9cm,trim=1cm 6.5cm 1.5cm 7cm,clip=true]{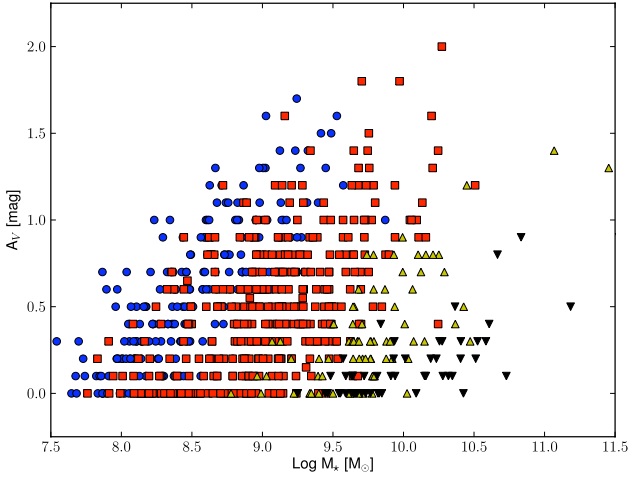}
	\caption{Relation between stellar mass and reddening for DEC+NEB model at $z\sim4$. Blue dots represent galaxies with median age $\leq10^7$ years; red squares: $10^7<$ age $\leq10^8$, yellow upward triangles: $10^8<$ age $\leq10^9$ and black downards triangles:  age $>10^9$ years.}
     \label{avsm2}
     \end{figure}

In Figure \ref{avsm}, we show the relation between the dust attenuation A$_\mathrm{V}$ and stellar mass
for a selected model set. For all models, a similar trend is found with the median $A_V$  which increases with galaxy mass,
and a wide range of attenuations that are allowed between 0 and $\sim1.5$ mag. Figure \ref{avsm2} helps to understand the correlation between stellar mass and dust reddening, since it shows that the dispersion comes mainly from age scatter. Indeed,  we find a clear trend of increasing extinction with increasing stellar mass for a given range of age.
     This trend has already been highlighted at lower redshifts \citep[eg.][]{buatetal2005,buatetal2008,burgarellaetal2007,daddietal2007,reddyetal2006,reddyetal2008,sawicki2012,dominguezetal2013},
     and it also seems to be observed at higher redshift \citep{yabeetal2009,bouwensetal2009,schaerer&debarros2010}. 
     This is clearly compatible with our results but with large uncertainties in both studies.
     While this trend could be explained by the age--reddening degeneracy, fixing age at a given value leads to no change.
The most likely natural explanation of this trend is probably that the dust attenuation is related to the stellar mass--metallicity relation \citep[cf.][]{tremontietal2004,erbetal2006,finlatoretal2007,maiolinoetal2008}.

          \begin{figure}[htbf]
     \centering
               \includegraphics[width=7.9cm,trim=7cm 6.75cm 7.5cm 7.5cm,clip=true]{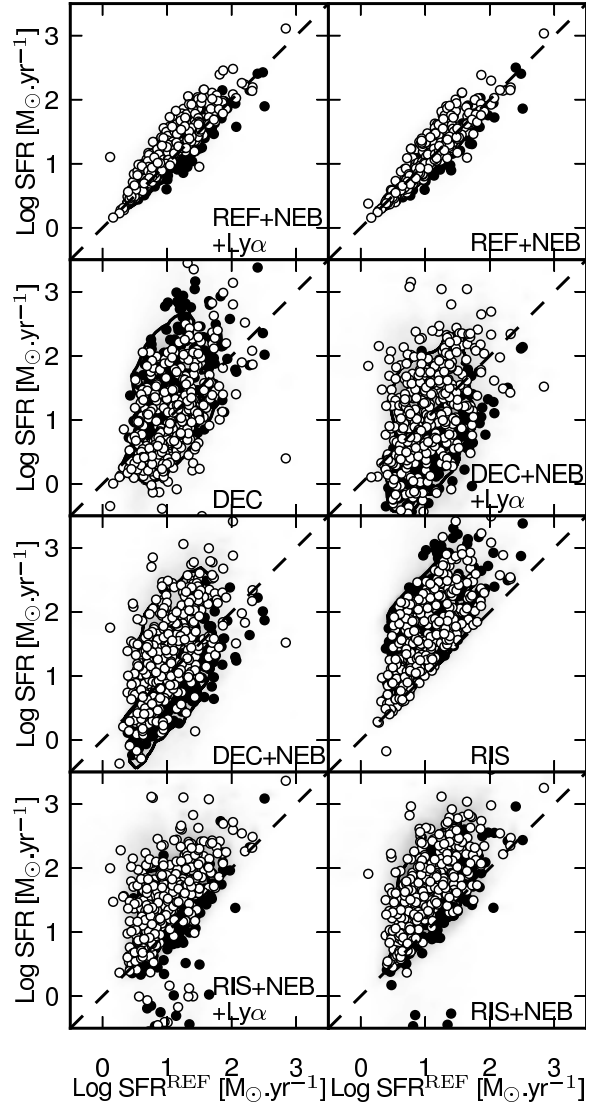}
     \caption{Same as Figure \ref{agecomp} for SFR.}
     \label{sfrcomp}
     \end{figure}   
          
     \subsubsection{Star formation rate}
     \label{sfr}
     
     \begin{figure}[htbf]
     \centering
               \includegraphics[width=7.9cm,trim=7cm 6.75cm 7.5cm 7.5cm,clip=true]{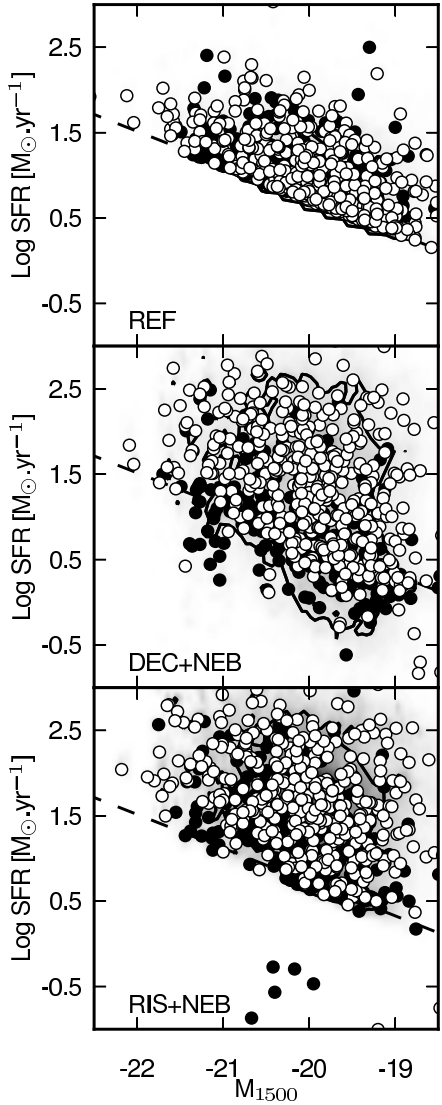}
     \caption{Composite probability distribution of $M_{1500}$ and SFR for the REF (top), DEC+NEB (centre) and RIS+NEB model (bottom) for the sample at $z \sim 4$ as determined for each galaxy from our 1000 Monte Carlo simulations. The  overlaid points show the median value properties for each object in the sample with black dots for ``weak" nebular emitters and white dots for ``strong" nebular emitters. The overlaid contours indicate the 68\% integrated probabilities on the ensemble properties measured from the centroid of the distribution. The dashed line represents the Kennicutt relation \citep{kennicutt1998}.}
     \label{sfr_muv}
     \end{figure}
     
     \begin{figure}[htbf]
     \centering
               \includegraphics[width=7.9cm,trim=6cm 6.75cm 6cm 7.5cm,clip=true]{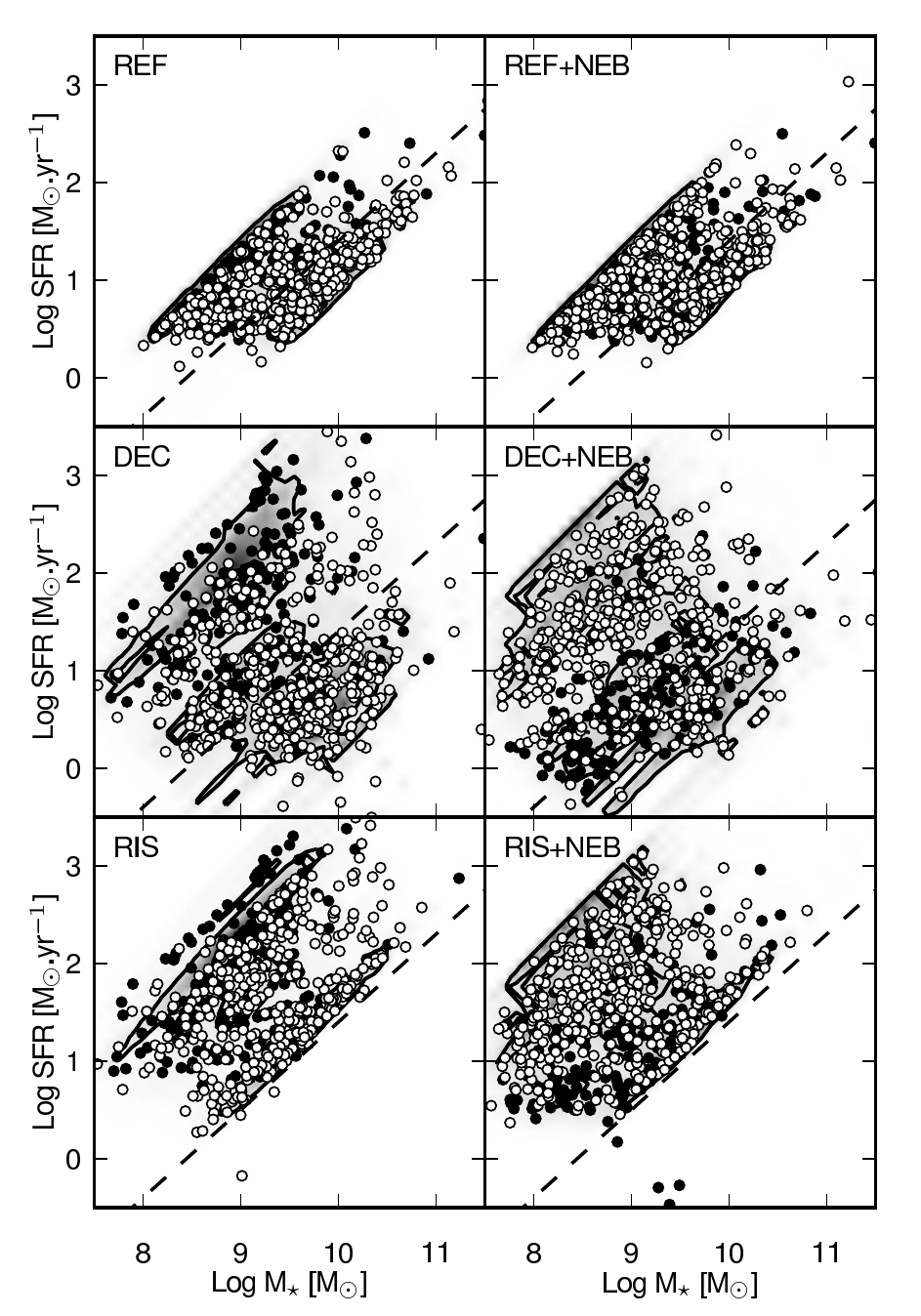}
     \caption{Same as Figure \ref{agesm} for $M_{\star}$ and SFR. The dashed line represents the SFR-$M_{\star}$ relation found in \cite{daddietal2007} at $z\sim2$.}
     \label{sfrsm}
     \end{figure}
     
     \begin{figure}[htbf]
     \centering
     \includegraphics[width=9cm,trim=5cm 0cm 5cm 1.5cm,clip=true]{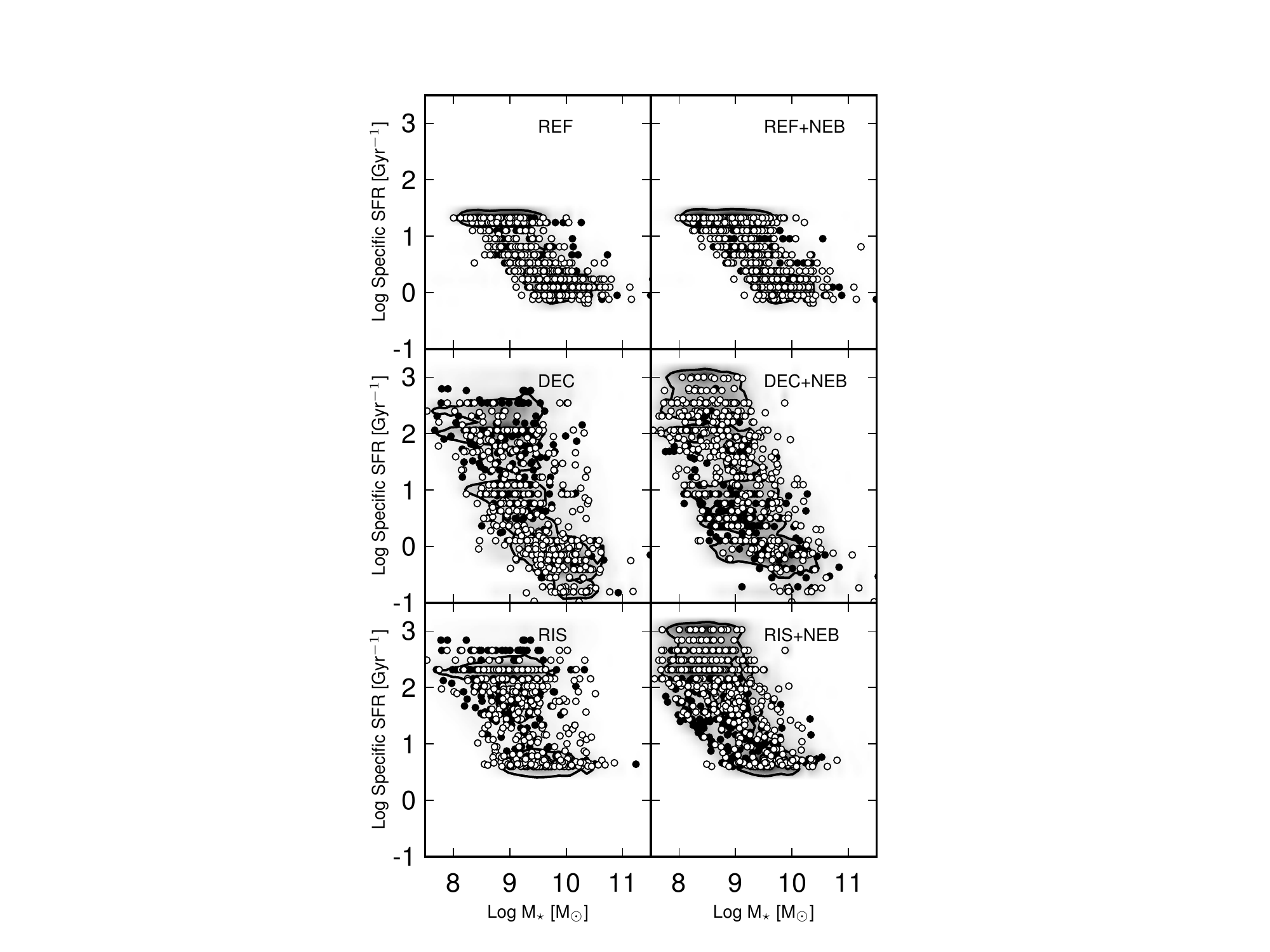}
     \caption{Same as Figure \ref{agesm} for specific SFR.}
     \label{ssfrsm}
     \end{figure}
     
     \begin{figure}[htbf]
     \centering
     \includegraphics[width=9cm,trim=0cm 0cm 0cm 0cm,clip=true]{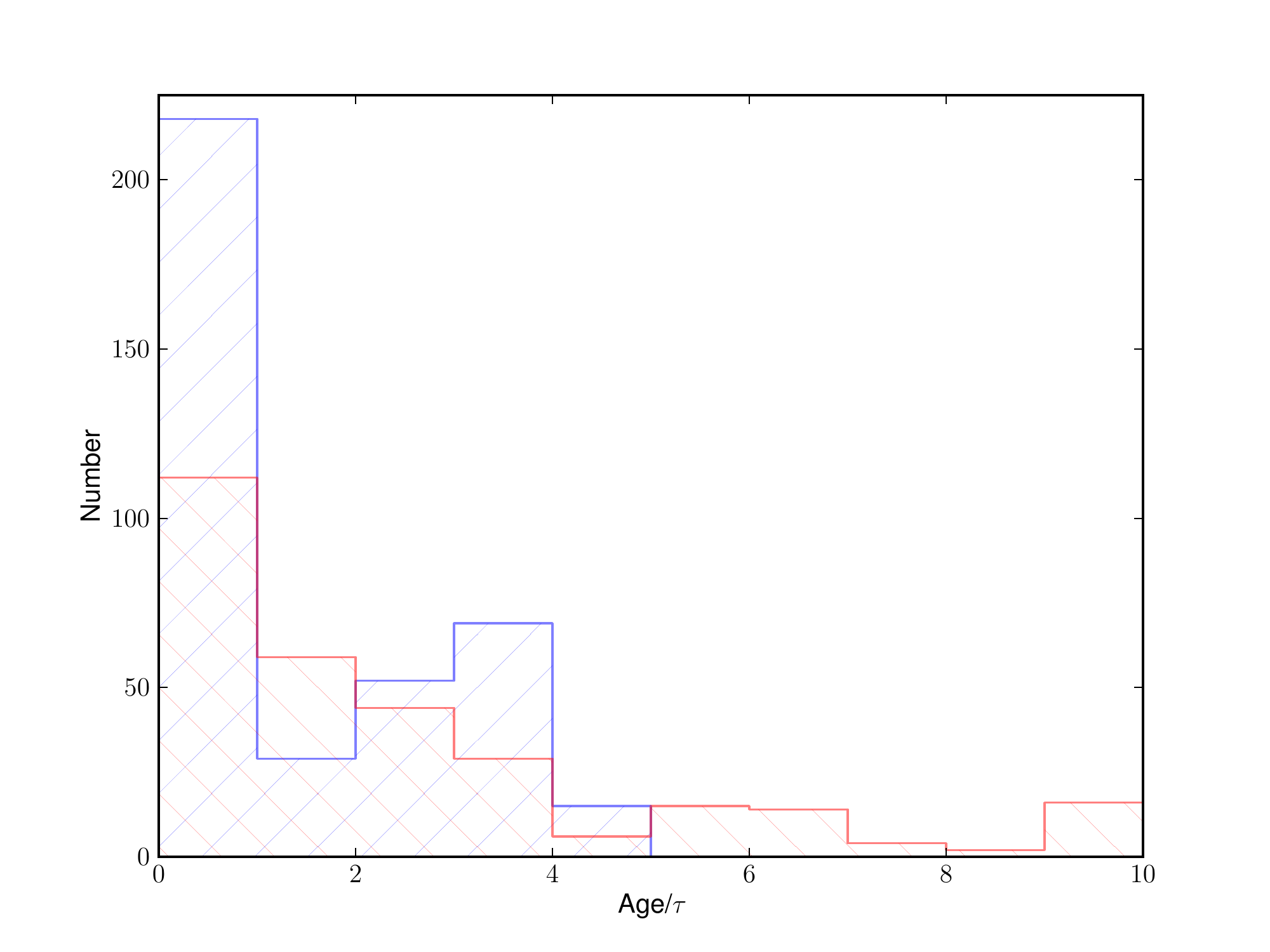}
     \caption{Distribution of the ratio $t/\tau$ at $z\sim4$, with objects with age$\le50$ Myr in blue, and objects with age$>50$ Myr in red.}
     \label{agetauratio}
     \end{figure}
     
     The star formation rate (defined here as the instantaneous value at the age $t$) depends strongly on the model assumptions,
as illustrated in Fig.\ \ref{sfrcomp}.
For the REF model (constant SF and age$>50 \mathrm{Myr}$), the inclusion of nebular emission leads to higher SFRs on average due to
the younger age, which requires a higher attenuation.
The largest differences (up to $\sim1$ dex) with respect to constant SFR models are obtained with DEC models. The reason for such differences is obviously due to the variations in the UV output with time and young ages ($< 50 \mathrm{Myr}$), which also imply a higher attenuation on average (cf. above).

An interesting feature of the declining SF histories
is that it also allows for SFRs which are lower than those derived using the canonical calibrations, assuming constant SFR.  Typically, galaxies with $t/\tau\lesssim1$ have higher SFR (up to 1 dex), if the timescale is short enough to diverge significantly from REF models. The range of $\tau$ values in our sample is given in Table \ref{tabmagall_dec}, while Figure \ref{agetauratio} shows the range of ratio $t/\tau$ for the DEC model at $z\sim4$.
Galaxies with $t/\tau\gtrsim2$ are  more quiescent \footnote{By quiescent galaxies we mean objects here with lower SFR and not galaxies with SFR$\sim0$.} and have lower SFR (up to 1 dex). Lastly, intermediate galaxies have SFRs that are consistent with results from the REF model.  This larger ``dynamic range''
may well be physical, as indicated by the existence of ``strong'' and ``weak'' nebular emitters, as we discuss below. 
Models with rising star formation histories lead to the highest SFRs, since their SED is always dominated by young stars.
This implies a narrower range of UV-to-optical fluxes, hence requiring a higher attenuation on average  than for 
other SF histories \citep[cf.\ above,][]{SP05}. For rising SF with ages above $\sim 10^8$ yr, all galaxies follow the canonical relation \citep{kennicutt1998}. Below this age, the SFR estimated by SED fitting is higher for the same reason as for decreasing SF (regardless of dust reddening).

We determine that nebular emission does not lead to any significant changes in the median SFR for the REF model, while  the median SFRs are lower for decreasing SF (mainly for +NEB+Ly$\alpha$) or equal (mainly for +NEB). For rising SF, the median SFRs are systematically lower, typically by a factor $\sim 2$. 
From $z\sim3$ to 5, this effect is due to the difference in dust reddening and age estimations, and to the contribution from Ly$\alpha$ line in the case of +NEB+Ly$\alpha$ due, which can decrease the UV flux necessary to fit the measured fluxes. 

  Relying on our previous identification of ``strong" nebular emitters and ``weak" nebular emitters (Section \ref{fitq}), we are able to check the consistency of star formation rate estimation. Since emission lines are produced by the strong UV flux from OB stars in H~{\sc II} regions, we should find a higher SFR for ``strong" nebular emitters in comparison with ``weak" nebular emitters, for a given stellar mass. As shown in Figure \ref{sfrcomp}, the REF model does not reproduce such a separation between ``strong" and ``weak" emitters, since SFRs are roughly similar for both populations or even showing an opposite trend to what is expected. The inclusion of nebular emission in the REF model does not lead to a significant difference on median SFR estimations between the two populations.
Other models without nebular emission (DEC and RIS) do not provide a better result than the REF model, since they also provide an opposite trend to what is expected -- a higher median SFR at a given stellar mass for ``weak" nebular emitters. 
On the other hand, the two populations are naturally separated in terms of SFR,
as shown in Fig.\ \ref{sfrcomp} when nebular emission is included. This reveals the ``strong'' nebular emitters as objects with a strong ongoing star formation episode, and ``weak'' emitters
as a more quiescent population. The capacity of both the declining and rising star formation histories to distinguish these populations
can be easily understood, since young ages ($<50$ Myr) lead to deviate from the canonical UV to SFR relation \citep{reddyetal2012a}.
Separating the two LBG populations, we find that the median SFR is higher by $\sim0.6$ dex (up to 0.75) for the ``strong" nebular emitters from $z\sim3$ to $z\sim5$ compared
to the ``weak'' emitters, although the typical uncertainty is relatively large ($\sim0.5$ dex).
Since the stellar mass is not significantly different  between ``weak" and ``strong" emitters, the specific SFR (SFR/\mstar) of the ``strong'' is higher than that 
of the ``weak".
At $z\sim6$, only DEC/RIS+NEB+Ly$\alpha$ models lead to the expected trend (separation between ``weak" and ``strong" emitters in term of SFR-M$_{\star}$ relation), showing that the presence of just one emission line, such as Ly$\alpha$ can have a large impact on parameter estimation \citep{schaereretal2011}.
          
We now explore the SFR-$M_{1500}$ relation, as illustrated in Figure \ref{sfr_muv} for the $z\sim4$ sample using three different models sets.
     For the constant star formation (REF) model, the SFR-$M_{1500}$ match with the Kennicutt calibration \citep{kennicutt1998}, once accounting for the effect of dust extinction. As explained in \cite{kennicutt1998}, the relation is valid for galaxies with continuous star formation over time scales of $10^8$ years. The SFR/L$_{1500}$ will be significantly higher in bursty galaxies with a decreasing SF and a short timescale, or for simply galaxies younger than $10^8$ years \citep[cf.][]{reddyetal2012a,schaereretal2013}.
     This leads to significantly higher SFR than those given by the Kennicutt relation, regardless of dust reddening. The SFRs found cover a large range of possible values, which strongly depend on both SFH and whether they fit or not with nebular emission (see Table \ref{tabmagall_ref}, \ref{tabmagall_dec} and \ref{tabmagall_ris}). Again, only rising and declining SF with nebular emission are able to separate the two populations previously identified as ``strong" and ``weak" nebular emitters, since they 
naturally separate these groups into higher and lower  SFR galaxies at a given $M_{\rm UV}$.

The SFR as a function $M_{\star}$ is plotted in Figure \ref{sfrsm} for the $z\sim4$ sample. The figure shows that we find a relation compatible to that found at $z\sim2$ by \cite{daddietal2007} with a relatively small dispersion, and no significant difference if we consider nebular emission for constant star formation (REF model). For decreasing SF, our results remain compatible with the relation at $z\sim2$ but with a very large dispersion, which can be explained by the large range of timescale. For rising SF, the star formation rates are systematically higher than those expected from the SFR-mass relation derived at $z\sim2$. For rising and decreasing star formation, we note that the galaxies seem to be separated in two groups: actively star forming galaxies, showing higher SFRs than expected from the \cite{daddietal2007} relation, and a group of more quiescent galaxies, which are compatible with this relation. These groups correspond again  to those previously identified as ``weak" and ``strong" nebular emitters.

 The specific star formation rate ($\mathrm{sSFR}=\mathrm{SFR}/\mathrm{M}_{\star}$) is plotted for $z\sim4$ as function of stellar mass in Figure \ref{ssfrsm}.
For all models, it decreases on average with increasing $M_{\star}$ and with decreasing redshift. The relation is fairly similar among all models but decreasing and rising star formation histories
lead to higher sSFR values by more than 1 dex. This increase is significant compared to the typical errors, which range from $\sim0.2$ dex for models with constant SF to $\sim0.6$ for decreasing and rising SFHs. 
For decreasing and rising SF, the presence of Ly$\alpha$ leads to a slightly lower SFR and a higher $M_{\star}$ (except at  $z\sim6$ where this trend is reversed), which explains the lower sSFR when compared to models that assume no Ly$\alpha$ emission. By comparing declining star formation histories to others, we find that they yield lower sSFR for some galaxies. With both the DEC+NEB(+Ly$\alpha$) and RIS+NEB(+Ly$\alpha$) models, ``strong" nebular emitters have a slightly lower median $M_{\star}$, and a higher SFR than the ``weak" nebular emitters.
In other words, we find that ``strong" emitters show a higher sSFR than ``weak" nebular emitters at a given mass.

     \subsubsection{Metallicity}

    Metallicity is the least constrained parameter by our SED fits. For individual objects the 68\% confidence interval for all samples basically covers  the three metallicity values (0.02, 0.2, 1 Z$_{\odot}$)
     used here. Considering the median metallicity, there is a trend for  RIS+NEB(+Ly$\alpha$) and DEC+NEB(+Ly$\alpha$) models to show an increase in the metallicity with galaxy mass. 
However, the uncertainties are too large to provide firm conclusions. This is consistent with the well known fact that metallicity is poorly constrained by SED fitting.
          
     \begin{figure}[htbf]
     \centering
     \includegraphics[width=5cm,trim=8.5cm 6.75cm 9cm 7.5cm,clip=true]{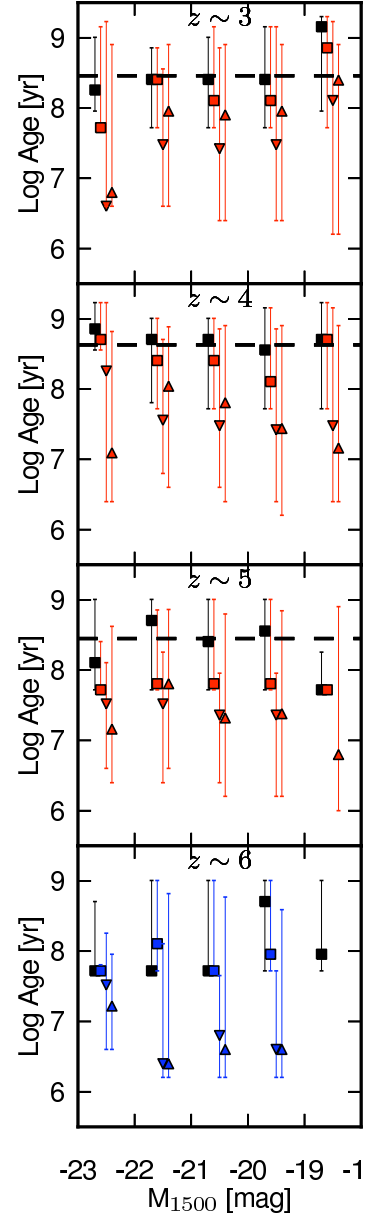}
\caption{Median ages are given in bin of absolute magnitude at 1500$\AA$ (no correction for dust) from $z\sim3$ to 6 for four models. Squares illustrate constant SFR with downward (upward) triangles decreasing (rising) SFHs. Black symbols stand for models without nebular emission; red symbols for models that include nebular emission (NEB), and blue symbols for NEB+\lya.
      The error bars correspond to the 68\% confidence limits of the probability distribution in each bin. Dashed lines show the amount of time spanned, since the previous redshift bin.}
      \label{muvage_all}
     \end{figure}
     
     \subsubsection{Physical properties: summary}
     
     Accounting for nebular emission in the SED fitting decreases the estimated age in most cases\footnote{The exception are the 
     ``weak emitters", making approximately 1/3 of the LBG population.}, since some strong lines can mimic a Balmer break, increase dust attenuation, and decrease stellar mass, as already found earlier \citep{schaerer&debarros2009,schaerer&debarros2010}. The extent of this impact strongly depends on assumptions on both star formation history and the allowed age range. An increasing SFH produces strong lines at any age, while constant and declining SFHs lead to a decreasing impact of emission lines with age (Figure \ref{color_ewha}). We find  that young ages ($<50$ Myr) are required to reproduce the most extreme observed (3.6-4.5)$\mu$m colours of Figure \ref{colortest}, which corresponds to 
     strong H$\alpha$ emission at $z$  $\epsilon \left[3.8,5\right]$ \citep{shimetal2011}.  Only DEC+NEB and RIS+NEB models are able to reproduce these colours.

From the best fit distribution (Sect.  \ref{s_sfh}), we identified two LBG populations, one which seems to show stronger nebular emission than the other, defined as ``strong" and ``weak" emitters (respectively). Using DEC+NEB and RIS+NEB models, this latter population is slightly more massive, less attenuated, and older than ``strong" emitters and also exhibits lower SFR for a given stellar mass, where the parameters are compatible with a more evolved population. While differences between these two populations are less important with REF+NEB model, it is still possible to statistically distinguish these two populations (e.g. Sect. \ref{av}).
     
     Differences in attenuation naturally lead  to a difference in SFR estimation, and a significant fraction of ``strong" emitters shows extremely young ages ($<50$ Myr), which leads to deviation from the SFR-UV standard relation \citep{kennicutt1998,madauetal1998}.
          
   \subsection{Evolution of the physical properties from $z\sim3$ to $z\sim6$}
   \label{evoz}
   
   We now examine the evolution of the median physical properties with redshift and discuss several implications.
   To allow a meaningful comparison that avoids variations of the completeness limit and the galaxy luminosity function with $z$, we make comparisons in bins of absolute UV magnitude. 
   To discuss the physical parameters derived from the models
   that include nebular emission, we choose the models with the \lya\ flux set to zero (+NEB option) for $z \sim$ 3--5, and the NEB+\lya\ option
   with maximum \lya\ emission for $z \sim 6$. Within the options discussed in this paper, this choice best describes  the 
   trend of increasing strength of \lya\ with redshift as observed from spectroscopic surveys \citep[eg.][]{andoetal2006,ouchietal2008,starketal2010},
   other studies \citep{blancetal2011,hayesetal2011}, and our models that allow for varying \lya\ strength \citep{schaereretal2011}.
   
In the following, we notice that there is a very small number of objects in the brightest bin at each redshift, while we plot parameters as a function of $M_{1500}$, in bins of magnitude. The faintest bin falls beyond the completeness limit. It is therefore more appropriate to mainly consider  the three intermediate bins to examine trends of the physical properties
with UV magnitude.

    \begin{figure}[htbf]
     \centering
          \includegraphics[width=5cm,trim=8.5cm 6.75cm 9cm 7.5cm,clip=true]{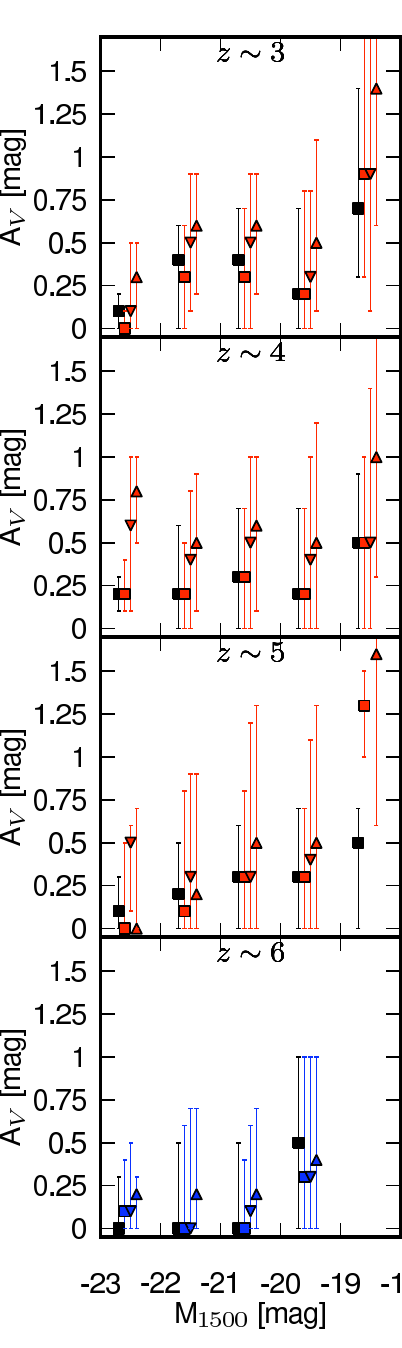}
     \caption{Same as Figure \ref{muvage_all} for $A_{\mathrm{V}}$.}
     \label{muvav_all}
   \end{figure}
      
   \subsubsection{Age}

The evolution of median ages with $z$ is shown in Figure \ref{muvage_all}. The median age increases with decreasing redshift for the four models, 
typically by one order of magnitude, with no significant trend with $M_{\mathrm{UV}}$.
Although declining and rising SFs lead to very similar ages at the highest redshift, they start to diverge with decreasing redshift. 
We found that the age differences are essentially driven by the group of quiescent galaxies with ``weak" nebular emission. This is explained  by both nebular emission and the intrinsically aging population with decreasing redshift. Quiescent galaxies with declining star formation prefer systematically short timescales $\tau$ (median at 10 Myr at each redshift) because the SED fitting minimize the contribution of nebular emission, which is achieved with ages significantly older than $\tau$ (at least $t/\tau \ga 2$), and produce a significant Balmer break.
On the other hand, the only way to minimize the contribution of nebular emission for rising SF (i.e.\ minimize the equivalent widths of lines) is by choosing relatively old ages.
Therefore, fitting SEDs that are both devoid of strong emission lines and with a strong Balmer break requires older ages than for DEC models.

Except for the model with constant SFR and no nebular emission (REF), Figure \ref{muvage_all} leads one to conclude that the bulk of the LBG populations at each redshift are dominated by young galaxies that were not present or visible at the previous redshift. For a constant (REF+NEB) and rising  (RIS+NEB) star formation, these results imply that an important number of galaxies formed in the interval from one redshift to another, in $\sim$ 300--400 Myr. For decreasing star formation, we can interpret this result with a scenario with episodes of active star formation which are followed by more quiescent episodes, because of the presence of actively star-forming and quiescent galaxies. In this scenario, age becomes a poorly constrained parameter since the underlying older stellar population, which is formed in previous episodes of strong activity, can be dominated by a newly emerged population.
Obviously, the {\em absolute} ages also depend on the SF timescale. For example, fixing the timescale to $\tau=100$ (300) Myr
for the DEC+NEB models leads to no significant differences on the median values of the stellar mass, SFR, or reddening, whereas the median age increases by a factor $\sim2$ (3)
for both active and quiescent galaxies.
     
   \subsubsection{Reddening}

  As shown in Fig.\ \ref{muvav_all}, the median dust attenuation decreases with increasing redshift for all the four models considered here. Inclusion of nebular emission for a constant star formation does not provide a significant difference, while decreasing and rising star formation histories lead  to +0.1--0.2 and +0.3--0.4 mag respectively in reddening compared to constant SFR. Many studies \citep[eg.][]{bouwensetal2009,castellanoetal2012} have shown that there is a trend of decreasing $\beta$ with increasing absolute UV magnitude, which corresponds to a similar trend on $M_{1500}-A_V$, which is not seen in Figure \ref{muvav_all}. This trend should be found for the actual $\beta-A_V$ relation, since both the classical relation \citep{bouwensetal2009} and the relation found at $z\sim4$ in this study (Section \ref{red_uvslope}) cannot change the sign of the slope. We attempt to retrieve this trend with our different models by fitting the entire composite probability distribution function A$_V$-M$_{1500}$. However, no model among the nine considered here has been able to reproduce the expected slope, while the $\beta$-M$_{1500}$ relation obtained for our $z\sim4$ sample is consistent with previous studies \citep[eg.][]{bouwensetal2009}. Important degeneracies and a high dispersion for each individual object can explain this effect.
  
As already mentioned in Sect.\ \ref{red_uvslope}, the median UV attenuation obtained from our models with rising and declining star formation histories
and nebular emission is higher than predicted from conventional methods that rely on the average UV slope. For all models, reddening evolves similarly with redshift: it increases with decreasing redshift. 

Unfortunately, while photometry in the filters commonly used to determine $\beta$ \citep{bouwensetal2012} is available for the whole sample at $z\sim4$, this is not the case at $z\sim5$ and 6, 
     where less than 20\% of the necessary fluxes are available for the $V$- and $i$-drop samples. 
     To circumvent this shortcoming, we use the fluxes predicted by the best fit model in these filters.
     At $z\sim5$,  we also find a deviation from the classical $A_V$--$\beta$ relation by using the equations of \cite{bouwensetal2012} to derive $\beta$, although it is less important than at $z\sim4$, which is an intermediate relation between the classical relation and the one we found at $z\sim4$. The result is consistent with equations \ref{betaavnoneb}, \ref{betaavneb} and \ref{betaavwlya} at $z\sim6$, but does not allow us to distinguish among them since they are very similar at low $A_{\mathrm{V}}$, and there are only few objects at $z\sim6$ with a significantly high extinction. We have to notice that we use a $J$-band from VLT/ISAAC, while \cite{bouwensetal2012} use data from Hubble/WFC3, which can lead to some differences
     for the $z\sim6$ objects.
        
     \begin{figure}[htbf]
     \centering
               \includegraphics[width=5cm,trim=8.5cm 6.75cm 9cm 7.5cm,clip=true]{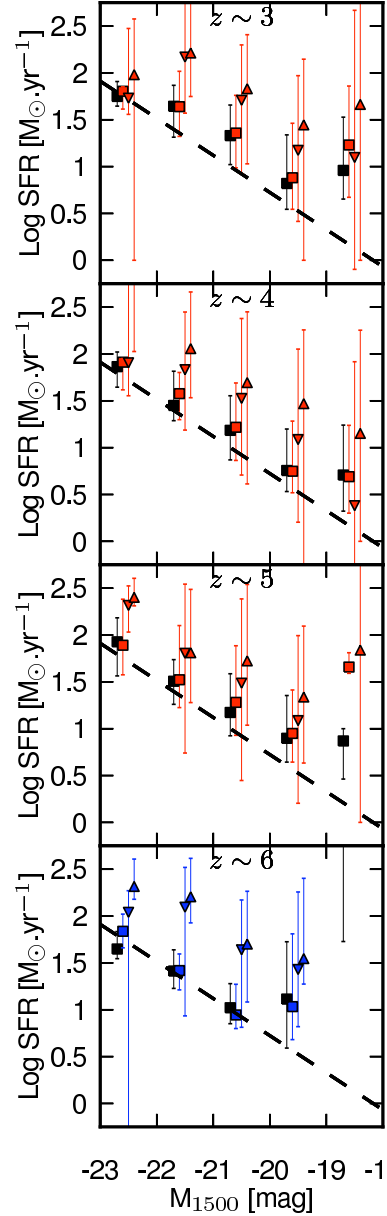}
          \caption{Same as Figure \ref{muvage_all} for SFR and for active  galaxies (i.e. ``strong" emitters). The dashed line shows the Kennicutt relation \citep{kennicutt1998}.}
     \label{muvsfr_neb}
     \end{figure}
     
     \begin{figure}[htbf]
     \centering
               \includegraphics[width=5cm,trim=8.5cm 6.75cm 9cm 7.5cm,clip=true]{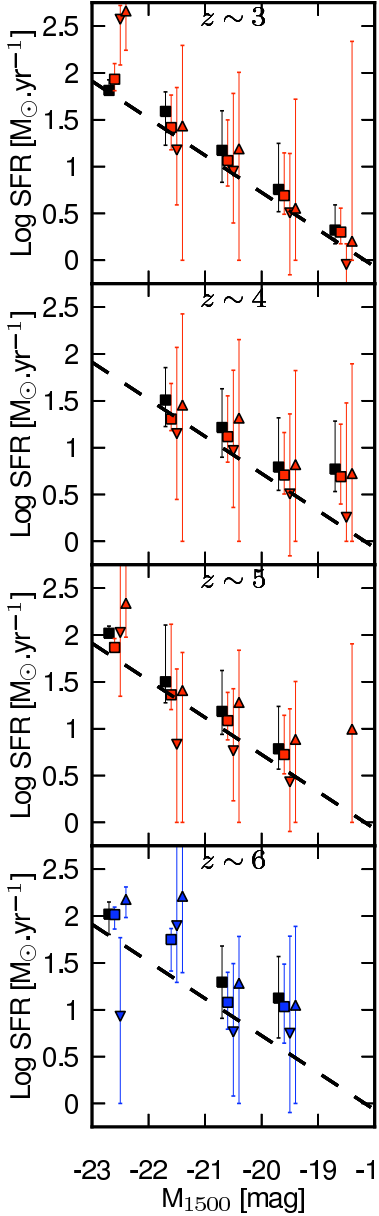}
     \caption{Same as Figure \ref{muvage_all} for SFR and for quiescent galaxies (i.e. ``weak" emitters). The dashed line shows the Kennicutt relation.}
    \label{muvsfr_noneb}
     \end{figure}   
     
      \subsubsection{Star-formation rate}
           
           The behaviour of the instantaneous SFR as a function of the UV magnitude and for all redshifts is shown in Figs.\ \ref{muvsfr_neb} and \ref{muvsfr_noneb},
where we have separated the sample in the two groups, LBGs with ``strong" and ``weak" nebular emission (also referred to as ``active" and ``quiescent" galaxies
here), as identified earlier. As previously stated, only DEC+NEB and RIS+NEB models are able to provide a consistent separation between these two populations in terms of SFR-M$_{\star}$ relation (see Section \ref{sfr}). While there is a large dispersion for individual objects, the median sSFR of ``strong" emitters at each redshift is larger by $\geq1$ dex ($\geq2$ dex) for DEC+NEB (RIS+NEB) models compared to REF models, whereas smaller differences( $<1$ or $<2$ dex, respectively) are found for ``weak" emitters.
For each group, we do not observe a significant change of the SFR with redshifts. The SFR--$M_{\mathrm{UV}}$ relation does not evolve with redshift, since the median age increases with decreasing redshift, while median reddening decreases, so the two effects cancel out on average.

Figure \ref{muvsfr_neb} and \ref{muvsfr_noneb}  clearly show the difference between ``active" and ``weak" emitters in terms of SFR: ``weak" emitters follow the Kennicutt relation for any model, while ``strong" emitters deviate from this relation for DEC+NEB and RIS+NEB models. This can be explained by a higher dust attenuation (since we use uncorrected M$_{1500}$) and/or by younger age ($<50$ Myr). While RIS+NEB leads to significant higher dust attenuation, differences in dust attenuation between models with constant SF and declining SF are not large (Sect. \ref{av}). Therefore, only young ages can explain the observed discrepancy between SFR(SED) and the Kennicutt relation. As shown in Section \ref{fitq}, these young ages are correlated with objects showing strong H$\alpha$ emission at $z\sim4$.

In the next section, we discuss some implications from these results on the star formation history.

     \begin{figure}[htbf]
     \centering
               \includegraphics[width=5cm,trim=8.5cm 6.75cm 9cm 7.5cm,clip=true]{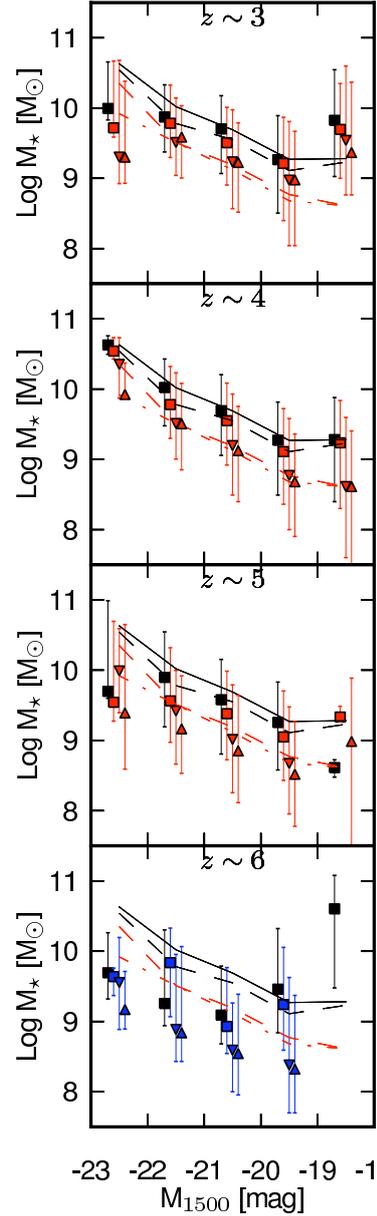}
     \caption{Same as Figure \ref{muvage_all} for stellar mass. In each panel, we plot the $M_{\star}$-$M_{1500}$ found at $z\sim4$ for more convenient comparison with other studies \citep{starketal2009,leeetal2011}. Black solid line: relationship for REF model, dashed black line: REF+NEB model, dashed red line: DEC+NEB, and dashed-dotted red line: RIS+NEB.}
     \label{muvsm_all}
     \end{figure}

   \subsubsection{Stellar mass and implications for star formation history}
   \label{sm_sfh}

\cite{starketal2009} suggest the exploration of the star formation history of LBGs through the evolution of the stellar mass in bins of UV magnitude, since the (observed/uncorrected) UV magnitude is the most reliable tool at a high redshift to track star formation, and since stellar mass is the most reliable parameter, because it is the least dependent on different assumptions \citep[SFH, metallicity, dust,][]{finlatoretal2007,yabeetal2009}. While this study challenges this last assumption (Section \ref{sm}), the $M_\star$--$M_{1500}$ relation remains a useful tool to test/falsify some scenarios, specifically constant star formation history. 
   As explained in \cite{starketal2009}, if the bulk of galaxies formed stars with a constant star formation rate over sufficiently long time, we would expect to see a systematic increase in the normalisation of the $M_{\star}$--$M_{1500}$ relation with cosmic time, and so with decreasing redshift. If no such change is observed, the SFR cannot be constant, at least not for more than $\Delta t \ga$ 300--400 Myr, which is the time corresponding
   to $\Delta z \approx 1$ between our samples. In other words, a non-evolution of the $M_\star$--$M_{1500}$ relation with redshift would require other star formation histories or the repeated emergence of new galaxies dominating the LBG population at each redshift.

The predicted relation between mass and UV magnitude, as obtained from all model sets and for all redshifts, is shown in Fig.\  \ref{muvsm_all}.
Although the absolute stellar masses depend on the model assumptions, all models yield essentially no evolution of the $M_\star$--$M_{1500}$ relation  
between $z \sim$ 5 to 3, but some shift between $z \sim$ 6 and 5.
In particular, models from redshift 5 to 3 with constant star formation (both with or without nebular emission, i.e.\ REF+NEB or REF) do not yield an evolution of stellar mass, 
which is inconsistent with the assumption
of constant SFR. By considering the median ages obtained from the fits (Figure \ref{muvage_all}), it seems difficult to reconcile the picture of constant star formation with the derived parameters. 
In contrast, the evolution of $M_{\star}$--$M_{1500}$ from $z\sim6$ to 5 for the REF model is fully compatible with what is expected from a constant star formation, while we still do not see the expected evolution for the same model
with nebular emission. However, median ages for this latter case allow us to assume that young LBGs dominate samples at each redshift.
We conclude that constant SF over long timescales is not compatible with the data.
   
   Assuming a constant star formation and no strong evolution of the dust contain or its geometric distribution with cosmic time, we expect galaxies to evolve at constant $M_{UV}$. Assuming this SFH, the evolution of UV luminosity function should be mainly due to the emergence of new galaxies (if we do not consider the effects of possible mergers). For $M_{UV}=-20.5$,  the number of galaxies per Mpc$^3$ increases by a factor $\sim3$ between $z\sim6$ and $z\sim4$ \citep{bouwensetal2007}, which means that at least one third of the LBGs seen at $z\sim4$ must have LBGs at $z\sim6$ as progenitors at this magnitude. Indeed, our sample shows one third of the LBGs having an age $>700$ Myr at $z\sim4$ and for $M_{UV}\simeq-20.5$ for REF+NEB model, which correspond to the time spans between $z\sim6$ to $z\sim4$. We predict the median stellar mass of these galaxies at $z\sim4$, using our parameter estimation at $z\sim6$: the median stellar mass should be $\simeq10^{10.11}$ \msun, while our sample of galaxies at $z\sim4$ that are sufficently older to be the descendants of $z\sim6$ galaxies have a median stellar mass $\simeq10^{9.55}$. This difference of $\sim0.6$ dex is three times larger than the typical uncertainty on stellar mass estimation with REF+NEB model, leading us to conclude that despite a better consistency of the data with constant SF, when accounting for nebular emission, there is still significant discrepancies.
      
      \begin{figure}[htb]
   \centering
        \includegraphics[width=9cm,trim=6cm 6.5cm 6cm 7.75cm,clip=true]{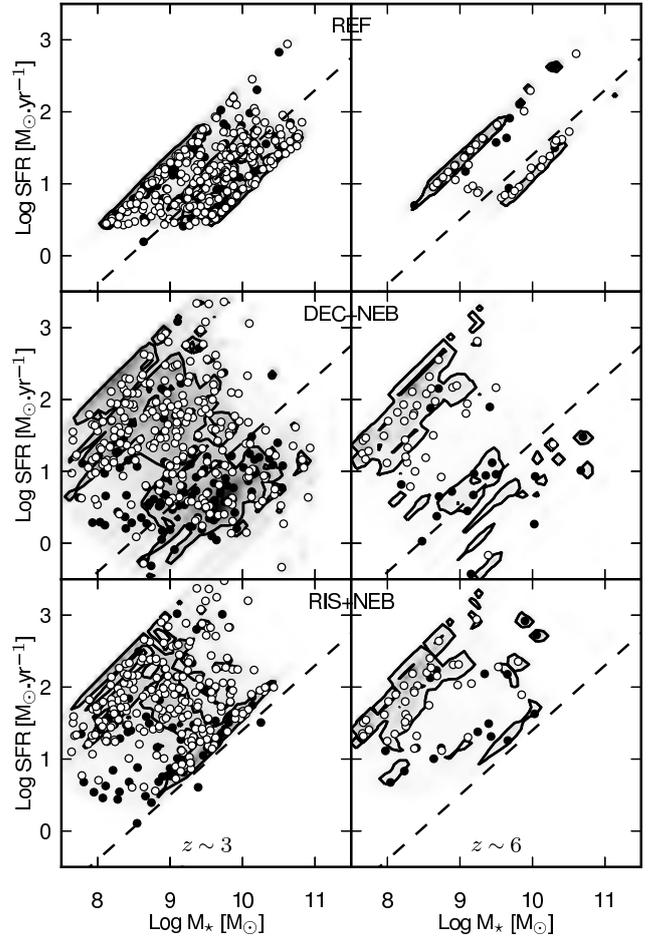}
   \caption{Same as Figure \ref{sfrsm} for three models (Top: REF, centre: DEC+NEB, bottom: RIS+NEB) at $z\sim3$ (Left) and $z\sim6$ (Right). For $z\sim6$, models with declining and rising SF include Ly$\alpha$. }
   \label{sfrsm_evol}
   \end{figure}      

 What about the rising star formation history adopted here? It is straightforward to compute how the stellar mass and SFR of each galaxy evolves with time, both forward and
 backward. Starting from the best fit of the 1000 MC simulations for each object, we determine how they would evolve in the future. Taking, for example LBGs at $z\sim5$, which have a median stellar mass of $10^{8.9}$ \msun\ and a median SFR $\sim30$ \msunyr, we find that they would have a median stellar mass of $10^{11.4}$ \msun\ and median SFR $\sim 2000$ \msunyr\ 
(corresponding to $M_{\mathrm{UV}}\simeq-26$ assuming the Kennicutt relation) after $\sim400$ Myr at $z\sim4$.
Obviously, such extreme objects are not seen, certainly not in large numbers! To hide most of them from the LBG selection, a strong increase in dust attenuation with time would
be necessary, and in this case, it should be fairly easy to find this large populations as IR/sub-mm galaxies. More likely, this discrepancy implies that the 
average rising star formation history adopted from the simulations of  \cite{finlatoretal2011} are not representative for typical $z \la 6$ galaxies:
that is, the growth does not continue significantly beyond the current ages of the LBGs, or the growth is too fast, or that a combination
of these arguments apply\footnote{Rising SFHs with varying timescales have been explored in \cite{schaereretal2013}.}.
In any case, the study of \cite{papovichetal2011} shows that  to have an observed number counts at $z \sim$ 3--8 compatible with a cosmologically average rising star formation history,
the growth of the SFR has to be fairly slow. Their SFR$(t) \propto (t/\tau)^\alpha$ with $\alpha=1.7\pm0.2$ and $\tau=180\pm40$ Myr 
corresponds to an increase in the SFR by a factor $\sim 2$ per $\Delta z =1$ from redshift 6 to 3, which is much slower in growth than predicted during the
first $\sim$ 100--400 Myr of the star formation history of \cite{finlatoretal2011}. Hence, applying the average rising SFH of \cite{papovichetal2011}
to our individual galaxies\footnote{It must be recognized that the average SFH derived by \cite{papovichetal2011} corresponds to a cosmologically averaged
history, which a priori does not apply to individual galaxies.} should yield results fairly close to those obtained from our models with constant SFR. This includes their inability to
reproduce the observed diversity of emission line strengths traced by the (3.6-4.5)\micron\ colour at $z\sim$ 3.8--5.
We therefore conclude that both rapidly and slowly rising star formation histories {\em over long time scales} ($\Delta t\ga$ 100--200Myr) are not appropriate to describe individual galaxies. Some mechanism turning off star formation or leading to episodic 
phases appears to be required.

For declining SF or other  episodic star formation histories, it becomes very difficult to connect galaxy populations at different redshifts and to draw conclusions 
from such a comparison. This is, of course, due to strong changes in UV luminosity with time and it would lead also to incorrect estimates of some parameters due to
the ``outshining" effect, where the glare of young massive stars can hide the properties of older stellar population. 
For example, \cite{papovichetal2001} estimates that an hypothetical old stellar population could contain up to $\sim3-8$ times the stellar mass of the young stars that dominate the observed SED. 
   
   However, \cite{starketal2009} predict the presence of massive objects with low star-forming activity under the assumption of an episodic star formation model. Since active and quiescent galaxies have similar distribution in stellar mass from $z\sim5$ to $z\sim3$ and a significant difference in median SFR ($\sim0.6$ dex), we can interpret this result as a confirmation of this  prediction.
   
   A consequence of episodic SF is that age becomes a poorly constrained parameter. It is also difficult to place constraints on the timescale of activity and inactivity and, indeed, to determine if the estimated parameters at different redshift are consistent with this scenario. \cite{leeetal2009} provide a higher limit on duration of star formation activity of 350 Myr based on observed UV luminosity function and clustering at $z\sim4$ and $z\sim5$. At high-redshift ($z\gtrsim6$) \cite{wyithe&loeb2011} found that the starburst timescale is set by the lifetime of massive stars, by comparing different assumptions on supernova feedback and observed evolution in galaxy size and UV luminosity function. These two results promote short timescales, which seems to be confirmed by our results, while uncertainties remain large. Furthermore, the result of \cite{wyithe&loeb2011} implies decreasing timescale with increasing redshift.
   Figure \ref{sfrsm_evol} show the difference in SFR-$M_{\star}$ relation between $z\sim3$ and $z\sim6$. At $z\sim6$, active and quiescent galaxies form two clearly separate groups while at $z\sim3$, the two groups populate all the intermediate states. We can interpret this result as an evolving star formation timescale, which is shorter 
   at $z\sim6$ than at $z\sim3$.
   
Under the assumption of episodic star formation, age becomes a tracer of the most recent star formation episode, since the youngest stars dominate the observed fluxes. This parameter does not allow us to determine if observed high redshift LBGs are progenitors of the observed low redshift LBGs. In this scenario, the number of free parameters 
(including duration of star-formation activity, duration of inactivity, the possibility for LBGs to evolve into a state with no star formation activity) are too large to conclude.

Several studies \citep{boucheetal2010,wuytsetal2011} suggest other SFHs, such as exponentially increasing or delayed star formation. These scenarios have been studied in \citet{schaereretal2013}.

           \begin{figure}[htbf]
     \centering
     \includegraphics[width=8.5cm,trim=0cm 0.5cm 2cm 1cm,clip=true]{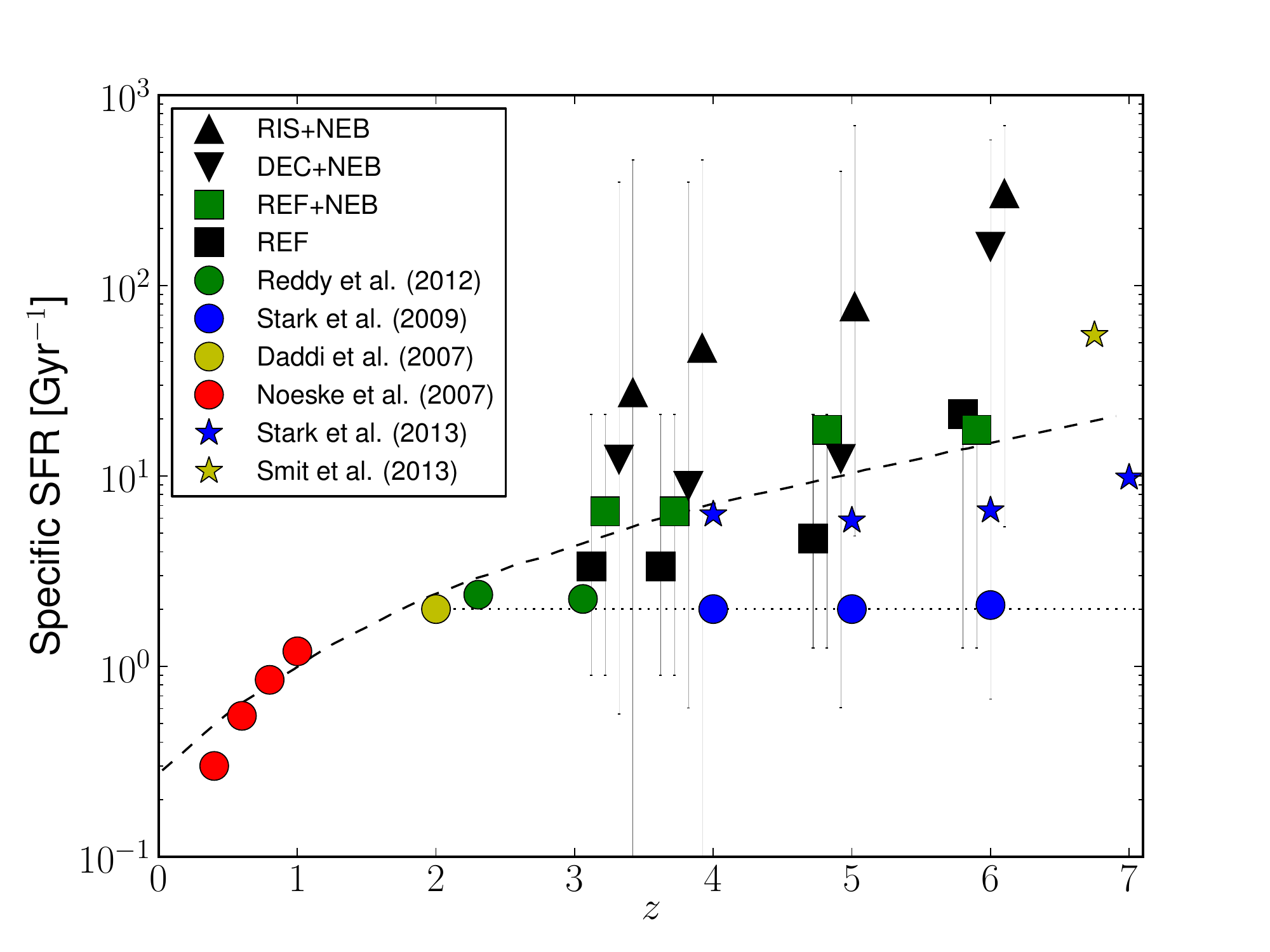}
     \caption{Median specific star formation rate as a function of redshift for four models with a 68\% confidence limit based on the whole probability distribution function with comparison of results from different studies
     %at $M_{\star}=10^{9.5}$
     \citep{noeskeetal2007,daddietal2007,starketal2009,reddyetal2012a} and results from studies accounting for nebular emission effect \citep{starketal2013,smitetal2013}. At $z\sim6$, we show results with +NEB+Ly$\alpha$ option for declining and rising SFHs. Typical errors are $\sim0.3$ dex. The dashed line shows the relation expected from \cite{boucheetal2010} for an exponentially increasing star formation at a fixed $M_{\star}=10^{9.5} \msun$. The dotted line given by sSFR=2 Gyr$^{-1}$is shown to guide the eye.}
     \label{ssfr_z}
\end{figure} 
   
   \subsection{Specific star formation rate}

        Since our study gives some elements supporting episodic star formation histories at high redshift, we compare the evolution of sSFR with results from other studies in Fig.\ \ref{ssfr_z}, using the compilation from \cite{gonzalezetal2010}, which is given at fixed stellar mass $M_{\star}=5\times 10^9 \msun$ \citep{noeskeetal2007,daddietal2007,starketal2009}. Since our different models lead to significant different median stellar masses (for e.g., $10^{9.6} \msun$ for REF model at $z\sim3$ and $10^{8.6} \msun$ for DEC+NEB+Ly$\alpha$ model at $z\sim6$), we compare the median sSFR for entire samples at each redshift. Typically, other studies found no significant change in median stellar mass with redshift and the median value of sSFR at $M_{\star}=5\times  10^9 \msun$ is near the median value for the whole sample \citep[e.g.][]{gonzalezetal2010}. Due to incompleteness, our values can be considered as a lower limit, since we find a trend of increasing sSFR with decreasing stellar mass for all models, albeit this trend is moderate with constant star formation (see Figure \ref{ssfrsm}). However, we cannot exclude the presence at these redshift of star forming galaxies enshrouded with dust, so our results have to be considered with appropriate caution. The impact of nebular emission on the sSFR evolution with redshift is also considered in \cite{starketal2013} and \cite{smitetal2013}, and while these studies conclude that the sSFR is higher than in previous studies not accounting for nebular emission, the exact evolutionary trend remains very uncertain. However, these studies confirm our main result: nebular emission can have a significant impact on stellar mass and star formation rate estimation at high-redshift.
        
        While our results for constant star formation seem to be consistent with those of \cite{starketal2009} for $z\sim4$ and $z\sim5$, we find a higher median sSFR at $z\sim6$. 
        Looking at IRAC channels, galaxies at $z\sim6$ with high sSFR ($\sim20$ Gyr$^{-1}$) have on average 2 to 3 channels with no detection (or no data), while galaxies with lower sSFR (median $\sim1.7$ Gyr$^{-1}$) have typically no more than 1 channel with no detection. No detection in the (rest-frame) optical bands leads to lower stellar mass estimation ($\sim1$ dex), while the estimated SFR stays similar. This explains the higher sSFR found for the REF models at $z\sim6$ compared to  \cite{starketal2009}.
        It is difficult to conclude if it is an effect due only to the lack of IRAC detection, or if it is physical. However, the difference of IRAC detections among the objects should correlate with physical differences in stellar mass. The precise extent of these differences is more difficult to constrain; additional data are needed.
The consideration of nebular emission with a constant star formation (REF+NEB) leads to higher sSFRs, and an evolution compatible with the trend from \citep{boucheetal2010}. This increase in sSFR is mainly due to a slight decrease in stellar mass estimation with a redshift due to an increase in emission lines strength with redshift (see Table \ref{ewha}).
         
        Our results with declining and rising star formation, which includes nebular emission, differ significantly from previous studies, with a higher median sSFR and a trend of increasing sSFR with redshift. Large confidence limits are due to a large dispersion of individual objects. For DEC+NEB and RIS+NEB, this dispersion is larger due to the two different LBG populations of ``weak" and ``strong" emitters. While studies neglecting nebular emission lead to the conclusion that star formation seems to be driven by different principles below and above $z\sim2$, the different assumptions used here and the results provided by DEC+NEB and RIS+NEB models highlight the possibility to reconcile theoretical expectations with inferred physical parameters.  
   
        \section{Discussion}
        \label{discussion}
        
        \subsection{Do we obtain realistic ages?}
        
        Since we have not imposed a lower limit for age estimation for both declining and rising SF, both models lead to a significant number of galaxies with an age below a typical dynamical timescale, especially for active galaxies (i.e. ``strong" emitters). Indeed, the median ages for this latter population are almost always close to our dynamical time estimate (Sect.\ \ref{s_tau}). Since the age estimation depends on star formation timescale for declining SF, we test some fixed values of $\tau$ to examine the effect on age and other parameters estimation. With $\tau=100$ Myr, we do not observe any significant change in median values of the different parameters, except for age, which increases by a factor $\sim2.5$ for both active and quiescent galaxies (i.e. ``weak" galaxies) at each redshift. However, if we look at the age distribution, there is still a significant number of galaxies with age $t<t_{\mathrm{dyn}}$. Imposing a lower limit of $\sim40$ Myr leads to an explanation of the previous result: nebular emission of active galaxies seems to be correctly fit only with very young ages. Using our sample at $z$ $\epsilon$ [3.8,5] with 3.6$\mu$m and 4.5$\mu$m fluxes measured, both declining and rising models are then not any more able to produce a significantly better fit of 3.6$\mu$m-4.5$\mu$m colour (in comparison with the REF model), which is correlated with EW(H$\alpha$).
        
        While this discrepancy between a significant fraction of our estimated ages and dynamical timescale may be a concern for our study, we are reminded of two elements. First, declining and rising models lead to higher uncertainty on age, which are typically by a factor $\sim3$ when nebular emission is considered and can allow us to reconcile almost 80\% of our samples with the dynamical timescale. Second, there is no measured velocity dispersion of nebular lines at $z>3$, since this measurement is confronted to the limit of current facilities.
Furthermore, we use typical velocity dispersion measurements at $z\sim2$ \citep[$\overline{\sigma}=129\pm50$ km s$^{-1}$, with a 7 km s$^{-1}$ error in the mean,][]{forsterschreiberetal2009}, which lead to a $\sim40$\% possible variation in the dynamical timescale of individual galaxies for a given typical r$_{hl}$.
        
We conclude that apparent discrepancies between dynamical timescale and estimated ages in this study are considerably reduced when uncertainties on estimated age and typical dispersions on velocity dispersion are taken into account. For rising and declining SFH, the estimated age of a fraction of LBGs ($\sim20$ \%) are inconsistent with dynamical timescale, mainly because of the strength of some lines (e.g. H$\alpha$), which can be reproduced by our SED fitting code with this set of assumptions only with extremely young ages.

        \subsection{Comparison with other studies}
        \label{compothst}
        
\begin{table*}[htdp]
\centering
\caption{Summary of other studies in the literature as compared with our results. Col.\ 3 indicates assumed star formation histories, col.\ 4 the extinction law, col.\ 5 the number of galaxies, and col.\ 6 indicates our model used for comparison.} 
\begin{tabular}{lccccc}
\hline
Redshift & Authors$^{\mathrm{a}}$ & SFH$^{\mathrm{b}}$ & ext.\ law & $N$ & Comparison model \\
\hline
$z\sim3$ & S01    & constant SFR                             & Calzetti                & 74   &  REF \\
$z\sim3$ & P01    & constant/exp.\ declining                           & Calzetti                & 33   &  DEC \\
$z\sim4$ & PG07 & exp.\ declining                           & Calzetti \& SMC & 47    &  DEC \\
$z\sim4$ & L11    & constant SFR                              & Calzetti               & 6$^{\rm c}$ & REF \\
$z\sim5$ & S07    & SSP/constant/exp.\ declining & Calzetti                & 14    &  DEC \\
$z\sim5$ & V07    & constant SFR                             & SMC                     & 21   & REF \\
$z\sim5$ & Y09    & constant SFR                             & Calzetti                & 105 & REF \\
$z\sim6$ & Y06    & SSP/constant SFR                    & Calzetti$^{\mathrm{d}}$                & 53 & DEC \\
$z\sim6$ & E07    & SSP/constant/exp.\ declining  & Calzetti                & 17  & DEC \\ 
\hline
\end{tabular}
\begin{list}{}{}
\item[$^{\mathrm{a}}$ S01: \cite{shapleyetal2001}, P01: \cite{papovichetal2001}, PG07: \cite{pentericcietal2007}, L11: \cite{leeetal2011}, S07: \cite{starketal2007},] 
\item[V07: \cite{vermaetal2007}, Y09: \cite{yabeetal2009},Y06: \cite{yanetal2006}, and E07: \cite{eylesetal2007}.]
\item[$^{\mathrm{b}}$ SSP: single stellar population, burst.]
\item[$^{\rm c}$: stack of 1913 objects in 6 UV magnitude bins.]
\item[$^{\mathrm{d}}$ \cite{yanetal2006} assume $A_V=0$. For a fair comparison, we run DEC model with the same assumption.]
\end{list}
\label{t_comparison}
\end{table*}

	During the last decade, various papers presented analyse of LBG properties at high redshift. Since we use a large range of star formation histories, we are able to compare our results to several of them, which are carried out for several LBG samples between $z\sim3$ and 6. The main assumptions made for the SED fits in these papers, size of their galaxy samples, and 
our corresponding models for comparison are listed in Table \ref{t_comparison}. To allow straightforward comparisons, we use the median/mean values derived from our probability distribution functions. Except stated otherwise, we determine the mean over our entire samples.
Although derived from the same pdfs, the values from our models quoted here do therefore not correspond to values listed in Tables \ref{tabmagall_ref}--\ref{tabmagall_ris} (which are {\it median} values, not mean values).

        \citet[][hereafter S01]{shapleyetal2001} and \citet[][hereafter P01]{papovichetal2001} have studied 74 and 33 LBGs respectively, at $z\sim3$. 
        The S01 sample is restricted to brighter objects ($M_{\mathrm{UV}} \lesssim -20.7$) than the sample studied here; thus, we use a subsample of our $z\sim3$ LBGs with the same magnitude limit.
Note that both studies (S01 and P01) include photometry up to the K-band and not longward.
        For both studies, we find very similar results for median ages (S01 obtain $\sim350$ Myr and P01 $\sim70$ Myr) and age distributions when corresponding
        models without nebular emission are used. This is not surprising since age is mainly constrained by the presence of the Balmer Break, which both studies can constrain from K$_{\mathrm{S}}$  and J (or H) band data. We also find a similar median stellar mass compared to P01 ($3\times 10^9$ $M_{\odot}$), while we obtain mean stellar mass slightly lower by $\sim0.2$ dex compared to S01 ($2\times 10^{10}$ $M_{\odot}$).
        Since  P01 provide best fit parameters for their sample, we can see that significant discrepancies appear on dust attenuation estimation. For both studies extinction factors are $\sim2$ times larger compared to our results. A possible explanation is that we can better constrain this parameter with IRAC data since longer wavelengths are less sensitive to reddening. Furthermore, S01 use a BC96 population synthesis models, while we use BC03 models. \cite{pentericcietal2010} have used BC03 models to study a sample of LBGs at $z\sim3$, and they found a similar result: same ages and lower dust attenuation in comparison with S01, while they used an exponentially declining SFH. These differences with S01 can come from the use of IRAC data but also from the difference in SFH.
        %It seems these differences with S01 come from the use of IRAC data.
%%%%%%%%%%%%%%%%%

	\citet[][hereafter PG07]{pentericcietal2007} and \citet[][hereafter L11]{leeetal2011} provide analysis of 47 and 1913 LBGs at $z\sim4$ with indidivual SED fittings in the first case, while L11 uses a stacking procedure in UV magnitude bins. The large sample of L11 has an upper limit $M_{UV}=-21.43$, at $z\sim3$, so we take care to do an appropriate comparison.
	Our results are similar with those from L11 in stellar mass, age, and dust extinction, although we have only 17 LBGs with $M_{UV}\leq-21.43$. A discrepancy appears with the results from PG07, since we find a lower mean stellar mass by $\sim0.4$ dex, while age, dust extinction, and SFRs are similar.

%%%%%
Stellar masses at $z \sim$ 5 and 6 are also in good agreement with previous studies, when the same/similar assumptions are used.
 Concretely, V07 found a median stellar mass of $2\times 10^9 \msun$ ($3.6\times 10^9 \msun$ for our study), and Y09 and S07 found a mean stellar mass of $4.1\times 10^9$ and $7.9\times 10^9 \msun$ respectively at $z\sim5$ (while we find $9.9\times 10^9$ and $1.2\times 10^{10} \msun$ with REF and DEC models respectively). At $z\sim6$, Y06 and E07 found $9.6\times 10^9$ and $1.6\times 10^{10}\msun$ respectively, as compared to our masses of $9.9\times 10^9$ and $1.1\times 10^{10} \msun$. The consistency of our results with other studies, by considering typical uncertainty of $\sim0.15$ dex for REF model and $\sim0.2$ dex for DEC model,  confirm that stellar mass is the most reliable parameter, since different assumptions on star formation history and also extinction law lead to similar 
 results \citep[cf.][]{papovichetal2001,vermaetal2007,yabeetal2009}.
 
Ages of $z \sim$ 5--6 LBGs obtained in the literature agree overall with our results, except for the study of \cite{vermaetal2007}.
Compared to their young median age of 25 Myr, we find 255 Myr, a difference which cannot be explained by uncertainty ($\sim0.15$ dex for REF model and $\sim0.4$ dex for DEC model). 
A possible explanation for the discrepancy with V07 could be the age-reddening degeneracy. However, V07 found a median $A_V=0.3$ mag, while we find 0.2 mag, which cannot explain the difference on the median age estimate. The origin of this difference remains unclear. On the other hand, Y09 and S07 found median ages of 25 Myr and 288 Myr (compared to 52 Myr and 320 Myr from our study) at $z\sim5$, and Y06 and E07 obtained 290 Myr and 400 Myr (262 Myr and 190 Myr)  at  $z \sim 6$, which are values in good agreement with our results.	
        
         Other differences appear on the reddening estimates: while Y09 found a mean $A_V \sim 0.9$ mag at $z \sim 5$, we obtain 0.4 mag for the corresponding constant SFR model (typical uncertainty $\sim0.1$ mag).
 At $z\sim6$, E07 found no reddening on average, while we obtain $A_V=0.4$ for the DEC model (typical uncertainty $\sim0.25$ mag). Our result at $z\sim5$ with the DEC model (mean $A_V=0.4$) is consistent with S07, since they found the same mean reddening. The differences on dust reddening estimation can have several explanations: the limited size of the samples compared to ours, or the lack of deep NIR and IRAC photometry which are able to put stronger constraints both on reddening and age. However, these latter arguments cannot be used to explain the discrepancy with Y09, and we do not find a satisfactory explanation.
         
        These discrepancies on reddening must obviously lead to differences on the derived star formation rates. Indeed, Y09 found a mean SFR of $141 \msunyr$ and V07 found a median SFR of 40 $\msunyr$ at $z\sim5$, while we find  $\sim50 \msunyr$ and $15 \msunyr$ from the models that best correspond to their assumptions.
        The difference on reddening estimation with Y09 explains the difference on SFR estimation, while the result of V07 can be explained by their very young median age, which leads to a significant deviation from the Kennicutt relation \citep{kennicutt1998}, and thus to a higher SFR. The sample from S07 does not provide SFR estimations but Y09 fits the parameters of S07: a median value of $20\msunyr$ was found, which is consistent with our result. However, these refitted parameters differ significantly from the results of S07 for both stellar mass and age, which casts some  doubt on the homogeneity of this comparison. For $z\sim6$, Y06 and E07 found a mean SFR $\lesssim10\msunyr$. While our results are consistent with Y06 ($7 \msunyr$), our SFR estimation differs significantly from E07 with an higher SFR ($\sim80 \msunyr$). This latter difference can be explained by the higher estimated dust reddening, since E07 estimates that $z\sim6$ galaxies are mainly dust free.
        
        Overall, we find good agreement with other studies in general for identical assumptions. Exceptions are the young ages of V07 and high dust extinction of Y09. When nebular emission is included, however the results change significantly, as we have shown in this study.
        
\subsubsection{Evolution of the mass -- UV magnitude relation?}
\cite{starketal2009} and \cite{gonzalezetal2011} cover a range of redshift between $z\sim4$ to $z\sim6$ (up to $z\sim7$ for the latter). Both studies present the $M_{\star}$--$M_{\mathrm{UV}}$ relation, which can be directly compared with our results in Figure \ref{sm_muv_z4} and \ref{muvsm_all}.  
As already discussed in Sect.\ \ref{sm}, we find similar results as \cite{gonzalezetal2011}, although our models for exponentially declining and for the rising star formation history 
suggest lower masses at high UV luminosities. Overall our relations remain within the scatter of $\pm0.5$ indicated by their work. 

\begin{figure}[htbf]
     \centering
          \includegraphics[width=9cm,trim=3cm 6.5cm 4.25cm 7.25cm,clip=true]{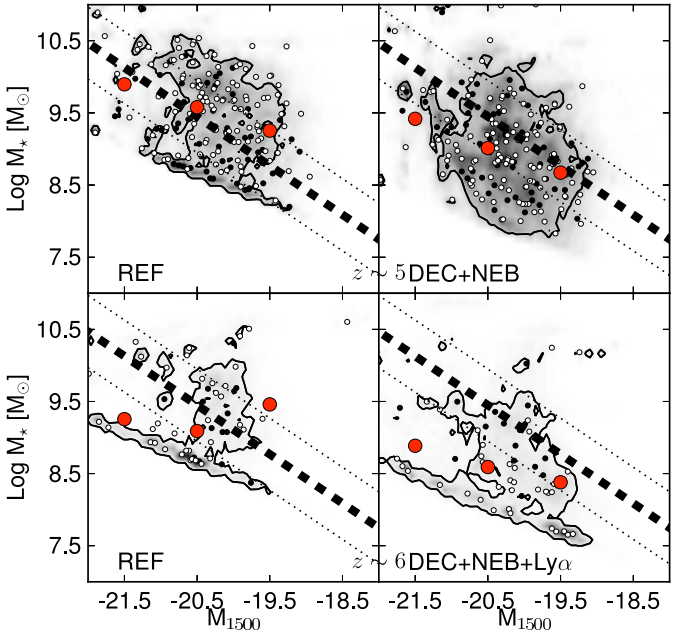}
     \caption{Composite probability distribution of $M_{1500}$ and $M_{\star}$ for REF and DEC+NEB(+Ly$\alpha$) models at $z \sim 5$ and $z\sim6$. The black dashed line represents the $M_{\star}$-$M_{1500}$ trend found by \cite{gonzalezetal2011} and black dotted lines shows a scatter of $\pm$0.5 dex. The points overlaid show the median value properties for each object in the sample: black dots for ``weak" nebular emitters, and white dots for ``strong" nebular emitters. The overlaid contour indicates the 68\% integrated probabilities on the ensemble properties measured from the centroid of the distribution. Red dots are median values of $M_{\star}$ in $M_{\mathrm{UV}}$ bins.}
     \label{sm_muv_diff}
\end{figure}

Both \cite{starketal2009} and \cite{gonzalezetal2011} find no evolution of the mass -- UV magnitude relation with redshift.
In contrast, our results seem to indicate a change of the $M_{\star}$--$M_{\mathrm{UV}}$ relation between $z \sim 5$ and 6,
as shown in Fig.\ \ref{sm_muv_diff}. (cf.\ also Sect.\ \ref{sm_sfh}). The main reason for this change is due to our finding of
relatively young ages for $z \sim 6$ galaxies, which implies a lower $M/L_{\rm UV}$ ratio.  This effect was also noticed 
by \cite{mclureetal2011} for their $z \sim$ 6--8 sample. A reanalysis of their sample with the same method and assumptions
used in the present paper confirms our finding from the $z \sim 6$ sample \citep[][in preparation]{schaerer&debarros2014}.

 % % % % % % % % % % % % % % % % % % % % 
 \subsubsection{LBGs with strong emission lines}
 An important finding from our quantitative models of LBGs with spectral models that include nebular lines 
 is the distinction of two separate categories of galaxies identified as "strong" and ``weak'' nebular emitters
 (cf.\ Sect.\ \ref{fitq}) Our work has revealed these categories from comparisons of the fit quality for
 models with and without nebular emission, and interestingly we found approximately the same
 fraction of objects ($\sim$ 2/3 strong versus 1/3 weak emitters) at each redshift.
 
As already mentioned above,  two other studies of $z\sim$ 4--5 LBGs have previously found similar objects with
strong H$\alpha$ emission identified by their excess in the 3.6 \micron\ filter with respect to 4.5 \micron.
From a sample of 74 LBGs with spectroscopic redshifts between 3.8 and 5, \cite{shimetal2011} found 
at least 65\% of galaxies that show a 3.6 \micron\ excess attributed to \ha\
(cf.\ also Stark et al. 2013 for a recent confirmation).
Earlier, \cite{yabeetal2009} had already noted a 3.6 \micron\ excess for 70\% of their $z \sim 5$ LBG sample
of $\sim$ 100 galaxies, which attributed to the same effect.
Obviously our study finds the same result and a very similar percentage of galaxies. Our work carries this
result further by revealing the existence of these two LBG categories among all the samples studied here,
ranging from U-drops to i-drops (i.e.\ $z \sim$ 3--6), and extending this result to fainter objects,
for which no spectroscopy is currently available.

Besides this important agreement, our results differ, however, on severals points from those of \cite{shimetal2011}.
For example, these authors conclude that 60\% of their so-called \ha\ emitters are forming stars at a relatively constant
rate by comparing estimated \ha\ equivalent width (obtained from the 3.6 \micron\ excess) 
and ages from broadband SED fits, whereas the rest prefer a more ``bursty" star formation. Our models yield nearly opposite results (cf.\ Sect.\ \ref{s_sfh}).
Since the effects of nebular emission affect the estimated ages in particular, the models of \cite{shimetal2011}, which neglect these
effects in their SED modelling, cannot give consistent physical parameters for their strong \ha\ emitters.
Similarly, whereas \cite{shimetal2011} prefer the SMC extinction law, our calculations for 
the B-drop sample using this law show that the vast majority of objects is a better fit with the Calzetti attenuation
law, when accounting for nebular emission.
Finally, we see no need for a top-heavy IMF or extremely low metallicities from our models, as suggested as possible
causes of the strong \ha\ emission by \cite{shimetal2011}.
In any case, the \ha\ star formation rates and the \ha\ equivalent widths derived in \cite{shimetal2011} are
comparable to ours. 	

 \subsection{Remaining uncertainties}
 
 \subsubsection{Uncertainties affecting individual objects}
Our study illustrates the way various physical parameters depend on model assumptions
 made for the broadband SED fits in detail, and a large range of parameter space has been explored here.
 Despite these extensive investigations, the impact of some assumptions have not been explored in 
 depth, and several uncertainties remain.
 
 The impact of different extinction laws, for example, has hardly been discussed here. Several papers
 presenting SED fits of LBGs at different redshifts have examined the differences obtained
 with the Calzetti attenuation law (adopted here) and the SMC extinction law \citep{prevotetal1984,bouchetetal1985} .
 Among them are the work of  \cite{papovichetal2001}, \cite{vermaetal2007}, and \cite{yabeetal2009} who study 
 LBG samples at $z \sim$ 3 and 5. A comprehensive comparison of the impact of different extinction
 laws (including SMC, LMC, Galactic, and the Calzetti law) is presented in \cite{yabeetal2009}.
 To examine how different extinction laws modify the results from SED fitting models that include nebular emission,
 we have modeled the B-drop sample with the declining star formation  history (DEC+NEB model) by adopting 
 the SMC law of \citep{prevotetal1984}. Qualitatively, our results show the same trends as found by \cite{yabeetal2009};
 that is, there is no big difference on stellar masses, lower $A_V$, older ages, and lower SFR. However, we find
 that the Calzetti attenuation law provides a better fit for the vast majority of galaxies. 
 Except for a few special cases such as the lensed galaxies, cB58 and the Cosmic Eye, which are studied in 
 detail and at different wavelengths but also include the IR \citep[See][but cf.\ Sklias et al. 2013]{sianaetal2008,wuytsetal2011}
 %(See \citep{sianaetal2008,wuytsetal2011} but cf.\ \citep{skliasetal2013}), where the 
 SMC law appears to be favoured from the measurement of the IR/UV luminosity and the 
 UV slope. It is generally thought that the Calzetti attenuation law is applicable (at least on average)
 to high redshift star forming galaxies \citep[but see][]{shimetal2011,oeschetal2013}.
     
Our models that include the effects of nebular emission assumed case B recombination to compute the strength
of the hydrogen recombination lines (and empirical line ratios for other lines from He and metals) and of 
nebular continuum emission. In particular we thereby assume that all Lyman continuum photons contribute to
nebular emission, which neglects therefore possible losses of ionizing photons due to dust inside \hii\ regions or 
to escape of Lyman continuum photons. Furthermore, we assume the same extinction/attenuation for stellar and 
nebular emission, whereas observations of nearby galaxies indicate that emission lines usually suffer from a 
higher extinction than the stellar continuum \citep[cf.][]{calzettietal2000}.
In this sense, our models maximize the effects of nebular emission, whereas the lines could be weaker 
than predicted by our models in reality. On the other hand, we have adopted empirical line ratios compiled by
\cite{anders&fritz2003} from nearby galaxies, whereas the conditions in distant galaxies could be different, which
leads to higher excitation or stronger lines \citep{erbetal2010,kewleyetal2013b,nakajimaetal2013}.

These uncertainties are currently difficult to quantify. In any case, the broadband SEDs clearly reveal
the presence of nebular lines, and our models include a wide range between maximum and no nebular emission.
Future spectroscopic observations or narrow/intermediate band imaging should try to provide 
more detailed and accurate observational constraints on the emission line strengths of LBGs at high redshift.
Detailed predictions from our models that regard individual lines will be presented elsewhere and can be made
available on request.

As clearly shown by our study, assumptions on the star formation history have a significant impact on the
estimated physical parameters when broadband SED fits are used. Although certain star formation histories
are found to provide better fits than others (cf.\ Sect.\ \ref{s_sfh}), it is obvious that the simple parametrisations
commonly adopted in literature and in our study can only be very crude representations of the true
SF histories of individual galaxies. 
In a companion paper \citep{schaereretal2013}, we have explored two additional families of star formation histories, a so-called delayed
star formation ($\mathrm{SFR} \propto t \exp(-t/\tau)$ and exponentially rising histories with variable timescales.
Using arbitrary star formation histories is, however, not practical, given the limited number of observational constraints.
Another approach has been taken by \cite{finlatoretal2007}, who have used the star formation history from 
their cosmological hydrodynamic simulations to fit observed high-$z$ galaxies. However, such studies
have so far been limited to a very small number of galaxies. Future improvements on this issue from both observations and 
simulations are certainly needed. For example, \ \citet{reddyetal2012a} test different SF histories using
IR and UV observations of $z \sim 2$ LBGs. Along similar lines, we show in \cite{schaereretal2013} how SFHs 
can be distinguished by measuring their dust emission with future ALMA observations and/or with measurements of 
emission lines. 
\citet{skliasetal2013} have carried out such an analysis for 7 lensed galaxies using recent Herschel
observations.

\subsubsection{Biases and selection effects}
For obvious reasons, selection effects and various biases affect studies in general of galaxy populations,
and these need to be taken into account to compare different samples in the analysis
of apparent correlations between derived physical parameters and in other contexts.
         
To allow meaningful comparisons with other studies, we have presented our detailed results in bins
of UV magnitude (see data in Tables \ref{tabmagall_ref}, \ref{tabmagall_dec}, and \ref{tabmagall_ris}).
Especially in the brightest and faintest bins, the number of galaxies is low, so that these
results should be taken with care. Of course, the number of galaxies at different redshifts varies strongly, affecting the accuracy of the median physical properties (and confidence range)
derived here.

As in most literature studies, selection effects and biases have not been treated here.
The impact of biases on the determination of physical parameters of LBGs from broadband SED fitting
has been studied by \cite{leeetal2009b}, who constructs mock galaxy catalogs
from semi-analytical models, which are then used to fit the simulated galaxies with a standard SED fitting tool.
They find that stellar masses can be recovered well, whereas single-component SED fitting methods
underestimate SFRs and overestimate ages. The differences are attributed in part to a ``mismatch" of 
star formation histories between their fitting tool (which assumes exponentially declining SF) and those 
predicted by their galaxy models (which are often rising).
A similar ``template mismatch" was also identified as the main cause for differences in the comparison of 
$z \sim$ 1.5--3 merger simulations with SED fitting results carried out by \cite{wuytsetal2009}.
Our models are also prone to such biases/uncertainties, and the role of the assumed SF histories on the derived
physical parameters has already been discussed above. However, since we have covered a wider
range of SFHs including rising SF, our results may suffer less from this potential problem. The results
from SED fits with additional SFHs (including exponentially declining histories with adjustable timescales)
are presented in \cite{schaereretal2013}.

Various correlations have been found between physical parameters as the stellar mass, the star formation
rate, age, and others in both the literature and in our study.
Although our work emphasizes the way the physical parameters depend on various model assumptions, it is important to be aware of the selection effects that may significantly affect
such correlations. \cite{stringeretal2011} have recently examined the behaviour of the SFR and the specific
SFR with stellar mass, two important quantities discussed extensively in the recent literature.
From their simulations of mock galaxies to which observational selection criteria and ``standard" analysis
are applied, they show how true underlying trends can be misrepresented. This study also echoes the 
caution expressed by \cite{dunneetal2009} on the apparent sSFR--mass relation, which they urge, could 
be severely affected by selection biases. \cite{reddyetal2012a} also discussed how the Malmquist bias in UV-selected sample can affect the SFR-M$_{\star}$ relation, and thus the sSFR determination.
To the best of our knowledge, the effect and extent of selection effects and biases on apparent correlation between extinction and mass, and age and mass (cf.\ Sect.\ \ref{sm}) has not yet been addressed. A quantitative study of these effects is clearly beyond the scope of the present publication.

%______________________________________________________________
 \section{Summary and conclusions}
 \label{conclusions}
 
 We present a homogeneous study of a sample of $\sim1700$ LBGs at $z\sim3-6$ from the GOODS-MUSIC catalogue \citep{santinietal2009} with deep photometry from the $U$ band to 8 $\mu$m. Using a modified version of the {\em HyperZ} photometric redshift code that takes into account nebular emission \citep{schaerer&debarros2009}, we explore a range of star formation history (constant, exponentially decreasing, and rising). We explore a wide parameter space in redshift, metallicity, age, and extinction as described by the Calzetti law \citep{calzettietal2000} by 
 varying e-folding timescales for star formation and determining whether or not nebular emission is included.
The main  physical parameters derived from our models are the stellar mass, age, reddening, star formation rate SFR, and specific SFR. Furthermore, our models also 
provide information on the characteristic SF timescale.

Our method and the selected sample has been described in Sects.\ \ref{data} and \ref{method}. The detailed model results concerning the physical parameters,
correlations among them, and the redshift evolution of the galaxy properties have been presented in Sect.\ \ref{results}.
Our main results can be summarized as follows:

 \begin{itemize}
 
 	\item Independent of the adopted star formation history, we find that $\sim65\%$ of the galaxies are better fit with nebular emission and $\sim35\%$ without (Fig.\ \ref{histcolor}) at all redshifts. According to the Akaike Information Criterium, models that include nebular emission are 5-10 times more likely to be better model than models without nebular emission for the first group (i.e $\sim65\%$ of the galaxies), while the rest shows a probability two times larger to be best fit by models without nebular emission. For galaxies with $z$ $\epsilon$ [3.8,5], these two groups are clearly identified by their 3.6$\mu$m-4.5$\mu$m colour (Fig.\ \ref{chi2_nebvsstd}), which is correlated with strong H$\alpha$ emission \citep[cf.][]{shimetal2011}. Furthermore, this colour cannot be reproduced if we impose an age limitation $>50$ Myr (Fig.\ \ref{colortest}). This observed colour distribution clearly 
indicates the existence of galaxies with strong emission lines and others with few or no discernible signs of emission lines (see Table \ref{ewha}). Our SED modelling reveals the presence of two LBG groups
(dubbed  ``strong" and ``weak" emitters respectively) at all redshifts studied here ($z \sim$ 3--6) from U-drops to i-drops.
	
\item Models that include the effects of nebular emission and account for variable (declining or rising) star formation histories naturally separate the two LBG groups
according to current star formation rate, where the group of ``strong" emitters show a larger SFR than the ``weak" emitters (Fig.\ \ref{sfrsm}). In a scenario of declining star formation histories,
these groups could be seen as starbursts and ``post-starbursts" with age differences compatible with this suggestion. Indeed, properties of ``weak" emitters are compatible with slightly more evolved population with older ages, lower dust attenuation, and slightly lower stellar mass and SFR, when compared to ``strong" emitters.
Models with constant SFR, nebular emission, and age $>50$ Myr cannot reproduce the observed range of 3.6$\mu$m-4.5$\mu$m colours of $z \sim$ 3.8--5 galaxies.
	
	\item Independent of the star formation history, the inclusion of nebular emission leads to younger ages on average (Fig.\ \ref{agecomp}), since nebular lines in optical (rest-frame) are able to mimic a Balmer break. This confirms our earlier findings \citep{schaerer&debarros2009,schaerer&debarros2010} for larger samples and over a wider redshift range.
	
	\item We find that the derived dust attenuation mainly depends on the assumed star formation history and that the treatment of nebular emission does
	not lead to a general systematic shift. Discrepancies found between our results with decreasing SFH and those from other studies at $z\sim2$ \citep[e.g.][]{shapleyetal2005,erbetal2006b,reddyetal2012a} seem to come from modelling assumptions on both age and metallicity. The largest attenuation is found with rising star formation histories, since these always predict very recent star formation and hence
	UV emission, as already discussed by \cite{SP05}.
	In this case, the inclusion of nebular emission decreases the average attenuation, whereas the attenuation increases for declining SFHs, and 
	remains unchanged for constant SFR.

	\item Based on our SED fits of 700 $z\sim4$ LBGs, we propose a new average relation between the observed UV slope $\beta$ and the 
	attenuation $A_V$. Our relation deviates from the classical relation \citep{meureretal1999}, which assumes constant SFR, ages $\ga 100$ Myr and solar metallicity, and
	leads to a higher attenuation for a given $\beta$ slope.
	
		\item Considering nebular emission, the stellar masses derived from the SED fits decrease by $\sim$ 0.4 on average and by larger amounts ($\sim0.4-0.9$ dex) for LBGs from the ``strong" emitter group (Fig.\ \ref{smcomp}). We find a trend of increasing dust attenuation with stellar mass (Fig.\ \ref{avsm}), as already suggested earlier for $z \sim$ 6--8 galaxies 
		\citep{schaerer&debarros2010}, and a trend of increasing age with galaxy mass  (Fig.\ \ref{agesm}).
		
		\item Given the large scatter found in the SFR--$M_{\star}$ relation for all models with variable star formation histories, we also find a large scatter for the specific star formation rate sSFR
		with stellar mass and at all redshifts. Our favoured models show a higher median sSFR at $z \sim 3-6$ than derived by previous studies \citep[e.g.][]{gonzalezetal2011} .
		Our results tend to indicate an increase of the median sSFR with redshift, as advocated by several theoretical galaxy formation and evolution models		
		\citep{boucheetal2010,duttonetal2010,weinmannetal2011}

		\item While uncertainties on $\tau$ remain large, our SED fits favour short median star formation timescales ($\lesssim300$ Myr). Furthermore, we find
		tentative evidence of decreasing values of $\tau$ with decreasing UV luminosity among the sample of ``strong" emitters. 
	
	\item As already shown in \cite{starketal2009}, constant star formation seems to be irreconcilable with the non-evolution of the $M_{\star}-M_{1500}$ relation between
	$z \sim$ 5 to 3 and with the UV luminosity function.
	
		\item The rising average star formation of \cite{finlatoretal2011} used in this study cannot represent the typical history of many individual LBGs over long time,
			since the predicted increase if continuing into the future, is too fast.
			This would lead to a strong increase of median stellar masses and the median SFR from one redshift to another, which would also imply an increase in dust attenuation to hide these objects from the LBG selection, since this very massive and strongly star-forming galaxies are not seen in the expected numbers
			\citep[cf.][]{reddyetal2012a}.
	
		\item Two groups of LBGs identified as active and more quiescent galaxies (respectively ``strong" and ``weak" emitters, Figure \ref{sfrsm}) are present. Our finding of best SED fits with declining star formation histories
		and the consistency of these results with constraints on duty cycles from clustering studies and other theoretical arguments \citep{leeetal2009,wyithe&loeb2011} all
		concur to consider episodic star formation as the scenario that best fits observations of LBGs at high redshift.
\end{itemize}

Our systematic and homogeneous analysis casts new light on the physical properties of LBGs from $z \sim$ 3 to 6
and possibly to higher redshift \citep[cf.][]{schaerer&debarros2010}.
Obviously, our results have a potentially important impact on a variety of questions, and several implications need
to be worked out. On the other hand, our study also calls for new observations and tests.

For example, both our preferred models (variable star formation histories with nebular emission) imply a higher UV attenuation 
than what is currently derived, using the observed UV slope. At $z \sim 4$, our models typically predict an increase
by a factor $\la$ 3 but smaller changes at higher redshift. 
Implications on the cosmic star formation history and related topics will be worked out in a separate publication.
If correct, a higher UV attenuation should lead to measurable changes in the IR emission of LBGs.
A stacking analysis of the LBGs studied in this paper shows that the models presented here are all compatible
with the current IR, sub-mm, and radio observations (Greve et al.\ 2013, in preparation). Detailed predictions
of the IR-mm emission from our galaxies are presented in \citet{schaereretal2013}. More sensitive 
observations in the future with ALMA should be able to detect individual LBGs over a wide redshift range, to determine
their attenuation, and, hence, to also distinguish different star formation histories \citep{shimetal2011,schaereretal2013}.

An important implication from our study is that the idea of a simple, well-defined ``star formation main sequence"
with the majority of star forming galaxies that show tight relation between SFR and \mstar,
as suggested from other studies at lower redshift \cite[$z \sim$ 0--2, cf.][]{daddietal2007,elbazetal2007,noeskeetal2007}, may not be appropriate 
at high redshift ($z \ga 3$). Indeed, a relatively small scatter is only obtained assuming constant star formation
over long enough timescales ($\ga 50$ Myr), whereas our models clearly provide indications for variable star formation
histories and episodic star formation  (cf.\ Sects.\ \ref{s_tau}, \ref{s_sfh}). A significant scatter is also obtained for 
models assuming rising or delayed star formation histories \citep[see Sect.\ \ref{sfr} and][]{schaereretal2013},
which are often suggested in the literature. If the scatter found from our models in the SFR--\mstar\ relation decreases
with decreasing redshift, convergence towards the results from other studies remains to be seen.
However, a smaller scatter is naturally found since constant star formation is usually {\em assumed} in
most models for establishing the standard SFR calibrations used in the literature \citep{daddietal2007,elbazetal2007,gonzalezetal2011,wuytsetal2011} or when analyzing stacked data, which naturally smoothes out
any possible variation \citep{leeetal2010}.
Establishing more precisely the attenuation, current SFR and star formation histories of galaxies is therefore 
crucial to shed more light on these questions.

Related to the above mentioned scatter is also the behaviour of the specific star formation rate (sSFR) with both galaxy mass and on average with redshift. Basically, the same questions and uncertainties concerning the
SFR-\mstar\ relation apply here. In any case, it must be recognized that the sSFR is strongly dependent on model
assumptions and seems to show a strong dependence on the galaxy mass. Whereas earlier determinations
of the sSFR at $z >3$ were considered in conflict with recent galaxy evolution models 
\citep{boucheetal2010}, our results are clearly in better agreement with the high sSFR values and the redshift evolution
predicted by these models. A detailed confrontation of our results with such models and more refined ones
will hopefully provide further insight into galaxy formation and evolution models at high redshift.

Finally, it is clear that our study reveals new aspects on the possibly complex and variable star formation histories 
of high redshift galaxies. Whereas different arguments are found in the literature that favour short duty cycles and episodic star 
formation \citep[cf.][]{sawicki&yee1998,vermaetal2007,starketal2009,leeetal2009,wyithe&loeb2011} based on SED studies, LBG clustering, and other arguments,
other studies favour long star formation timescales \citep[][]{shapleyetal2001,leeetal2011,shimetal2011}.
Our study uses quantitatively features probing emission lines for the first time, whose strength is naturally
sensitive to relatively rapid variations in the recent SFR. Although such variations have been more difficult to
uncover before, this may therefore not be surprising. Direct observations of the emission lines in high-z LBGs
should be very useful to test our models and provide more stringent constraints on the importance of nebular 
emission and on the star formation histories of distant galaxies \citep{schaereretal2013}.
Other studies providing new measurement of LBG clustering at $z>4$, searches for
passive galaxies at high redshift, or theoretical studies on star formation and regulation processes 
can also help to improve our understanding of these important issues closely related to key questions on 
galaxy formation and evolution.

\begin{acknowledgements}
We would like to thank numerous colleagues who have contributed to interesting discussions and raised useful questions during the time
where this work has been carried out.  We thank them here collectively.
We acknowledge the GOODS-MUSIC collaboration. We thank the referee for useful comments and suggestions that helped improve the quality and presentation of this paper. The work of SdB and DS is supported by the Swiss National Science Foundation. DPS is supported by an STFC postdoctoral research fellowship. SdB wants to dedicate this work to the memory of Jean Faure-Brac.	
\end{acknowledgements}

\bibliographystyle{aa}
\bibliography{ref}

%%%%%%%%%%%%%%%%%

\appendix

\section{}
  
\begin{table*}[htbf]
\centering
\caption{Galaxy properties over $3\leq z\leq6$ for constant star formation and solar metallicity set of models in $M_{1500}$ bins. For each parameter, we give the median value and 68\% confidence limits derived from the probability distribution function.}
\scalebox{0.85}{\begin{tabular}{ccccccc}
\hline
$M_{\mathrm{UV}}$ & Num & Age & $M_{\star}$ & SFR & $A_{\mathrm{V}}$ & sSFR \\
 (mag) & & (Myr) & ($10^8 \mathrm{M}_{\odot}$) & ($\mathrm{M}_{\odot}.\mathrm{yr}^{-1}$) & (mag) & (Gyr$^{-1}$) \\
\hline
\hline
\multicolumn{7}{c}{REF model}\\
\multicolumn{7}{c}{$U$ drops ($z\sim3$)}\\
\hline
-22.5 &    5 &   180.5 (90.5-1015.2) &    98.6 (67.7-452.9) &    64.3 (48.6-81.2) &     0.1 (0.0-0.2) &     6.5 (1.2-12.6) \\
-21.5 &   28 &   255.0 (52.5-718.7) &    74.6 (23.7-213.9) &    40.7 (19.1-68.7) &     0.4 (0.0-0.6) &     4.7 (1.7-21.1)  \\
-20.5 &  146 &   255.0 (52.5-1015.2) &    50.4 (11.7-149.8) &    19.1 (8.7-43.3) &     0.4 (0.0-0.7) &     4.7 (1.2-21.1)  \\
-19.5 &  176 &   255.0 (52.5-1434.0) &    18.4 (3.2-77.8) &     6.3 (3.4-20.7) &     0.2 (0.0-0.7) &     4.7 (0.9-21.1) \\
-18.5 &   24 &  1434.0 (90.5-2000.0) &    66.8 (10.6-350.8) &     7.6 (3.3-31.1) &     0.7 (0.3-1.4) &     0.9 (0.7-12.6) \\
\hline
\multicolumn{7}{c}{$B$ drops ($z\sim4$)}\\
\hline
-22.5 &    3 &   718.7 (360.2-1700.0) &   424.1 (313.5-572.8) &    73.2 (44.2-104.9) &     0.2 (0.1-0.3) &     1.7 (0.8-3.4)\\
-21.5 &   62 &   508.8 (64.1-1015.2) &   105.3 (30.0-268.1) &    29.2 (18.0-67.6) &     0.2 (0.0-0.6) &     2.4 (1.2-17.5)  \\
-20.5 &  255 &   508.8 (52.5-1015.2) &    49.2 (11.4-160.4) &    15.6 (7.6-37.9) &     0.3 (0.0-0.7) &     2.4 (1.2-21.1)  \\
-19.5 &  335 &   360.2 (52.5-1434.0) &    18.7 (3.1-65.3) &     5.9 (3.4-17.2) &     0.2 (0.0-0.7) &     3.4 (0.9-21.1) \\
-18.5 &   47 &   508.8 (52.5-1700.0) &    19.1 (2.5-75.8) &     5.2 (2.2-17.4) &     0.5 (0.0-0.9) &     2.4 (0.8-21.1)  \\
\hline
\multicolumn{7}{c}{$V$ drops ($z\sim5$)}\\
\hline
-22.5 &    3 &   127.8 (52.5-1015.2) &    49.9 (39.7-970.8) &   104.4 (38.0-131.1) &     0.1 (0.0-0.3) &     9.1 (1.2-21.1)  \\
-21.5 &   23 &   508.8 (52.5-1015.2) &    78.8 (15.6-350.6) &    32.0 (18.3-69.6) &     0.2 (0.0-0.5) &     2.4 (1.2-21.1)  \\
-20.5 &   89 &   255.0 (52.5-1015.2) &    37.9 (6.4-142.0) &    15.0 (8.5-39.9) &     0.3 (0.0-0.6) &     4.7 (1.2-21.1)  \\
-19.5 &   67 &   360.2 (52.5-1015.2) &    18.0 (3.8-67.6) &     7.3 (4.1-21.5) &     0.3 (0.0-0.7) &     3.4 (1.2-21.1) \\
-18.5 &    1 &    52.5 (52.5-180.5) &     4.1 (3.0-5.3) &     7.4 (2.9-10.0) &     0.5 (0.0-0.7) &    21.1 (6.5-21.1)  \\
\hline
\multicolumn{7}{c}{$i$ drops ($z\sim6$)}\\
\hline
-22.5 &    5 &    52.5 (52.5-508.8) &    49.3 (20.9-183.2) &    57.8 (37.2-125.9) &     0.0 (0.0-0.3) &    21.1 (2.4-21.1)  \\
-21.5 &   10 &    52.5 (52.5-1015.2) &    18.0 (8.7-201.1) &    28.3 (17.0-68.5) &     0.0 (0.0-0.5) &    21.1 (1.2-21.1)  \\
-20.5 &   30 &    52.5 (52.5-1015.2) &    12.3 (4.8-61.4) &    11.4 (7.4-27.1) &     0.0 (0.0-0.5) &    21.1 (1.2-21.1) \\
-19.5 &   11 &   508.8 (52.5-1015.2) &    28.8 (6.9-210.1) &    13.2 (4.6-48.0) &     0.5 (0.0-1.0) &     2.4 (1.2-21.1)  \\
-18.5 &    1 &    90.5 (52.5-1015.2) &   400.6 (30.0-1199.0) &   642.2 (53.5-1019.0) &     2.7 (2.0-3.5) &    12.6 (1.2-21.1)  \\
\hline
\multicolumn{7}{c}{REF+NEB+Ly$\alpha$ model}\\
\multicolumn{7}{c}{$U$ drops ($z\sim3$)}\\
\hline
-22.5 &    6 &   127.8 (64.1-1434.0) &    62.7 (42.6-350.1) &    61.2 (41.9-85.9) &     0.1 (0.0-0.3) &     9.1 (0.9-17.5)  \\
-21.5 &   27 &    90.5 (52.5-508.8) &    57.7 (13.2-155.3) &    45.1 (16.5-104.2) &     0.4 (0.0-0.7) &    12.6 (2.4-21.1)  \\
-20.5 &  114 &    90.5 (52.5-718.7) &    32.8 (11.3-90.5) &    23.8 (9.4-60.5) &     0.5 (0.1-0.8) &    12.6 (1.7-21.1)  \\
-19.5 &  199 &    90.5 (52.5-1015.2) &    14.4 (3.6-61.0) &     9.4 (3.7-31.3) &     0.4 (0.0-0.9) &    12.6 (1.2-21.1) \\
-18.5 &   39 &   180.5 (52.5-1700.0) &    33.3 (3.6-156.5) &    10.9 (3.2-55.9) &     0.9 (0.3-1.7) &     6.5 (0.8-21.1)  \\
\hline
\multicolumn{7}{c}{$B$ drops ($z\sim4$)}\\
\hline
-22.5 &    3 &   360.2 (180.5-1700.0) &   299.5 (188.5-524.3) &    98.2 (41.1-123.9) &     0.3 (0.1-0.5) &     3.4 (0.8-6.5)  \\
-21.5 &   59 &   127.8 (52.5-718.7) &    56.1 (20.0-187.8) &    39.1 (18.1-82.0) &     0.3 (0.0-0.6) &     9.1 (1.7-21.1)  \\
-20.5 &  247 &   180.5 (52.5-1015.2) &    34.0 (11.1-112.3) &    18.6 (8.4-53.6) &     0.4 (0.0-0.8) &     6.5 (1.2-21.1)  \\
-19.5 &  346 &    90.5 (52.5-1015.2) &    12.3 (2.9-49.2) &     6.9 (3.6-22.1) &     0.3 (0.0-0.8) &    12.6 (1.2-21.1) \\
-18.5 &   48 &   180.5 (52.5-1434.0) &    12.8 (2.3-60.2) &     5.4 (2.1-16.6) &     0.5 (0.1-1.0) &     6.5 (0.9-21.1)  \\
\hline
\multicolumn{7}{c}{$V$ drops ($z\sim5$)}\\
\hline
-22.5 &    3 &   127.8 (52.5-180.5) &    94.0 (32.6-402.4) &    86.9 (68.2-330.3) &     0.0 (0.0-0.8) &     9.1 (6.5-21.1)  \\
-21.5 &   20 &    90.5 (52.5-718.7) &    54.5 (11.5-229.8) &    28.2 (17.0-150.7) &     0.1 (0.0-0.9) &    12.6 (1.7-21.1)  \\
-20.5 &   92 &    90.5 (52.5-1015.2) &    25.1 (5.9-99.8) &    15.2 (7.9-48.3) &     0.3 (0.0-0.8) &    12.6 (1.2-21.1)  \\
-19.5 &   70 &    90.5 (52.5-1015.2) &    13.8 (3.0-53.6) &     8.1 (4.0-28.4) &     0.3 (0.0-0.9) &    12.6 (1.2-21.1)  \\
-18.5 &    1 &   180.5 (127.8-360.2) &     5.5 (4.0-6.9) &     2.9 (2.2-4.9) &     0.2 (0.0-0.5) &     6.5 (3.4-9.1)  \\
\hline
\multicolumn{7}{c}{$i$ drops ($z\sim6$)}\\
\hline
-22.5 &    3 &    52.5 (52.5-64.1) &    43.5 (23.4-57.5) &    88.0 (46.1-113.2) &     0.1 (0.0-0.4) &    21.1 (17.5-21.1)  \\
-21.5 &    8 &   127.8 (52.5-1015.2) &    68.7 (11.7-213.2) &    28.1 (18.8-86.3) &     0.0 (0.0-0.6) &     9.0 (1.2-21.1)  \\
-20.5 &   30 &    52.5 (52.5-1015.2) &     8.5 (3.5-58.4) &     9.1 (6.3-20.9) &     0.0 (0.0-0.4) &    21.1 (1.2-21.1)  \\
-19.5 &   16 &    90.5 (52.5-1015.2) &    17.4 (3.9-113.4) &    10.8 (4.5-39.7) &     0.3 (0.0-1.0) &    12.6 (1.2-21.1) \\
-18.5 &    0 & -& -& -& -& -\\
\hline
\multicolumn{7}{c}{REF+NEB model}\\
\multicolumn{7}{c}{$U$ drops ($z\sim3$)}\\
\hline
-22.5 &    6 &    52.5 (52.5-1434.0) &    52.2 (38.6-466.0) &    69.7 (43.4-107.6) &     0.0 (0.0-0.1) &    21.1 (0.9-21.1)  \\
-21.5 &   29 &   255.0 (52.5-718.7) &    60.7 (19.8-212.0) &    35.7 (16.0-68.6) &     0.3 (0.0-0.6) &     4.7 (1.7-21.1)  \\
-20.5 &  134 &   127.8 (52.5-1434.0) &    32.1 (8.0-102.2) &    17.1 (7.6-45.4) &     0.3 (0.0-0.7) &     9.1 (0.9-21.1) \\
-19.5 &  185 &   127.8 (52.5-1434.0) &    16.3 (2.5-73.8) &     6.2 (3.3-25.4) &     0.2 (0.0-0.8) &     9.1 (0.9-21.1)  \\
-18.5 &   30 &   718.7 (52.5-2000.0) &    49.6 (10.0-223.0) &    10.5 (3.4-65.8) &     0.9 (0.3-1.8) &     1.7 (0.7-21.1) \\
\hline
\multicolumn{7}{c}{$B$ drops ($z\sim4$)}\\
\hline
-22.5 &    3 &   508.8 (360.2-1700.0) &   347.4 (268.0-538.8) &    81.7 (41.4-89.3) &     0.2 (0.1-0.4) &     2.4 (0.8-3.4)  \\
-21.5 &   56 &   255.0 (52.5-1015.2) &    60.2 (19.9-209.7) &    29.8 (16.4-61.4) &     0.2 (0.0-0.5) &     4.7 (1.2-21.1) \\
-20.5 &  267 &   255.0 (52.5-1015.2) &    35.5 (8.3-121.6) &    14.8 (7.2-42.3) &     0.3 (0.0-0.7) &     4.7 (1.2-21.1)  \\
-19.5 &  331 &   127.8 (52.5-1434.0) &    12.9 (2.3-52.7) &     5.4 (3.2-17.3) &     0.2 (0.0-0.7) &     9.1 (0.9-21.1)  \\
-18.5 &   46 &   508.8 (52.5-1700.0) &    17.2 (2.0-68.7) &     4.7 (2.0-17.0) &     0.5 (0.0-1.0) &     2.4 (0.8-21.1)  \\
\hline
\multicolumn{7}{c}{$V$ drops ($z\sim5$)}\\
\hline
-22.5 &    4 &    52.5 (52.5-255.0) &    34.9 (18.8-494.7) &    73.6 (39.6-182.0) &     0.0 (0.0-0.5) &    21.1 (4.7-21.1) \\
-21.5 &   23 &    64.1 (52.5-718.7) &    36.5 (9.4-208.2) &    26.4 (16.2-126.5) &     0.1 (0.0-0.8) &    17.5 (1.7-21.1) \\
-20.5 &   89 &    64.1 (52.5-1015.2) &    24.0 (5.3-96.6) &    15.3 (8.1-53.3) &     0.3 (0.0-0.8) &    17.5 (1.2-21.1)  \\
-19.5 &   70 &    64.1 (52.5-1015.2) &    11.2 (2.7-50.5) &     7.2 (3.9-23.2) &     0.3 (0.0-0.7) &    17.5 (1.2-21.1) \\
-18.5 &    1 &    52.5 (52.5-52.5) &    21.5 (18.5-30.6) &    45.5 (39.0-64.7) &     1.3 (1.0-1.5) &    21.1 (21.1-21.1)  \\
\hline
\multicolumn{7}{c}{$i$ drops ($z\sim6$)}\\
\hline
-22.5 &    3 &    52.5 (52.5-508.8) &    41.2 (34.7-143.5) &    72.7 (33.0-85.3) &     0.0 (0.0-0.1) &    21.1 (2.4-21.1) \\
-21.5 &   10 &    52.5 (52.5-718.7) &    16.8 (8.0-140.3) &    30.1 (15.8-66.0) &     0.0 (0.0-0.5) &    21.1 (1.7-21.1)  \\
-20.5 &   32 &    52.5 (52.5-718.7) &     8.6 (4.2-50.4) &    10.0 (6.7-19.6) &     0.0 (0.0-0.4) &    21.1 (1.7-21.1)  \\
-19.5 &   12 &    64.1 (52.5-1015.2) &    16.1 (2.9-174.6) &    10.2 (4.7-42.0) &     0.3 (0.0-1.0) &    17.5 (1.2-21.1)  \\
-18.5 &    0 & -& -& -& -& -\\
\hline
\hline
\end{tabular}}
\label{tabmagall_ref}
\end{table*}

%%%%%%%%%%%%%%

\begin{table*}[htbf]
\centering
\caption{Same as Table \ref{tabmagall_ref} for decreasing star formation history with variable timescale and metallicity Z.}
\scalebox{0.85}{\begin{tabular}{ccccccccc}
\hline
$M_{UV}$ & Num & Age & $M_{\star}$ & SFR & $A_{\mathrm{V}}$ & sSFR & Z & $\tau$ \\
 (mag) & & (Myr) & ($10^8 \mathrm{M}_{\odot}$) & ($\mathrm{M}_{\odot}.\mathrm{yr}^{-1}$) & (mag) & (Gyr$^{-1}$) & ($\mathrm{Z}_{\odot}$) & (Myr) \\
\hline
\hline
\multicolumn{9}{c}{DEC model}\\
\multicolumn{9}{c}{$U$ drops ($z\sim3$)}\\
\hline
-22.5 &    6 &     2.5 (1.0-508.8) &   233.3 (154.8-398.4) &  7717.0 (21.7-25670.0) &     1.2 (0.0-1.3) &   350.2 (0.9-1000.0) &  0.02 (0.02-1.00) & 70 (10-$\infty$)\\
-21.5 &   26 &    30.0 (6.3-508.8) &    67.8 (22.3-178.5) &    69.2 (11.8-508.5) &     0.6 (0.1-1.2) &    17.0 (0.7-114.7) &   0.20 (0.02-1.00) & 100 (10-$\infty$)\\
-20.5 &  140 &    52.5 (6.3-718.7) &    40.1 (8.6-134.9) &    17.9 (4.1-175.0) &     0.5 (0.0-1.1) &     4.1 (0.4-115.0) &    1.00 (0.02-1.00) & 70 (10-3000)\\
-19.5 &  184 &    45.0 (6.3-1015.2) &    15.2 (2.7-81.6) &     6.9 (1.9-64.9) &     0.4 (0.0-1.0) &     5.1 (0.3-115.6) &    0.20 (0.02-1.00) & 70 (10-$\infty$)\\
-18.5 &   28 &   360.2 (15.1-1700.0) &    57.8 (8.5-437.9) &     2.2 (0.0-34.1) &     0.6 (0.0-1.6) &     0.3 (0.0-45.1) &  0.20 (0.02-1.00) & 70 (10-1000)\\
\hline
\multicolumn{9}{c}{$B$ drops ($z\sim4$)}\\
\hline
-22.5 &    3 &   180.5 (6.3-718.7) &   404.9 (129.8-510.8) &    38.0 (0.7-1387.0) &     0.2 (0.1-1.4) &     0.9 (0.0-115.6) &    1.00 (0.02-1.00) & 300 (10-700)\\
-21.5 &   55 &    64.1 (6.3-1015.2) &    81.6 (19.7-254.6) &    30.6 (9.0-281.3) &     0.4 (0.0-1.0) &     3.2 (0.4-114.7) &    0.20 (0.02-1.00) & 70 (10-3000)\\
-20.5 &  260 &    33.0 (6.3-1015.2) &    32.4 (7.5-131.3) &    20.1 (4.2-228.8) &     0.5 (0.0-1.2) &     8.5 (0.4-160.0) &    0.20 (0.02-1.00) & 50 (10-3000)\\
-19.5 &  343 &    30.0 (6.3-1015.2) &    13.1 (2.6-61.1) &     5.7 (1.5-89.9) &     0.3 (0.0-1.1) &     5.8 (0.4-159.6) &    0.20 (0.02-1.00) & 30 (10-3000)\\
-18.5 &   44 &    90.5 (6.3-1434.0) &    13.0 (1.7-61.0) &     2.2 (0.0-40.8) &     0.4 (0.0-1.2) &     0.9 (0.0-114.7) &  1.00 (0.02-1.00) & 30 (10-1000)\\
\hline
\multicolumn{9}{c}{$V$ drops ($z\sim5$)}\\
\hline
-22.5 &    3 &     4.0 (2.5-255.0) &    74.8 (31.8-634.1) &   700.5 (42.2-2661.0) &     0.8 (0.1-0.9) &   205.1 (0.7-350.2) &   1.00 (1.00-1.00) & 10 (10-100)\\
-21.5 &   25 &    52.5 (2.5-718.7) &    66.9 (16.8-295.7) &    38.2 (5.8-1048.0) &     0.5 (0.0-1.1) &     3.2 (0.2-350.2) &    0.02 (0.02-1.00) & 10 (10-1000)\\
-20.5 &   91 &    90.5 (4.0-1015.2) &    31.7 (5.8-132.2) &     7.0 (0.3-188.9) &     0.1 (0.0-1.0) &     1.3 (0.0-205.0) &    0.20 (0.02-1.00) & 30 (10-1000)\\
-19.5 &   67 &    52.5 (6.3-1015.2) &    14.6 (3.7-75.2) &     4.0 (0.5-55.9) &     0.2 (0.0-0.9) &     1.6 (0.2-144.5) &   1.00 (0.02-1.00) & 10 (10-1000)\\
-18.5 &    1 &    11.5 (2.5-26.3) &     3.9 (2.1-17.2) &    16.8 (3.0-615.5) &     0.8 (0.3-1.7) &    59.9 (8.5-350.2) &    0.02 (0.02-1.00) & 10 (10-30)\\
\hline
\multicolumn{9}{c}{$i$ drops ($z\sim6$)}\\
\hline
-22.5 &    7 &     6.3 (2.5-127.8) &    43.4 (19.8-178.7) &   498.4 (19.6-1227.0) &     0.3 (0.0-0.9) &   114.7 (0.1-350.2) &   0.02 (0.02-1.00) & 10 (10-70)\\
-21.5 &   12 &     2.5 (2.5-255.0) &     8.8 (3.6-115.6) &   143.5 (7.2-384.6) &     0.1 (0.0-0.5) &   350.2 (0.6-350.3) &   0.02 (0.02-0.20) & 10 (10-$\infty$)\\
-20.5 &   28 &    22.9 (2.5-508.8) &    14.2 (4.1-60.0) &    20.1 (3.3-177.4) &     0.2 (0.0-0.9) &    22.7 (0.6-350.2) &   0.02 (0.02-1.00) & 10 (10-3000)\\
-19.5 &    8 &    39.0 (4.0-718.7) &    15.4 (3.6-170.8) &     5.3 (0.0-96.1) &     0.2 (0.0-0.9) &     3.2 (0.0-251.8) &   0.20 (0.02-1.00) & 10 (10-500)\\
-18.5 &    0 & -& -& -& -& -& -& -\\
\hline
\multicolumn{9}{c}{DEC+NEB+Ly$\alpha$ model}\\
\multicolumn{9}{c}{$U$ drops ($z\sim3$)}\\
\hline
-22.5 &    4 &    64.1 (52.5-90.5) &   140.3 (120.8-165.9) &     3.1 (0.2-9.8) &     0.2 (0.0-0.3) &     0.2 (0.0-0.6) &  0.02 (0.02-0.02) & 10 (10-10)\\
-21.5 &   27 &    45.0 (20.0-180.5) &    44.0 (15.7-192.2) &    14.5 (1.8-110.5) &     0.4 (0.1-0.9) &     2.3 (0.2-35.0) &  1.00 (0.02-1.00) & 10 (10-100)\\
-20.5 &  126 &    39.0 (11.5-127.8) &    29.8 (9.1-92.4) &     7.4 (0.7-64.6) &     0.5 (0.0-1.1) &     2.3 (0.2-59.8) &     1.00 (0.02-1.00) & 10 (10-70)\\
-19.5 &  192 &    45.0 (11.5-508.8) &    12.9 (3.1-60.4) &     2.8 (0.2-30.3) &     0.4 (0.0-1.0) &     1.3 (0.1-59.8) &   0.20 (0.02-1.00) & 10 (10-500)\\
-18.5 &   37 &    30.0 (2.5-1434.0) &    23.3 (5.8-275.9) &    12.1 (0.2-454.6) &     1.2 (0.2-1.8) &     5.8 (0.0-350.2) &   1.00 (0.02-1.00) & 30 (10-$\infty$)\\
\hline
\multicolumn{9}{c}{$B$ drops ($z\sim4$)}\\
\hline
-22.5 &    2 &    36.0 (30.0-360.2) &   370.9 (249.4-429.9) &   150.2 (40.2-215.5) &     1.2 (0.2-1.2) &     4.3 (1.0-5.8) &  0.02 (0.02-1.00) & 10 (10-500)\\
-21.5 &   58 &    45.0 (20.0-255.0) &    55.6 (17.8-179.0) &    22.1 (2.9-112.3) &     0.5 (0.0-0.9) &     3.1 (0.6-25.1) &  0.20 (0.02-1.00) & 10 (10-700)\\
-20.5 &  249 &    39.0 (10.0-255.0) &    24.5 (6.0-88.1) &     9.4 (1.3-76.6) &     0.5 (0.0-1.1) &     3.2 (0.6-86.3) &  0.20 (0.02-1.00) & 10 (10-1000)\\
-19.5 &  340 &    33.0 (6.3-255.0) &     8.3 (1.9-37.3) &     4.3 (0.6-47.2) &     0.4 (0.0-1.1) &     5.8 (0.6-115.6) &  0.20 (0.02-1.00) & 10 (10-1000)\\
-18.5 &   56 &    22.9 (2.5-360.2) &     4.7 (0.8-21.9) &     3.8 (0.1-84.9) &     0.8 (0.1-1.5) &    16.9 (0.3-350.2) &  1.00 (0.02-1.00) & 10 (10-$\infty$)\\
\hline
\multicolumn{9}{c}{$V$ drops ($z\sim5$)}\\
\hline
-22.5 &    3 &    64.1 (1.6-127.8) &   122.2 (46.4-504.0) &     8.3 (2.2-2998.0) &     0.2 (0.0-0.7) &     0.6 (0.1-630.8) &  0.20 (0.02-1.00) & 10 (10-3000)\\
-21.5 &   20 &    45.0 (17.4-180.5) &    50.8 (16.8-125.4) &    10.5 (1.0-79.6) &     0.3 (0.0-1.0) &     1.4 (0.2-29.7) &  0.02 (0.02-1.00) & 10 (10-70)\\
-20.5 &   88 &    30.0 (4.0-127.8) &    15.6 (2.7-69.6) &    11.0 (1.2-121.9) &     0.4 (0.0-1.3) &     8.5 (0.6-251.9) &  0.02 (0.02-1.00) & 10 (10-500)\\
-19.5 &   73 &    26.3 (4.0-127.8) &     7.7 (1.2-36.2) &     8.2 (1.0-60.2) &     0.5 (0.0-1.2) &    12.2 (0.6-251.3) &  0.20 (0.02-1.00) & 10 (10-1000)\\
-18.5 &    3 &    30.0 (4.0-90.5) &     5.5 (2.3-12.3) &     6.4 (0.0-49.5) &     0.8 (0.0-1.3) &     8.5 (0.0-205.1) &  1.00 (0.02-1.00) & 10 (10-500)\\
\hline
\multicolumn{9}{c}{$i$ drops ($z\sim6$)}\\
\hline
-22.5 &    3 &    33.0 (4.0-180.5) &    35.8 (7.7-156.7) &    42.3 (0.2-160.3) &     0.1 (0.0-0.5) &     4.3 (0.0-204.4) & 0.20 (0.02-1.00) & 10 (10-100)\\
-21.5 &    8 &     2.5 (1.6-127.8) &     7.7 (2.7-90.0) &   120.2 (1.8-341.6) &     0.0 (0.0-0.7) &   350.1 (0.2-582.4) &  0.02 (0.02-0.20) & 10 (10-500)\\
-20.5 &   24 &     6.3 (1.6-45.0) &     3.9 (1.0-18.3) &    21.3 (3.3-124.0) &     0.1 (0.0-0.6) &   143.4 (2.3-582.3) &  0.02 (0.02-1.00) & 10 (10-1000)\\
-19.5 &   22 &     4.0 (1.6-52.5) &     2.4 (0.5-42.2) &    22.4 (3.2-150.2) &     0.3 (0.0-1.0) &   234.9 (1.3-582.3) &  0.02 (0.02-0.20) & 10 (10-700)\\
-18.5 &    0 & -& -& -& -& -& -& -\\
\hline
\multicolumn{9}{c}{DEC+NEB model}\\
\multicolumn{9}{c}{$U$ drops ($z\sim3$)}\\
\hline
-22.5 &    4 &     4.0 (4.0-1700.0) &    20.0 (8.4-476.8) &   234.4 (41.6-517.8) &     0.1 (0.0-0.5) &   204.6 (0.8-251.2) &  0.20 (0.02-1.00) & 700 (10-$\infty$)\\
-21.5 &   30 &    30.0 (4.0-360.2) &    32.3 (11.0-139.0) &    46.3 (8.7-345.0) &     0.5 (0.1-0.9) &    12.6 (0.7-241.3) &  0.20 (0.02-1.00) & 30 (10-$\infty$)\\
-20.5 &  145 &    26.3 (2.5-718.7) &    17.1 (3.7-94.5) &    26.6 (5.0-154.0) &     0.5 (0.0-0.9) &    16.1 (0.7-350.0) &  1.00 (0.02-1.00) & 30 (10-$\infty$)\\
-19.5 &  174 &    30.0 (2.5-1434.0) &     9.4 (1.1-64.2) &     6.8 (1.6-65.0) &     0.3 (0.0-0.8) &     8.5 (0.5-350.1) &  1.00 (0.02-1.00) & 30 (10-$\infty$)\\
-18.5 &   31 &   127.8 (1.6-1700.0) &    34.6 (5.8-395.7) &     7.4 (0.6-424.9) &     0.9 (0.1-1.8) &     1.4 (0.0-582.1) & 1.00 (0.02-1.00) & 300 (10-$\infty$)\\
\hline
\multicolumn{9}{c}{$B$ drops ($z\sim4$)}\\
\hline
-22.5 &    3 &   180.5 (2.5-1700.0) &   224.6 (74.2-538.8) &    80.4 (35.8-2606.0) &     0.6 (0.1-1.0) &     3.4 (0.8-350.2) &  0.20 (0.02-1.00) & 70 (10-$\infty$)\\
-21.5 &   59 &    36.0 (6.3-508.8) &    32.4 (10.1-170.3) &    39.3 (6.3-218.6) &     0.4 (0.0-0.8) &     5.8 (0.7-159.6) &  0.20 (0.02-1.00) & 30 (10-$\infty$)\\
-20.5 &  260 &    30.0 (4.0-718.7) &    15.7 (3.1-85.0) &    18.3 (3.7-166.9) &     0.5 (0.0-1.0) &    10.7 (0.7-251.9) &  0.20 (0.02-1.00) & 10 (10-$\infty$)\\
-19.5 &  339 &    26.3 (2.5-718.7) &     5.9 (1.0-38.0) &     5.6 (1.0-77.0) &     0.4 (0.0-1.0) &    12.2 (0.6-350.2) &  0.20 (0.02-1.00) & 10 (10-$\infty$)\\
-18.5 &   44 &    30.0 (2.5-1434.0) &     4.1 (0.4-39.5) &     2.1 (0.0-64.2) &     0.5 (0.0-1.4) &     8.5 (0.2-350.3) & 0.20 (0.02-1.00) & 10 (10-$\infty$)\\
\hline
\multicolumn{9}{c}{$V$ drops ($z\sim5$)}\\
\hline
-22.5 &    4 &    33.0 (4.0-127.8) &    97.4 (24.8-390.6) &   149.0 (24.4-702.4) &     0.5 (0.1-0.6) &    12.1 (1.3-205.1) &  1.00 (0.02-1.00) & 10 (10-100)\\
-21.5 &   23 &    33.0 (2.5-180.5) &    26.2 (4.6-99.2) &    21.2 (3.3-218.3) &     0.3 (0.0-0.9) &     8.5 (0.6-251.9) &  0.20 (0.02-1.00) & 10 (10-3000)\\
-20.5 &   92 &    22.9 (2.5-90.5) &    10.3 (1.8-61.8) &    12.8 (2.2-160.6) &     0.3 (0.0-1.2) &    19.5 (0.6-398.2) &  0.20 (0.02-1.00) & 10 (10-3000)\\
-19.5 &   69 &    22.9 (1.6-90.5) &     4.7 (0.9-29.9) &     7.0 (1.1-75.0) &     0.4 (0.0-1.1) &    16.9 (0.6-582.3) & 1.00 (0.02-1.00) & 10 (10-$\infty$)\\
-18.5 &    0 & -& -& -& -& -& -& -\\
\hline
\multicolumn{9}{c}{$i$ drops ($z\sim6$)}\\
\hline
-22.5 &    6 &    22.9 (4.0-30.0) &    17.7 (7.1-32.7) &    48.5 (10.4-221.6) &     0.0 (0.0-0.1) &    16.9 (5.8-251.4) &  0.02 (0.02-1.00) & 10 (10-30)\\
-21.5 &    9 &    30.0 (10.0-180.5) &    13.3 (4.7-107.4) &     8.6 (2.7-23.8) &     0.0 (0.0-0.6) &     5.8 (0.4-77.0) &  0.20 (0.02-1.00) & 10 (10-300)\\
-20.5 &   31 &    30.0 (6.3-52.5) &     6.4 (2.3-22.6) &     3.9 (2.0-22.3) &     0.0 (0.0-0.4) &     5.8 (1.3-159.6) & 0.20 (0.02-1.00) & 10 (10-70)\\
-19.5 &   11 &    30.0 (1.6-180.5) &     9.6 (2.2-121.3) &     6.1 (0.8-169.2) &     0.3 (0.0-1.2) &     5.8 (0.4-582.2) &  0.20 (0.02-1.00) & 10 (10-500)\\
-18.5 &    0 & -& -& -& -& -& -& -\\
\hline
\hline
\end{tabular}}
\label{tabmagall_dec}
\end{table*}

%%%%%%%%%%%%%%%%%%

\begin{table*}[htbf]
\centering
\caption{Same as Table \ref{tabmagall_ref} for rising star formation history with variable metallicity Z.}
\scalebox{0.85}{\begin{tabular}{cccccccc}
\hline
$M_{UV}$ & Num & Age & $M_{\star}$ & SFR & $A_{\mathrm{V}}$ & sSFR & Z  \\
 (mag) & & (Myr) & ($10^8 \mathrm{M}_{\odot}$) & ($\mathrm{M}_{\odot}.\mathrm{yr}^{-1}$) & (mag) & (Gyr$^{-1}$) & ($\mathrm{Z}_{\odot}$) \\
\hline
\hline
\multicolumn{8}{c}{RIS model}\\
\multicolumn{8}{c}{$U$ drops ($z\sim3$)}\\
\hline
-22.5 &    5 &     2.5 (1.0-700.0) &   272.8 (218.9-319.9) & 10600.0 (121.3-31920.0) &     1.3 (0.6-1.4) &   456.5 (4.4-1064.0) & 0.02 (0.02-0.20) \\
-21.5 &   28 &    37.0 (6.3-730.0) &    59.7 (24.5-160.5) &   200.1 (38.3-868.7) &     0.9 (0.4-1.3) &    50.1 (4.3-207.8) &  0.20 (0.02-1.00) \\
-20.5 &  148 &    47.5 (6.3-770.0) &    27.2 (8.3-92.4) &    71.0 (9.8-406.9) &     0.8 (0.4-1.3) &    41.3 (4.0-207.5) &   0.20 (0.02-1.00) \\
-19.5 &  171 &    64.1 (6.3-801.0) &    11.0 (2.6-54.3) &    24.8 (0.0-149.5) &     0.7 (0.2-1.2) &    32.9 (0.0-207.4) &   0.20 (0.02-1.00) \\
-18.5 &   33 &   730.0 (6.3-1015.2) &    45.1 (9.8-292.1) &    16.6 (0.0-316.3) &     1.2 (0.4-2.2) &     4.2 (0.0-207.4) &  0.20 (0.02-1.00) \\
\hline
\multicolumn{8}{c}{$B$ drops ($z\sim4$)}\\
\hline
-22.5 &    2 &     6.3 (6.3-770.0) &   152.6 (142.1-475.2) &  2935.0 (143.1-3174.0) &     1.4 (0.8-1.5) &   207.9 (4.0-208.6) &  0.20 (0.02-1.00) \\
-21.5 &   59 &    90.5 (6.3-770.0) &    54.9 (20.9-216.5) &   143.4 (16.9-623.7) &     0.8 (0.3-1.2) &    25.2 (4.0-207.5) &    0.20 (0.02-1.00) \\
-20.5 &  274 &    31.0 (6.3-801.0) &    23.8 (7.3-97.4) &    79.0 (0.0-484.9) &     0.8 (0.3-1.3) &    56.9 (0.0-207.8) &   0.20 (0.02-1.00) \\
-19.5 &  324 &    34.0 (6.3-801.0) &    10.8 (2.7-42.1) &    29.7 (0.0-219.2) &     0.7 (0.1-1.3) &    53.1 (0.0-207.8) &  0.20 (0.02-1.00) \\
-18.5 &   46 &   420.0 (6.3-801.0) &    12.3 (3.0-47.1) &    12.6 (0.0-157.4) &     0.9 (0.1-1.5) &     7.0 (0.0-207.4) &  0.20 (0.02-1.00) \\
\hline
\multicolumn{8}{c}{$V$ drops ($z\sim5$)}\\
\hline
-22.5 &    3 &     4.0 (2.5-10.0) &    75.9 (35.7-232.0) &  3275.0 (1078.0-3566.0) &     0.9 (0.8-1.4) &   305.6 (144.1-456.5) & 1.00 (0.20-1.00) \\
-21.5 &   27 &     6.3 (2.5-801.0) &    50.2 (15.7-161.6) &   270.8 (0.0-2021.0) &     0.7 (0.2-1.3) &   207.3 (0.0-456.5) & 0.20 (0.02-1.00) \\
-20.5 &   93 &    64.1 (4.0-801.0) &    24.6 (6.1-97.0) &    38.9 (0.0-512.9) &     0.6 (0.1-1.3) &    32.9 (0.0-305.6) & 0.20 (0.02-1.00) \\
-19.5 &   61 &    55.0 (4.0-801.0) &    13.0 (3.5-54.4) &    27.5 (0.0-320.8) &     0.7 (0.2-1.3) &    36.9 (0.0-305.5) & 0.20 (0.02-1.00) \\
-18.5 &    1 &     6.3 (2.5-40.0) &     3.5 (2.3-17.8) &    35.2 (12.5-858.0) &     0.8 (0.5-1.7) &   207.4 (47.1-456.6) & 0.02 (0.02-1.00) \\
\hline
\multicolumn{8}{c}{$i$ drops ($z\sim6$)}\\
\hline
-22.5 &    7 &     4.0 (2.5-19.1) &    39.3 (12.6-65.0) &   890.8 (200.9-1659.0) &     0.4 (0.0-1.0) &   305.5 (83.6-456.5) & 0.02 (0.02-1.00) \\
-21.5 &   14 &     2.5 (2.5-730.0) &    10.1 (3.8-63.9) &   205.1 (25.7-697.5) &     0.3 (0.0-0.8) &   456.4 (4.2-456.6) &  0.02 (0.02-1.00) \\
-20.5 &   27 &     6.3 (2.5-801.0) &    14.0 (4.1-51.6) &    87.7 (0.0-386.7) &     0.5 (0.1-1.1) &   207.3 (0.0-456.4) & 0.02 (0.02-0.20) \\
-19.5 &    7 &     6.3 (2.5-1015.2) &    35.9 (4.1-265.2) &    84.6 (0.0-3619.0) &     0.8 (0.0-2.2) &   207.3 (0.0-456.5) & 0.20 (0.02-1.00) \\
-18.5 &    0 & -& -& -& -& -& -\\
\hline
\multicolumn{8}{c}{RIS+NEB+Ly$\alpha$ model}\\
\multicolumn{8}{c}{$U$ drops ($z\sim3$)}\\
\hline
-22.5 &    4 &   360.0 (160.9-801.0) &    61.8 (39.8-132.3) &    45.3 (0.0-79.3) &     0.0 (0.0-0.2) &     8.1 (0.0-16.0) & 1.00 (0.02-1.00) \\
-21.5 &   28 &   250.0 (20.9-801.0) &    50.7 (13.6-166.5) &    49.4 (0.0-162.6) &     0.5 (0.1-0.9) &    11.1 (0.0-78.3) &  1.00 (0.20-1.00) \\
-20.5 &  125 &   180.5 (12.6-801.0) &    26.0 (7.6-79.9) &    34.4 (0.0-126.5) &     0.6 (0.2-1.0) &    14.5 (0.0-118.7) & 1.00 (0.20-1.00) \\
-19.5 &  182 &   180.5 (10.0-801.0) &     9.5 (2.0-40.1) &    10.4 (0.0-72.5) &     0.6 (0.1-1.1) &    14.5 (0.0-144.1) &   1.00 (0.20-1.00) \\
-18.5 &   45 &    47.5 (2.5-801.0) &    17.8 (3.9-135.8) &    46.4 (0.0-607.0) &     1.3 (0.6-2.0) &    41.3 (0.0-456.5) &  0.20 (0.02-1.00) \\
\hline
\multicolumn{8}{c}{$B$ drops ($z\sim4$)}\\
\hline
-22.5 &    2 &   227.3 (6.3-660.0) &   250.9 (65.8-287.2) &   283.2 (117.7-1281.0) &     0.8 (0.5-1.0) &    12.0 (4.6-207.8) &  0.20 (0.02-1.00) \\
-21.5 &   66 &   321.0 (27.5-801.0) &    54.1 (11.9-155.4) &    47.9 (0.0-132.1) &     0.5 (0.1-0.8) &     8.9 (0.0-62.5) & 1.00 (0.02-1.00) \\
-20.5 &  262 &   180.5 (11.0-801.0) &    21.1 (4.1-71.4) &    25.5 (0.0-118.6) &     0.6 (0.2-1.1) &    14.5 (0.0-133.3) &   1.00 (0.02-1.00) \\
-19.5 &  304 &    80.0 (6.3-801.0) &     7.4 (1.4-27.0) &    14.8 (0.0-91.0) &     0.6 (0.1-1.2) &    27.5 (0.0-208.6) &  1.00 (0.02-1.00) \\
-18.5 &   54 &    16.6 (2.5-801.0) &     3.5 (0.3-16.6) &    17.7 (0.0-114.8) &     1.0 (0.4-1.7) &    94.4 (0.0-456.4) &  1.00 (0.02-1.00) \\
\hline
\multicolumn{8}{c}{$V$ drops ($z\sim5$)}\\
\hline
-22.5 &    5 &    47.5 (1.6-630.0) &    44.9 (23.5-535.6) &    97.6 (71.6-3396.0) &     0.6 (0.0-0.8) &    41.0 (4.8-692.8) &  0.20 (0.02-1.00) \\
-21.5 &   15 &   227.3 (20.9-801.0) &    41.8 (5.2-110.0) &    25.6 (0.0-139.0) &     0.2 (0.0-0.9) &    12.0 (0.0-78.3) &  1.00 (0.20-1.00) \\
-20.5 &   90 &    55.0 (2.5-801.0) &    11.5 (1.5-50.0) &    23.9 (0.0-166.8) &     0.5 (0.0-1.3) &    36.7 (0.0-456.5) &  0.20 (0.02-1.00) \\
-19.5 &   68 &    37.0 (2.5-801.0) &     5.3 (0.8-25.7) &    15.0 (0.0-88.1) &     0.6 (0.0-1.4) &    49.7 (0.0-456.5) &  0.20 (0.02-1.00) \\
-18.5 &    2 &   127.8 (1.0-1015.2) &     5.0 (0.1-11.0) &     3.8 (0.0-73.8) &     0.9 (0.0-3.1) &    19.2 (0.0-1063.4) & 1.00 (0.20-1.00) \\
\hline
\multicolumn{8}{c}{$i$ drops ($z\sim6$)}\\
\hline
-22.5 &    2 &    16.6 (4.0-90.5) &    14.8 (7.8-51.5) &   172.7 (127.1-269.8) &     0.2 (0.0-0.3) &    93.9 (25.2-305.0) &  1.00 (0.20-1.00) \\
-21.5 &    9 &     2.5 (1.6-660.0) &     6.9 (2.7-116.4) &   160.4 (51.6-486.7) &     0.2 (0.0-0.7) &   456.4 (4.6-692.8) &  0.20 (0.02-1.00) \\
-20.5 &   20 &     4.0 (1.6-590.0) &     3.5 (0.9-24.4) &    41.8 (8.7-169.0) &     0.2 (0.0-0.7) &   304.9 (5.1-692.7) &  0.02 (0.02-1.00) \\
-19.5 &   25 &     4.0 (1.6-390.0) &     2.1 (0.5-23.5) &    30.3 (4.9-172.2) &     0.4 (0.0-1.0) &   305.0 (7.5-692.7) &  0.02 (0.02-1.00) \\
-18.5 &    0 & -& -& -& -& -\\
\hline
\multicolumn{8}{c}{RIS+NEB model}\\
\multicolumn{8}{c}{$U$ drops ($z\sim3$)}\\
\hline
-22.5 &    4 &     6.3 (4.0-801.0) &    20.0 (8.5-240.7) &   324.1 (0.0-633.1) &     0.3 (0.0-0.5) &   208.6 (0.0-305.6) & 0.20 (0.02-1.00) \\
-21.5 &   31 &    90.5 (4.0-801.0) &    38.6 (10.1-107.9) &    95.3 (0.0-467.4) &     0.6 (0.2-0.9) &    25.2 (0.0-305.0) &  1.00 (0.02-1.00) \\
-20.5 &  149 &    80.0 (2.5-801.0) &    16.8 (3.3-61.5) &    44.5 (0.0-209.7) &     0.6 (0.2-0.9) &    27.5 (0.0-456.4) & 1.00 (0.02-1.00) \\
-19.5 &  163 &    90.5 (2.5-801.0) &     9.5 (1.1-46.6) &    16.4 (0.0-109.9) &     0.5 (0.1-1.1) &    25.0 (0.0-456.5) &  0.20 (0.02-1.00) \\
-18.5 &   36 &   250.0 (1.6-801.0) &    23.3 (5.8-233.0) &    26.9 (0.0-534.3) &     1.4 (0.6-2.0) &    11.1 (0.0-692.6) &  0.20 (0.02-1.00) \\
\hline
\multicolumn{8}{c}{$B$ drops ($z\sim4$)}\\
\hline
-22.5 &    2 &    12.4 (2.5-660.0) &    83.9 (74.9-235.0) &   777.8 (106.2-3473.0) &     0.8 (0.5-1.0) &   191.6 (4.6-456.5) &  0.02 (0.02-0.20) \\
-21.5 &   66 &   110.0 (4.0-770.0) &    32.1 (7.1-120.6) &    77.6 (13.1-395.9) &     0.5 (0.1-0.9) &    21.5 (4.0-305.0) & 0.20 (0.02-1.00) \\
-20.5 &  277 &    64.1 (2.5-801.0) &    13.3 (2.5-56.7) &    36.4 (0.0-227.0) &     0.6 (0.1-1.0) &    32.9 (0.0-456.5) &  0.20 (0.02-1.00) \\
-19.5 &  308 &    27.5 (1.6-801.0) &     4.8 (0.8-22.9) &    17.0 (0.0-133.7) &     0.5 (0.0-1.2) &    62.5 (0.0-692.5) &  0.20 (0.02-1.00) \\
-18.5 &   47 &    14.5 (2.5-801.0) &     4.1 (0.3-25.4) &     9.6 (0.0-139.2) &     1.0 (0.3-1.8) &   105.2 (0.0-456.6) & 0.20 (0.02-1.00) \\
\hline
\multicolumn{8}{c}{$V$ drops ($z\sim5$)}\\
\hline
-22.5 &    5 &    14.5 (2.5-420.0) &    24.6 (3.9-445.1) &   245.4 (105.0-450.3) &     0.0 (0.0-0.7) &   105.8 (7.0-456.6) & 1.00 (0.20-1.00) \\
-21.5 &   25 &    64.1 (4.0-730.0) &    14.6 (3.4-82.8) &    37.0 (11.9-242.8) &     0.2 (0.0-0.9) &    32.9 (4.2-305.7) & 1.00 (0.02-1.00) \\
-20.5 &   92 &    20.9 (1.6-630.0) &     7.1 (1.3-43.9) &    34.8 (8.3-231.3) &     0.5 (0.0-1.3) &    78.3 (4.9-692.7) &  0.20 (0.02-1.00) \\
-19.5 &   58 &    24.0 (1.6-700.0) &     3.3 (0.6-18.7) &    16.1 (3.2-100.1) &     0.5 (0.0-1.3) &    70.3 (4.4-692.7) & 1.00 (0.02-1.00) \\
-18.5 &    2 &     6.3 (1.0-801.0) &     9.7 (0.2-76.9) &    14.5 (0.0-279.3) &     1.6 (0.6-3.6) &   207.4 (0.0-1063.7) & 1.00 (0.02-1.00) \\
\hline
\multicolumn{8}{c}{$i$ drops ($z\sim6$)}\\
\hline
-22.5 &    4 &     6.3 (2.5-80.0) &     9.7 (6.5-29.3) &   184.2 (88.2-330.4) &     0.0 (0.0-0.1) &   207.8 (27.7-456.5) &  0.20 (0.02-1.00) \\
-21.5 &    8 &    90.5 (11.0-770.0) &     7.1 (5.2-53.4) &    33.5 (6.4-79.0) &     0.0 (0.0-0.4) &    25.1 (4.0-132.2) & 0.20 (0.02-1.00) \\
-20.5 &   33 &    47.5 (4.0-560.0) &     3.7 (1.7-22.3) &    13.6 (6.5-45.6) &     0.0 (0.0-0.6) &    41.0 (5.4-305.1) &  1.00 (0.02-1.00) \\
-19.5 &   10 &    27.5 (1.6-520.0) &     5.6 (1.1-53.1) &    18.9 (4.8-365.7) &     0.5 (0.0-1.4) &    62.5 (5.8-692.7) &  0.20 (0.02-1.00) \\
-18.5 &    0 & -& -& -& -& -\\
\hline
\hline
\end{tabular}}
\label{tabmagall_ris}
\end{table*}

\end{document}